\documentclass[aps,prl,twocolumn,superscriptaddress,nobibnotes]{revtex4-2}
\usepackage{graphicx}
\usepackage{amsmath}
\usepackage{hyperref} 
\setcitestyle{super}
\usepackage{xcolor}
\newcommand{\revision}{\textcolor{black} } 
\begin{document}

\title{Atomic Alignment in PbS Nanocrystal Superlattices with Compact Inorganic Ligands via Reversible Oriented Attachment of Nanocrystals}

\author{Ahhyun Jeong}
\affiliation{Department of Chemistry, James Franck Institute, and Pritzker School of Molecular Engineering, University of Chicago, Chicago, Illinois 60637, United States}

\author{Aditya N. Singh}
\affiliation{Department of Chemistry, University of California, Berkeley, California 94720, USA}

\author{Josh Portner}
\affiliation{Department of Chemistry, James Franck Institute, and Pritzker School of Molecular Engineering, University of Chicago, Chicago, Illinois 60637, United States}

\author{Xiaoben Zhang}
\affiliation{Department of Chemistry, James Franck Institute, and Pritzker School of Molecular Engineering, University of Chicago, Chicago, Illinois 60637, United States}

\author{Saghar Rezaie}
\affiliation{Renewable and Sustainable Energy Institute, University of Colorado, Boulder, Colorado, USA}

\author{Justin C. Ondry}
\affiliation{Department of Chemistry, James Franck Institute, and Pritzker School of Molecular Engineering, University of Chicago, Chicago, Illinois 60637, United States}

\author{Zirui Zhou}
\affiliation{Department of Chemistry, James Franck Institute, and Pritzker School of Molecular Engineering, University of Chicago, Chicago, Illinois 60637, United States}

\author{Junhong Chen}
\affiliation{Department of Chemistry, James Franck Institute, and Pritzker School of Molecular Engineering, University of Chicago, Chicago, Illinois 60637, United States}

\author{Ye Ji Kim}
\affiliation{Department of Chemistry, James Franck Institute, and Pritzker School of Molecular Engineering, University of Chicago, Chicago, Illinois 60637, United States}

\author{Richard D. Schaller}
\affiliation{Center for Nanoscale Materials, Argonne National Laboratory, Argonne, Illinois 60439, United States}
\affiliation{Department of Chemistry, Northwestern University, Evanston, Illinois, USA}

\author{Youssef Tazoui}
\affiliation{Department of Chemistry, James Franck Institute, and Pritzker School of Molecular Engineering, University of Chicago, Chicago, Illinois 60637, United States}

\author{Zehan Mi}
\affiliation{Department of Chemistry, James Franck Institute, and Pritzker School of Molecular Engineering, University of Chicago, Chicago, Illinois 60637, United States}

\author{Sadegh Yazdi}
\affiliation{Renewable and Sustainable Energy Institute, University of Colorado, Boulder, Colorado, USA}

\author{David T. Limmer}
\affiliation{Department of Chemistry, University of California, Berkeley, California 94720, USA}
\affiliation{Chemical Sciences and Materials Sciences Divisions, Lawrence Berkeley National Laboratory, Berkeley, California 94720, USA}
\affiliation{Kavli Energy NanoSciences Institute, University of California, Berkeley, California 94720, USA}

\author{Dmitri V. Talapin}
\email[Corresponding author: ]{dvtalapin@uchicago.edu}
\affiliation{Department of Chemistry, James Franck Institute, and Pritzker School of Molecular Engineering, University of Chicago, Chicago, Illinois 60637, United States}
\affiliation{Center for Nanoscale Materials, Argonne National Laboratory, Argonne, Illinois 60439, United States}

\date{\today}

\begin{abstract}

Nanocrystals (NCs) serve as versatile building blocks for the creation of functional materials, with NC self-assembly offering opportunities to enable novel material properties. Here, we demonstrate that PbS NCs functionalized with strongly negatively charged metal chalcogenide complex (MCC) ligands, such as Sn\textsubscript{2}S\textsubscript{6}\textsuperscript{4-} and AsS\textsubscript{4}\textsuperscript{3-}, can self-assemble into all-inorganic superlattices with \emph{both} long-range superlattice translational and atomic-lattice orientational order. Structural characterizations reveal that the NCs adopt unexpected edge-to-edge alignment, and numerical simulation clarifies that orientational order is thermodynamically stabilized by many-body ion correlations originating from the dense electrolyte. Furthermore, we show that the superlattices of Sn\textsubscript{2}S\textsubscript{6}\textsuperscript{4-}-functionalized PbS NCs can be fully disassembled back into the colloidal state, which is highly unusual for orientationally attached superlattices with atomic-lattice alignment. The reversible oriented attachment of NCs, enabling their dynamic assembly and disassembly into effectively single-crystalline superstructures, offers a pathway toward designing reconfigurable materials with adaptive and controllable electronic and optoelectronic properties.

\end{abstract}

\maketitle



\section{Introduction}

Self-assembly of nanocrystals (NCs) enables the bottom-up design of a
wide range of structures from a library of NCs as building
blocks.\cite{1,2} Extensive work has established formation of
a library of different structures built from NCs, including close-packed
structures,\cite{3} multicomponent crystalline and
quasi-crystalline structures,\cite{4,5,6,7} and chiral
structures.\cite{8,9} When NCs organize themselves into
densely-packed superlattices with long-range order, they can exhibit
collective phenomena. For instance, electronic coupling within
superlattices can lead to miniband formation, enhanced charge carrier
mobility, and collective optical effects such as super-fluorescence and
superradiance.\cite{10} These emergent properties expand the
functional landscape of NC materials, opening avenues for advanced
optoelectronic applications, quantum technologies, and nanoscale
sensing.\cite{1,2}

Traditional approaches for NC self-assembly rely on surface
functionalization to control pair potentials between NCs, superlattice
phase behavior or to enable stimuli-responsive assembly and disassembly.
For example, length and coverage of organic ligands can be tuned to
access different superlattice phases,\cite{11,12,13,14,15} and
DNA-based functionalization enables self-assembly of NCs into
programmable architectures.\cite{16,17,18,19} Chemically- or
photo-responsive ligands further enable external control of assembly and
disassembly processes.\cite{8,20,21,22} While these ligands
maintain colloidal stability and direct assembly, they also act as
insulating barriers that suppress interparticle coupling in the
resulting superlattices.\cite{23} Switching to shorter
ligands or removing them entirely can facilitate direct NC contact and
strong coupling but often drives irreversible aggregation or oriented
attachment preventing reversibility of the assembly process and
restricting long-range ordering of NCs.\cite{24,25,26} These
challenges reveal the pressing need for alternative strategies that
enable both strong electronic coupling and structural adaptability in NC
superlattices.

\begin{figure*}[t] 
    \centering
    \includegraphics[width=.95\linewidth]{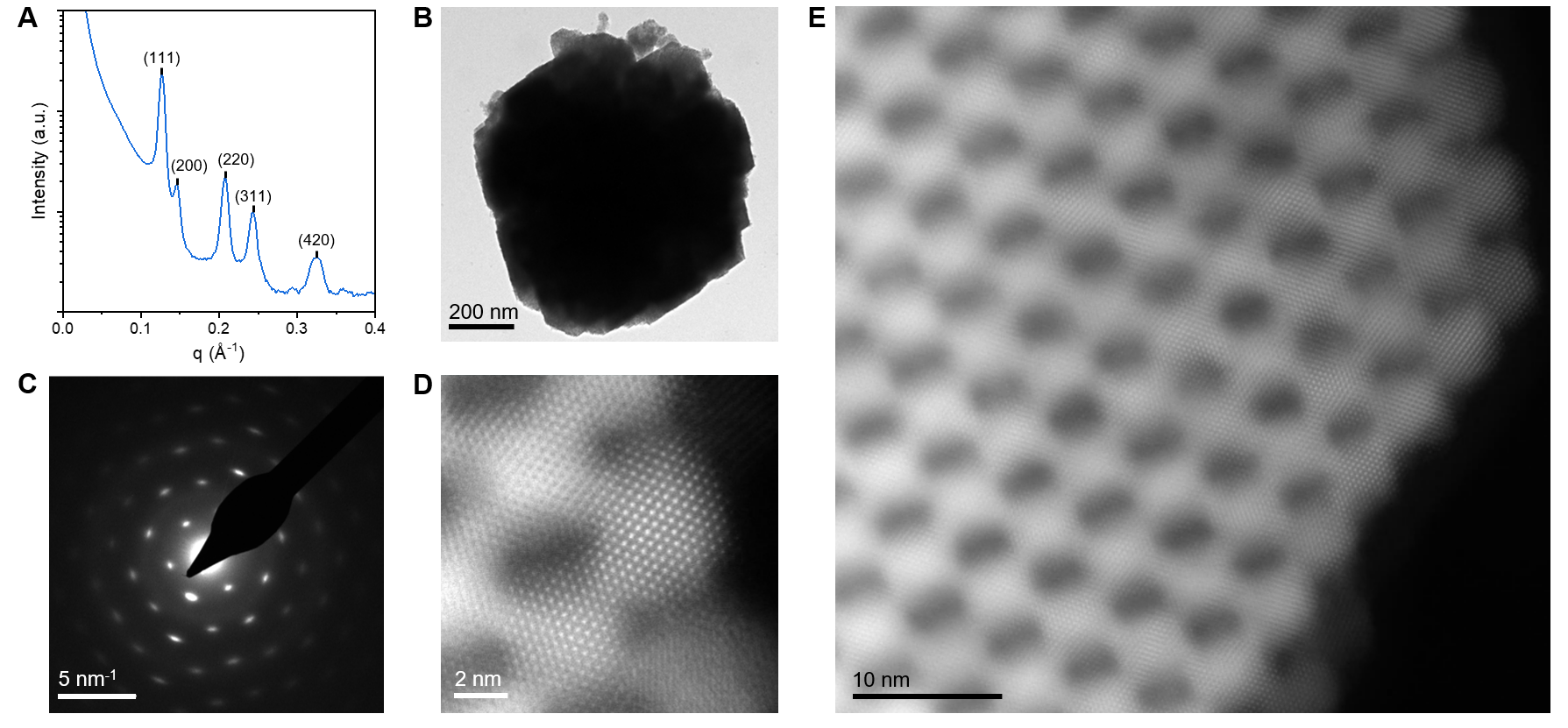}
    
    \caption{(A) Small-angle X-ray scattering (SAXS) pattern of 5.7 nm PbS-Sn\textsubscript{2}S\textsubscript{6}\textsuperscript{4-} nanocrystal (NC) superlattice. The peak assignment shows that the superlattice adopts a face-centered cubic (\emph{fcc}) structure. (B) Bright-field transmission electron microscopy (TEM) image of a 5.7 nm PbS-Sn\textsubscript{2}S\textsubscript{6}\textsuperscript{4-} NC superlattice domain. (C) Selected area electron diffraction (SAED) image of the entire grain. Distinct spot patterns indicate that the PbS NCs are atomically aligned. (D, E) Scanning transmission electron microscopy (STEM) high-angle annular dark-field (HAADF) image of a 5.7 nm PbS-Sn\textsubscript{2}S\textsubscript{6}\textsuperscript{4-} NC superlattice.}
    \label{fig:1} 
\end{figure*}

One promising direction is the use of short, compact inorganic ligands
to functionalize NCs, replacing their organic counterparts. These
inorganic ligands substantially reduce interparticle spacing, allowing
for enhanced electronic coupling and potentially new regimes of
collective behavior.\cite{27,28,29,30} However, unlike organic
ligands, inorganic species stabilize NCs predominantly through
electrostatic interactions, introducing a fundamentally different
mechanism for both colloidal stability and
self-assembly.\cite{31,32,33} For this reason, the
self-assembly of electrostatically stabilized NCs remains relatively
underexplored. A deeper understanding of this approach is critical, as
it holds promise for creating functional superlattices with robust
collective electronic, optical, and structural properties.

In this study, we demonstrate the self-assembly of PbS NCs with metal
chalcogenide complex (MCC) ligands (e.g.
Sn\textsubscript{2}S\textsubscript{6}\textsuperscript{4-},
AsS\textsubscript{4}\textsuperscript{3-}) into all-inorganic
superlattices exhibiting both long-range translational \emph{and}
orientational order, with individual NCs bridged with epitaxial
``necks''. We show that the self-assembly of these NCs exhibits
controllable reversibility, a counterintuitive feature for
orientationally attached NC superlattices with atomic-lattice alignment.
Structural characterizations reveal that the NCs adopt an edge-to-edge
alignment, a unique configuration facilitated by strong ionic
correlations in the dense electrolyte. The formation of densely packed
and oriented superlattices with controlled reversibility offers
possibilities to utilize the benefits of reversible self-assembly in
dynamic and adaptive material systems.

\begin{figure*}[t]
    \centering
    \includegraphics[width=\linewidth]{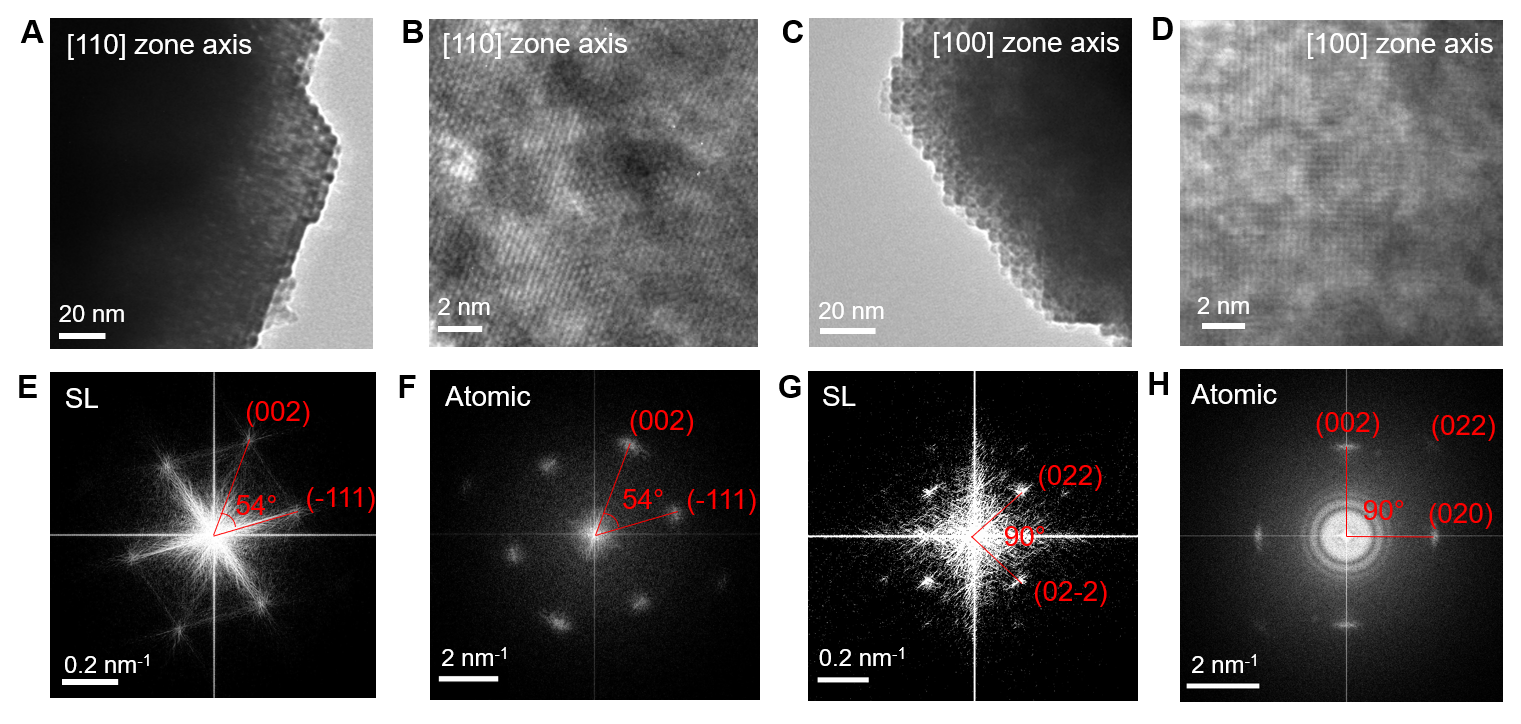}
    
    \caption{(A) Transmission electron microscopy (TEM) and
    (B) High-resolution TEM (HRTEM) image of the superlattice of
    5.7 nm PbS-Sn\textsubscript{2}S\textsubscript{6}\textsuperscript{4-}
    NCs, viewed along the (110) zone axis. (C) TEM and (D)
    HRTEM images of the superlattice of the same NC superlattice viewed
    along the (100) zone axis. (E, F) Fourier-transformed TEM
    (FT-TEM) images of the superlattice viewed along the (110) zone axis,
    with the distinct spots indicating highly ordered \emph{fcc} structures
    at both nanometer and atomic scales. (G, H) FT-TEM of the
    superlattice viewed along the (100) zone axis, also with distinct spots
    indicating a high degree of \emph{fcc} ordering.}
    \label{fig:2}
\end{figure*}

\section{Results and Discussion}

\subsection{Self-assembly of nanocrystals with charged surfaces into
superlattices featuring translational and atomic-lattice orientational
order} Monodisperse PbS NCs with various sizes (5.4±0.3, 5.7±0.3,
7.3±0.7, 8.3±0.5 nm) were synthesized from lead oleate and substituted
thioureas using the methods developed by the Owen group (Figure
S1).\cite{34} The surface oleate (OA) ligands of the PbS NCs
are then substituted with thiostannate ligands
(Sn\textsubscript{2}S\textsubscript{6}\textsuperscript{4-}) by stirring
them with a solution of
K\textsubscript{4}Sn\textsubscript{2}S\textsubscript{6} in
N-methylformamide (NMF). After stirring overnight, the PbS NCs transfer
to the NMF phase, indicating the substitution of OA ligands by
Sn\textsubscript{2}S\textsubscript{6}\textsuperscript{4-} ligands.
Small-angle X-ray scattering (SAXS) patterns of
Sn\textsubscript{2}S\textsubscript{6}\textsuperscript{4-}-capped PbS NCs
(PbS-Sn\textsubscript{2}S\textsubscript{6}\textsuperscript{4-} NCs), in
addition to oscillations originating from the spherical form factor of
individual NCs, exhibit a broad peak in the low-\emph{q} region
(\emph{q} \textless{} 0.1 $\mathring{\mathrm{A}}^{-1}$), indicative of long-range repulsive
forces between the particles due to the presence of highly charged
surface ligands (Figure S2).\cite{32} The absence of a
low-\emph{q} upturn confirms that the particles are free of aggregation.
Similar SAXS patterns were observed for MCC-capped PbS NCs of different
sizes, and with different charged surface ligands (Figure S2).

The self-assembly of
PbS-Sn\textsubscript{2}S\textsubscript{6}\textsuperscript{4-} NCs
colloidally dispersed in NMF was induced by adding a non-solvent and a
flocculant that screened the surface potential of the
PbS-Sn\textsubscript{2}S\textsubscript{6}\textsuperscript{4-} NCs,
reducing long-range electrostatic repulsion between NC
surfaces.\cite{30} As a flocculant, we used
K\textsubscript{3}AsS\textsubscript{4},
K\textsubscript{4}GeS\textsubscript{4} or
K\textsubscript{4}Sn\textsubscript{2}Se\textsubscript{6} that dissociate
in polar solvents greatly increasing the ionic strength. Additionally, a
non-solvent (e.g., DMF, MeCN) lowers the solvent dielectric constant
further facilitating self-assembly by decreasing the solubility of
PbS-Sn\textsubscript{2}S\textsubscript{6}\textsuperscript{4-} NCs. A
black solid precipitates out of the solution, leaving unassembled
PbS-Sn\textsubscript{2}S\textsubscript{6}\textsuperscript{4-} NCs
remaining as a brown supernatant. To wash as-prepared 5.7 nm
PbS-Sn\textsubscript{2}S\textsubscript{6}\textsuperscript{4-} NC
superlattice, the collected precipitate was suspended in acetonitrile,
followed by centrifugation to remove the solvent. The residual
acetonitrile was then allowed to evaporate, yielding washed superlattice
as a free-flowing black solid powder. The SAXS pattern of the
precipitated solid revealed that they consist of highly ordered
superlattices, evidenced by the emergence of sharp Bragg peaks (Figures
1A and S3), while the photoluminescence (PL) spectrum of the
superlattice confirms that the quantum confinement in the NCs is
preserved (Figure S4). Our findings suggest that various multivalent
salts can function as flocculants, and different solvents miscible with
NMF and having a lower polarity than NMF can act as the non-solvent
(Figure S3). Consistent Bragg peak patterns were observed when various
flocculants and non-solvents were used, indicating that these factors do
not substantially affect the superlattice crystal structure.
Additionally,
PbS-Sn\textsubscript{2}S\textsubscript{6}\textsuperscript{4-} NCs
ranging from 5.4 nm to 8.1 nm in size also self-assembled into
superlattices (Figure S5), with shifts in Bragg peaks along the
\emph{q}-axis corresponding to expanded lattice constants as NC diameter
increases. PbS NCs with another charged ligand
(AsS\textsubscript{4}\textsuperscript{3-}) were also able to be
self-assembled into an ordered superlattice under the same experimental
conditions (Figure S5).

Figure \ref{fig:1} demonstrates that
PbS-Sn\textsubscript{2}S\textsubscript{6}\textsuperscript{4-} NCs
self-assemble into superlattices exhibiting not only translational but
also orientational order of NCs. Figure \ref{fig:1}A presents a SAXS pattern of
washed 5.7 nm
PbS-Sn\textsubscript{2}S\textsubscript{6}\textsuperscript{4-} NC
superlattices. The Bragg peak assignment indicates that the superlattice
adopts a face-centered cubic (\emph{fcc}) structure (Figure S6). Figures
1B and 1C show transmission electron microscopy (TEM) and corresponding
selected area electron diffraction (SAED) pattern of washed 5.7 nm
PbS-Sn\textsubscript{2}S\textsubscript{6}\textsuperscript{4-} NC
superlattices, respectively. The SAED pattern consists of distinct spot
patterns that correspond to an \emph{fcc} atomic lattice viewed at (110)
zone axis (Figure S7). The absence of rings or unidentified spots in the
SAED data indicates that the PbS NCs maintain orientational order
throughout the entire superlattice. Additional TEM and SAED images can
be found in Figure S8. High-angle annular dark-field scanning
transmission electron microscopy (HAADF-STEM) imaging with atomic
resolution (Figures \ref{fig:1}D, \ref{fig:1}E, and S9) reveals the crystalline necks
between NCs formed across the superlattice. To rule out the possibility
of projection artifact coming from NCs in different layers, the
high-resolution STEM images were acquired with a relatively large
convergence angle (25 mrad), which provides depth sensitivity on the
order of a few nanometers, comparable to the NC size. If the particles
were at different heights, they would not both be in focus. Furthermore,
during STEM imaging of these superlattices (beam current less than 50
pA), we did not observe any changes that would suggest the electron beam
is inducing the crystalline connections (necks). Additional TEM images
of the superlattices show that spontaneous oriented self-assembly occurs
for PbS-Sn\textsubscript{2}S\textsubscript{6}\textsuperscript{4-} NCs of
different sizes (Figure S10), and similar behavior can be found from
PbS-AsS\textsubscript{4}\textsuperscript{3-} NCs and
PbSe-Sn\textsubscript{2}S\textsubscript{6}\textsuperscript{4-} NCs
(Figure S11). The coherence of the atomic lattice across the
superlattice is further supported by the Scherrer analysis of the
diffraction spots in SAED patterns -- the Scherrer size of atomic
crystalline domains in the superlattice exceeds both the Scherrer size
and SAXS diameter of individual PbS NCs (Figures S12-S16).

Some degree of orientational alignment is often observed in
self-assembled NC superlattices, typically associated with NC faceting
and parallel facet alignment in superlattices that maximizes NC packing
density.\cite{35} However, the soft nature of the organic
surface ligands cannot tightly lock the NC alignment, and SAED patterns
typically show diffraction arcs rather than spots typical for single
crystals.\cite{36} An important exception is the assembly of
colloidal NCs at the liquid/liquid interface, followed by the removal of
surface ligands---the oriented attachment of NCs can locally create a
comparable order of the atomic lattices through irreversible NC
necking.\cite{25,26} There, the liquid-liquid interface plays
an important role for in-plane alignment of NCs and gentle removal of
surface ligands to induce the oriented attachment. To the best of our
knowledge, the orientational ordering of sub-10 nm NCs in large 3D NC
superlattices shown in Figure \ref{fig:1} is rather unprecedented and requires
further investigation.

\begin{figure*}[t]
    \centering
    \includegraphics[width=.9\linewidth]{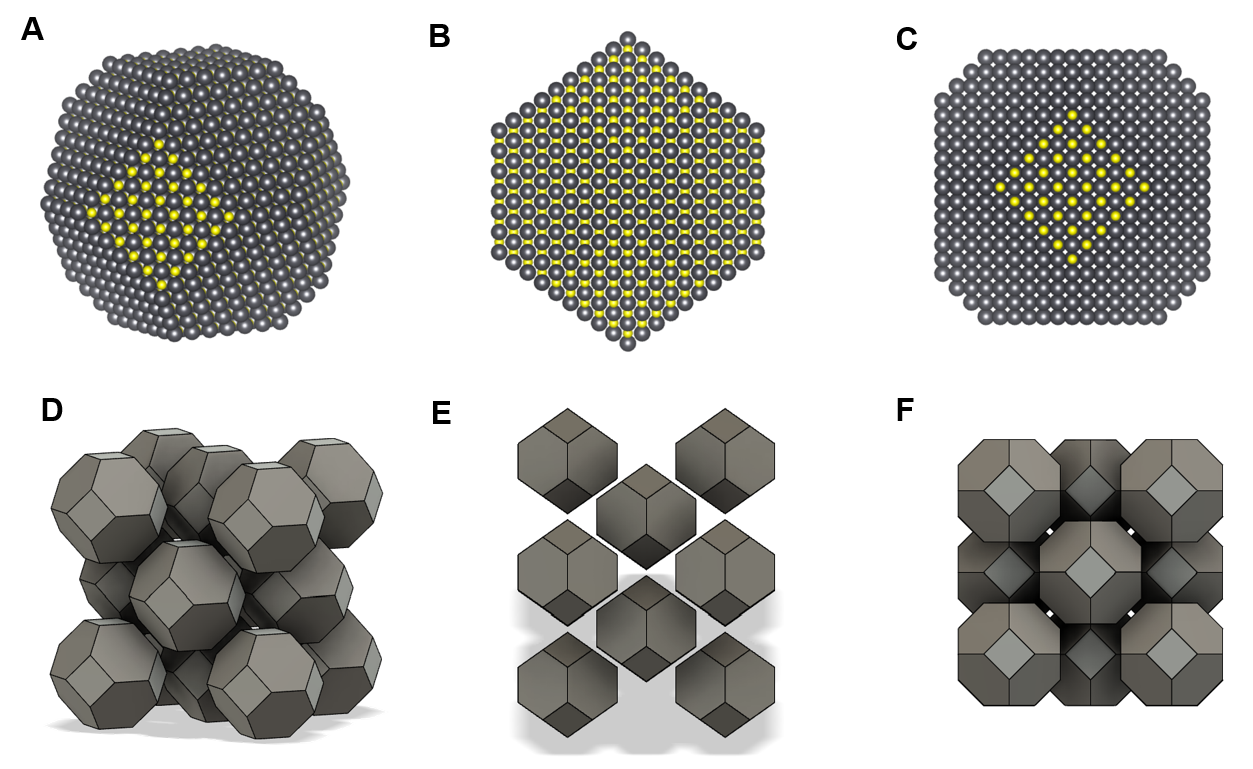}
    
    \caption{(A) Illustration of 5.7 nm
    PbS-Sn\textsubscript{2}S\textsubscript{6}\textsuperscript{4-} NC
    superlattice. Illustration of the superlattice viewed along (B)
    (110) zone axis and (C) (100) zone axis. (D)
    Illustration of the PbS NC viewed along (E) (110) zone axis and
    (F) (100) zone axis.}
    \label{fig:3}
\end{figure*}

\subsection{Structural analysis of
PbS-Sn\textsubscript{2}S\textsubscript{6}\textsuperscript{4-}
nanocrystal superlattices} The translational and orientational ordering
of the 5.7 nm
PbS-Sn\textsubscript{2}S\textsubscript{6}\textsuperscript{4-} NCs was
further examined using real and reciprocal space analysis of TEM images
of NC superlattices. Figure \ref{fig:2} presents TEM and Fourier-transformed (FT)
TEM images of 5.7 nm
PbS-Sn\textsubscript{2}S\textsubscript{6}\textsuperscript{4-} NC
superlattice. Figures \ref{fig:2}A and \ref{fig:2}B are the low-magnification TEM and
high-resolution TEM (HRTEM) images of the resulting superlattice, while
Figures \ref{fig:2}E and \ref{fig:2}F display the corresponding FT-TEM images, respectively.
The FT-TEM image in Figure \ref{fig:2}E reflects the periodicity of NC packing in
the superlattice, showing six spots arranged in an elongated hexagonal
pattern, where two adjacent spots form a 54° angle. This arrangement is
characteristic of the reciprocal-space representation of a face-centered
cubic (\emph{fcc}) crystal viewed along the (110) zone axis, which is
consistent with the SAXS pattern (Figure \ref{fig:1}A).\cite{37} The
four innermost spots arise from the (-111), (1-11), (-11-1), and (1-1-1)
sets of superlattice planes, while the two outer spots correspond to the
(002) and (00-2) planes. The calculation of the \emph{d}-spacing from
the spot-to-center distances yields a superlattice unit cell length of
8.7 nm, which is consistent with the SAXS measurement of 8.5 nm. The
FT-TEM image in Figure \ref{fig:2}F indicates a PbS atomic unit cell lattice
constant of 6.1 Å, which is in good agreement with the powder X-ray
diffraction (PXRD) measurement (5.9 Å, Figure S13). The TEM, HRTEM, and
the corresponding FT-TEM images of the 5.7 nm
PbS-Sn\textsubscript{2}S\textsubscript{6}\textsuperscript{4-} NC
superlattice viewed along (100) zone axis are shown in Figures \ref{fig:2}C, \ref{fig:2}D,
\ref{fig:2}G, and \ref{fig:2}H, respectively.

Analysis of PXRD patterns of 5.7 nm
PbS-Sn\textsubscript{2}S\textsubscript{6}\textsuperscript{4-} NCs
provides an average \emph{h}(111)/\emph{h}(100) Wulff construction ratio
of 1.02 (Figure S13), indicating that these NCs exhibit a truncated
octahedral shape (0.58 \textless{} \emph{h}(111)/\emph{h}(100)
\textless{} 1.15),\cite{38} consistent with previous
experimental and computational studies.\cite{38,39} TEM
images of colloidal
PbS-Sn\textsubscript{2}S\textsubscript{6}\textsuperscript{4-} NCs show
that the NCs are nearly spherical, which further supports this (Figure
S17). A model of a truncated octahedral NC is shown in Figure \ref{fig:3}A. The
FT-TEM images in Figures \ref{fig:2}E and \ref{fig:2}F demonstrate that both the NC
superlattice and the atomic lattices of individual NC are simultaneously
oriented along the (110) zone axis. This alignment is only achievable if
the PbS NCs are assembled edge-to-edge, with the closest nearly touching
edges being parallel to the \textless110\textgreater{} set of equivalent
lattice directions of the superlattice and NC atomic lattices, as
depicted in Figures \ref{fig:3}B and \ref{fig:3}E. Furthermore, the matching spot patterns
in the FT-TEM images in Figures \ref{fig:2}D and \ref{fig:2}H for the superlattice viewed
along the (100) zone axis further confirm the edge-to-edge alignment of
NCs within the superlattice, as depicted in Figures \ref{fig:3}C and \ref{fig:3}F.

This edge-to-edge alignment is highly unusual for NC assemblies, which
typically favor face-to-face orientations to maximize the cohesive
energy from van der Waals interactions. For NCs with steric colloidal
stabilization provided by ligands with long hydrocarbon chains, there
are only a few reported examples of superlattices supported by
vertex-to-vertex contacts,\cite{40} but such examples are far
rarer compared to the face-to-face assemblies. We propose that this
unique superlattice configuration plays a key role in enabling the
reversible nature of self-assembly processes in the absence of bulky
organic surface ligands---the traditional face-to-face packing would be
more prone to irreversible aggregation under strong van der Waals
forces. Furthermore, the narrow connecting NC edges may be instrumental
in establishing epitaxial necks with atomically precise order (Figure \ref{fig:1}D
and 1E) as discussed below.

\subsection{Theoretical insight into the origin of edge-to-edge orientation
in PbS-Sn\textsubscript{2}S\textsubscript{6}\textsuperscript{4-}
nanocrystal superlattices} In this Section, we introduce a theoretical
model helping to elucidate possible origin of the edge-to-edge
orientation in our NC superlattices. Such edge-to-edge orientation is
very unusual not only in experimental but also in computational studies
of NC self-assembly.\cite{41} The simulations of
self-assembly of nanoparticles with conventional bulky organic ligands
have shown superlattices with various kinds of orientational order, but
not edge-to-edge or vertex-to-vertex arrangements, because of the van
der Waals attraction strongly favoring face-to-face
contacts.\cite{35,42} Intriguingly, calculations using
combination of mean-field electrostatic repulsion and van der Waals
attraction (e.g. DLVO theory) predict that electrostatically stabilized
nanoparticles, e.g., truncated octahedra, also adopt face-to-face
orientation.\cite{43,44,45} The discrepancy between the
previous computational studies of NC self-assembly and the
experimentally observed edge-to-edge alignment in
PbS-Sn\textsubscript{2}S\textsubscript{6}\textsuperscript{4--}~NC
superlattices may arise from the fundamental limitations of DLVO theory,
which is strictly applicable to colloids suspended in dilute 1:1
electrolytes and ignores ionic correlations present in concentrated
electrolyte solutions, which are particularly strong for multivalent
ions.\cite{46,47}

\begin{figure*}[t]
    \centering
    \includegraphics[width=.85\linewidth]{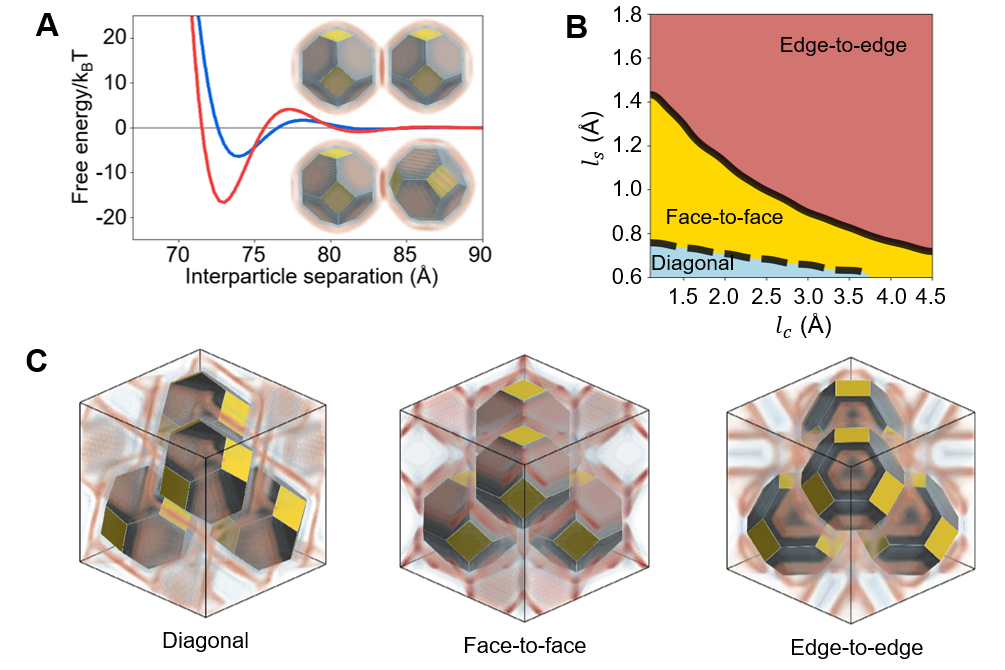}

    \caption{(A) Free energy difference between two aligned (red
line) and nonaligned (blue line) nanocrystals (NCs) as a function of the
distance of separation. Top and bottom inset show the asymmetric charge
density for the aligned and non-aligned NCs at the distance of minimum
free energy. B) Phase diagram for orientational order in a four-site
\emph{fcc} lattice as a function of characteristic length scales related
to ion and solvent size (\(l_{c}\)) and screening length (\(l_{s}\)).
The three phases correspond to diagonal alignment (blue), face-to-face
alignment (gold) and edge-to-edge alignment (red). (C) Representative
snapshots of the three phases along with the asymmetric charge density.
The snapshot consists of truncated octahedral PbS NCs and the charge
distributions around the NCs. Red indicates positive charge, blue
indicates negative charge.}
    \label{fig:4}
\end{figure*}

To address this limitation, we develop a theoretical model accounting
for ionic correlations in a system of highly charged colloidal particles
in a solution of multivalent ions. The theoretical framework combines
phenomenological Landau-Ginzburg field theory adapted for describing
strongly correlated ionic systems\cite{limmer2015interfacial, 48}, with explicit geometric descriptions
of truncated octahedral NCs and numerical solution of the resulting
Euler--Lagrange equations. Full technical details are provided in
Section 5 of Supporting Information. The effective Hamiltonian, which
serves as a basis for calculating the free energy of a system, accounts
for (i) short-range interactions including dispersion forces and steric
effects; (ii) long-range Coulomb interactions. The former term describes
local structure with the characteristic length scale \(l_{c}\)
determined by ion and solvent size, while the second contribution
accounts for the electrostatic screening with characteristic length
\(l_{s}\) closely related to the Debye screening length. This approach
has been successfully employed to rationalize interparticle forces in
highly charged colloidal systems in concentrated electrolytes such as
molten salts and ionic liquids.\revision{\cite{48,kamysbayev2019nanocrystals}}

This model predicts that the response of the electrolyte to a localized
charge, the susceptibility \(\chi(r)\), has the form of a damped
oscillation\revision{\cite{limmer2015interfacial, 48}}:
\begin{equation*}
    \chi(r) = \frac{A}{4\pi r}\ \exp\left( - \frac{\alpha}{r} \right)\cos\left( \frac{\gamma r}{\alpha} - \theta \right),
\end{equation*}

where \(A,\ \alpha\), \(\gamma\) and \(\theta\) are simple analytic
functions of \(l_{c}\) and \(l_{s}\):
\begin{align*}
\alpha &= 2l_{c}l_{s}^{1/2}\left( 2l_{c} - l_{s} \right)^{- 1/2},\quad\gamma = \sqrt{\frac{2 + l_{s}/l_{c}}{2 - l_{s}/l_{c}}},\notag \\
\quad\theta &= \sin^{- 1}(l_{s}/2l_{c}),\quad A = \frac{1}{al_{c}\sqrt{4l_{c}^{2} - l_{s}^{2}}}
\end{align*}

.

The plots of susceptibility \(\chi(r)\) in respect to interparticle
distance (\(r\)) are shown in Figure S18. The oscillatory component
arises from the formation of alternating layers of cations and anions in
the vicinity of a charged surface, which results from the competition
between ionic packing constraints and the requirement for local charge
neutrality.\revision{\cite{bazant2011double, smith2016electrostatic, kornyshev2007double, gebbie2013ionic}} The oscillation period is related to ion size and spacing,
and the oscillation amplitude decays away from the interface over a
distance \(\alpha\). These model parameters are directly tied to
familiar experimental tuning knobs. Thus, the electrolyte ionic strength
and ion valency control the screening length \(l_{s}\) (shorter at
higher ionic strength or higher ion charge). Solvent dielectric constant
and temperature: high-\(\varepsilon\) solvents and higher temperatures
lead to longer \(l_{s}\) and more extended ion layering. Ion size
controls the short-range correlation length \(l_{s}\), which sets the
length scale for layering. Ligand surface charge density determines the
magnitude of the electrolyte-mediated NC--NC interaction, which strongly
increases with more highly charged ligands. In this way, the model
provides a direct link between chemically controllable parameters (salt
concentration, solvent, ligand chemistry, NC size and shape) and the
emergent orientational order and interactions in NC superlattices
mediated by strong electrolytes.

The susceptibility \(\chi(r)\) is the kernel that generates the pair
potential \(W(r)\) between NCs. In the simplest approximation of NC as a
point, the pair potential between NCs is directly proportional to
\(- \chi(r)\). To account for the NC size and shape (truncated
octahedron), we introduce surface and volume form factors and derive
analytic expressions for these form factors\revision{\cite{wuttke2021numerically, lee1983fourier}}. Charged ligands at the NC
surface are modeled as an effective surface field with the strength
controlled by a parameter directly related to the surface charge density
of the ligands. This approach enables us to directly calculate the
solvent contribution to the total free energy for any given position and
orientation of the NCs.

Figure \ref{fig:4}A shows the free energy profile for two 5.7 nm truncated
octahedral NCs as a function of their separation distance. The red curve
represents the free energy when the two PbS NCs are orientationally
aligned, whereas the black curve illustrates the case where they are
tilted with respect to each other. In both configurations, we observe
short-ranged free energy oscillatory decay with increasing separation.
This behavior arises from the oscillatory charge profiles that surround
NCs in solution. As two NCs approach, the overlap of these profiles
results in constructive or destructive interference, resulting in minima
(in-phase) and maxima (out-of-phase) in the free energy landscape.
Additionally, we find that the faceted geometry of PbS NCs leads to
pronounced anisotropy in their interactions, evidenced by the strong
dependence of the free energy profile on relative orientation. These
findings suggest that the driving force for orientational ordering in
PbS NCs is rooted in the structured layering of ligands and cations
within the superlattice.

Next, we carry out Monte Carlo (MC) simulations to explore the
orientational alignment of four truncated octahedral NCs arranged on the
sites of an \emph{fcc} lattice with the lattice constant of 90 Å. By
performing iterative free energy calculations, we determined the most
energetically favorable configuration that minimizes the free energy
(Figures S19 and S20). The robustness of the model is further validated
using a larger supercell containing 32 NCs (Figure S20). These
simulations produced a detailed orientational phase diagram (Figure \ref{fig:4}B),
illustrating the relationship between the decay length ($\alpha$) and the
oscillation parameter ($\gamma$), which are set by \(l_{c}\) and \(l_{s}\). The
diagram reveals three distinct ordered states: diagonal alignment,
face-to-face alignment, and the experimentally observed edge-to-edge
alignment (Figure \ref{fig:4}C). This diverse orientational landscape arises from
charge layering effects in solutions containing multivalent ions, which
strongly influence free energy as a function of orientation. The
edge-to-edge configuration is thermodynamically preferred over a wide
parameter range and becomes exclusively favored as the decay length ($\alpha$)
increases. These findings highlight the crucial role of long-range
electrostatic interactions with charge layering effects in stabilizing
the edge-to-edge orientation.

\subsection{\emph{Reversible} oriented attachment of edge-to-edge assembled
PbS-Sn\textsubscript{2}S\textsubscript{6}\textsuperscript{4-}
nanocrystals} In this Section, we focus on the formation of epitaxial
``necks'' between
PbS-Sn\textsubscript{2}S\textsubscript{6}\textsuperscript{4-} NCs
(Figure \ref{fig:1}D). This process resembles oriented attachment of
NCs.\cite{24} Oriented attachment is common for nanoscale
crystals, while it is rarely observed for large crystals because joining
two crystals together requires rearranging a number of surface atoms
roughly proportional to the areas of merging surfaces. Oriented
attachment can be an ideal approach to establish strong electronic
coupling within NC superlattices.\cite{25,26} However, a
major limitation of oriented attachment as universal mechanism for
assembling nanoscale building blocks into meso- and macroscale objects
with near-atomic precision is the lack of micro-reversibility required
for healing occasional bonding mistakes. For example, a NC superlattice
in Figure \ref{fig:1}A contains about 3·10\textsuperscript{6} individual
PbS-Sn\textsubscript{2}S\textsubscript{6}\textsuperscript{4-} NCs. It is
highly likely that the formation of such a complex construct requires
numerous rearrangements. In traditional crystal growth, the reversible
bonding and surface mobility of added building blocks are critical for
maintaining structural coherence.\cite{49} However,
disassembly of a crystalline ``neck'' between NCs is highly improbable
because of a large total energy of all chemical bonds that need to be
broken, thus making traditional oriented attachment a ``forward-only''
process.

\begin{figure*}[t]
    \centering
    \includegraphics[width=\linewidth]{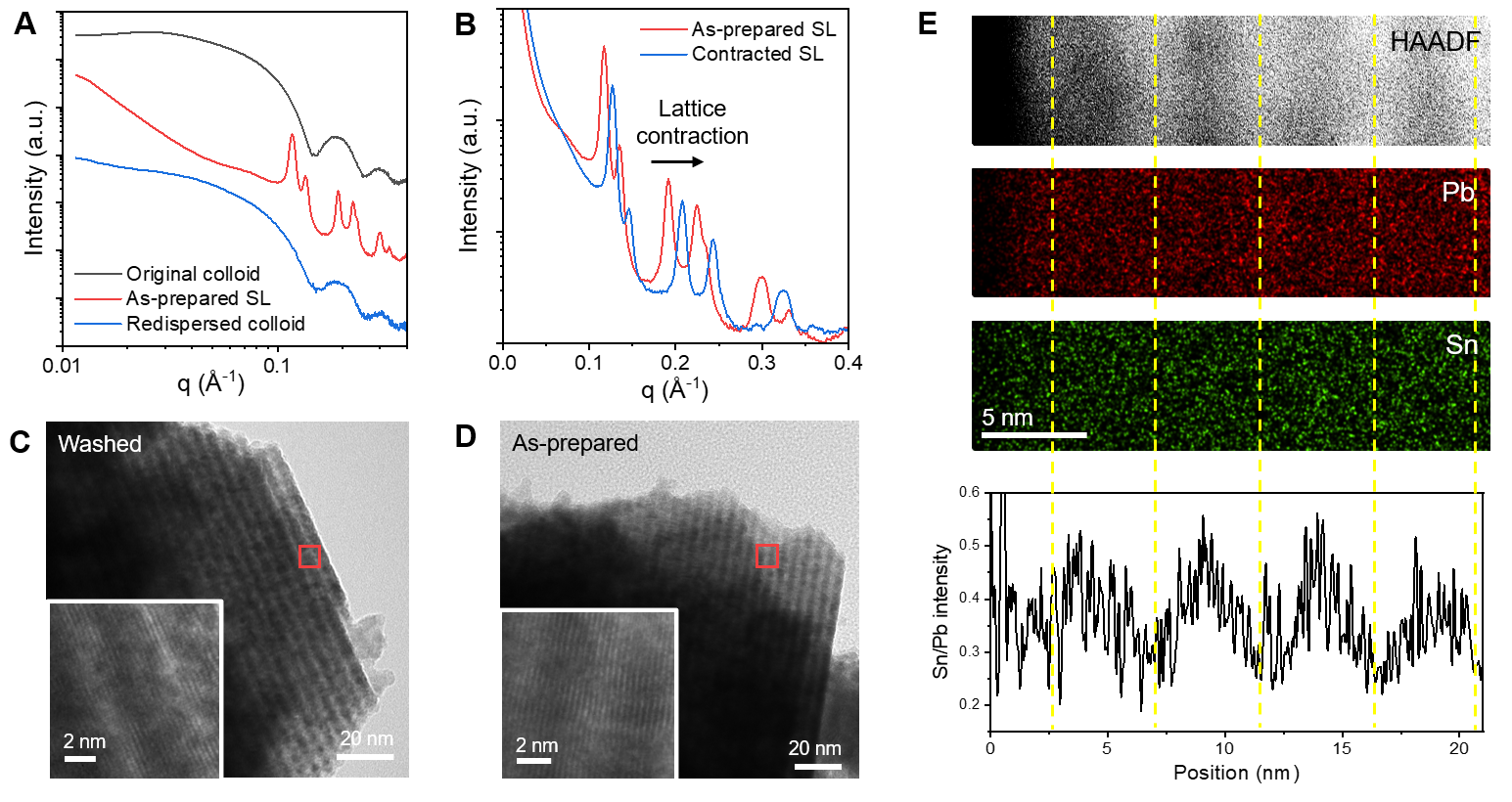}
    
    \vspace{1em} 
    
    \includegraphics[width=\linewidth]{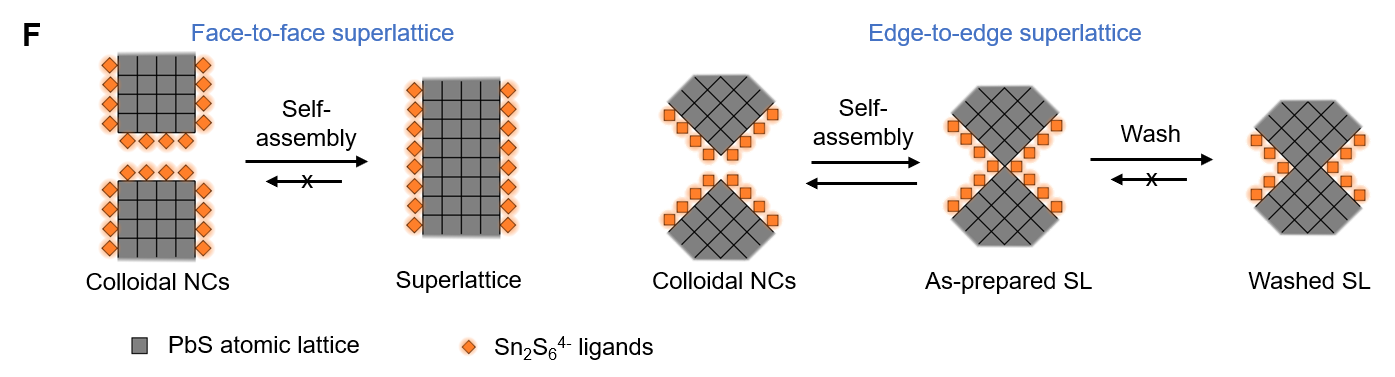}
    
    \caption{(A) Small-angle X-ray scattering (SAXS)
patterns of the original colloid solution, as-prepared superlattices
(SL) and redispersed colloid of 5.7 nm
PbS-Sn\textsubscript{2}S\textsubscript{6}\textsuperscript{4-}
nanocrystals (NCs). The y-axis is vertically offset for clarity. Absence
of Bragg peaks from the SAXS pattern of redispersed PbS NCs indicates
full disassembly of superlattices. (B) SAXS patterns of
as-prepared and washed 5.7 nm
PbS-Sn\textsubscript{2}S\textsubscript{6}\textsuperscript{4-} NC
superlattices. (C, D) Transmission electron microscopy (TEM)
images of as-prepared (D) and washed (E) 5.7 nm
PbS-Sn\textsubscript{2}S\textsubscript{6}\textsuperscript{4-} NC
superlattices. Insets: magnified images of the regions highlighted by
red squares. (E) HAADF-STEM image and STEM-EDS elemental
mapping of the superlattice. The plot below presents the ratio of the Sn
and Pb signal counts across the superlattice. (F) Schematic of
self-assembly, washing and redispersion processes.}
    \label{fig:5}
\end{figure*}

Counterintuitively, Figure \ref{fig:5} demonstrates that self-assembled
PbS-Sn\textsubscript{2}S\textsubscript{6}\textsuperscript{4-} NC
superlattices can be fully reversed to the colloidal state. The SAXS
pattern of a colloidal solution of 5.7 nm
PbS-Sn\textsubscript{2}S\textsubscript{6}\textsuperscript{4-} NCs is
shown as the black line in Figure \ref{fig:5}A. The red line corresponds to a
representative SAXS pattern of a superlattice formed using
dimethylformamide (DMF) as the non-solvent and
K\textsubscript{3}AsS\textsubscript{4} as the flocculant. After
centrifugation to remove solvents and unassembled NCs, the residual
solid was collected for SAXS measurements of the as-prepared
superlattices. The sharp Bragg peaks in the SAXS pattern confirm the
formation of a highly ordered \emph{fcc} superlattice with a unit cell
length of 9.3 nm (Figure S21). This
PbS-Sn\textsubscript{2}S\textsubscript{6}\textsuperscript{4-} NC
superlattice can be disassembled back into colloidal NCs by adding
highly polar NMF solvent (Figure S22). The SAXS pattern of the
redispersed
PbS-Sn\textsubscript{2}S\textsubscript{6}\textsuperscript{4-} NCs, shown
as the blue line in Figure \ref{fig:5}A, exhibits identical Bessel oscillations in
the high-\emph{q} region, (\emph{q} \textgreater{} 0.1
Å\textsuperscript{-1}) to those in the original colloidal solution, thus
confirming that the PbS NCs were fully recovered with no changes in size
or shape distribution. This reversible self-assembly remained effective
even after the superlattices were aged for over two days prior to
redispersion (Figure S23). TEM images of the original solution of
PbS-Sn\textsubscript{2}S\textsubscript{6}\textsuperscript{4-} NCs and
the solution obtained by redissolving a superlattice (Figure S24)
further confirm complete disassembly of superlattices into individual
NCs.

TEM and HRTEM images of as-prepared superlattices display continuous
atomic lattice fringes, indicating strong orientational alignment of PbS
NCs (Figure \ref{fig:5}C). Washing these superlattice with excess acetonitrile
does not change the qualitative appearance of the superlattices in TEM
and HRTEM images (Figure \ref{fig:5}D), while making them more robust and less
prone to electron beam damage. The washing causes a lattice contraction
of the 5.7 nm
PbS-Sn\textsubscript{2}S\textsubscript{6}\textsuperscript{4-} NC
superlattice from 9.3 nm to 8.5 nm, evidenced by the shift of Bragg
peaks in the SAXS patterns to lower \emph{q}-values after washing
(Figure \ref{fig:5}B). Immediately after washing with acetonitrile, the
superlattices can be redissolved into individual NCs, but lose
solubility upon aging in dry state. Attempts to redisperse washed and
aged PbS-Sn\textsubscript{2}S\textsubscript{6}\textsuperscript{4-} NC
superlattices in NMF (Figure S25) result in SAXS patterns with Bragg
peaks, confirming that the superlattices remain attached and do not
revert to the initial colloidal state.

To reconcile these observations, here we discuss the possibility of
\emph{reversible} oriented attachment (ROA) and how it may work on the
example of PbS-Sn\textsubscript{2}S\textsubscript{6}\textsuperscript{4-}
NC superlattices. An important prerequisite for achieving ROA is the
minimization of NC contact areas that touch to form a crystalline neck,
as large contact areas can make the orientational attachment more prone
to defects.\cite{50,51,52,53} Ideally, NCs should approach each
other\textquotesingle s edge-to-edge to create a line of contact that
further grows into a neck (Figures \ref{fig:1}D and 5F). Such alignment with
low-area contacts is indeed observed in
PbS-Sn\textsubscript{2}S\textsubscript{6}\textsuperscript{4-} NC
superlattices (Figures 1-3), stabilized by ion layering as described in
the previous Section (Figure \ref{fig:4}).

The second requirement for ROA can be related to the chemical nature of
inorganic surface ligands. Ideally, such ligands should guide the
oriented attachment of NC cores while enabling local rearrangements by
delaying formation of irreversible necks between NCs. To study the necks
formed between PbS NCs, we carried out STEM-EDS elemental mapping within
a 5.9 nm PbS-Sn\textsubscript{2}S\textsubscript{6}\textsuperscript{4-}
NC superlattice (Figure \ref{fig:5}E), along with the corresponding elemental line
scan across the superlattice (also see Figure S26). The accumulated EDS
spectrum can be found in Figure S27. The bright regions in the
HAADF-STEM image correspond to superlattice planes formed by PbS NCs,
while the darker regions represent the interparticle gaps with necks
connecting adjacent NCs. The STEM-EDS Pb signal intensity is higher in
the brighter regions of the HAADF-STEM image, confirming this
correspondence. Notably, the spatial distribution of tin (Sn) exhibits
an anti-phase relationship with that of lead (Pb), clearly indicated by
the plot of Sn/Pb ratio (Figures \ref{fig:5}E and S26). This indicates that the
interparticle regions are enriched in Sn, suggesting that the epitaxial
bridging of PbS NCs involves
Sn\textsubscript{2}S\textsubscript{6}\textsuperscript{4-} surface
ligands. Furthermore, the use of K\textsubscript{3}AsS\textsubscript{4}
as the flocculant can further promote reversibility at an early stage of
the oriented attachment because tetrahedrally bonded
AsS\textsubscript{4}\textsuperscript{3-} ions cannot epitaxially
integrate with PbS lattice. We propose that the reversible self-assembly
of 5.7 nm PbS-Sn\textsubscript{2}S\textsubscript{6}\textsuperscript{4-}
NCs is enabled by the incorporation of excess flocculants
(K\textsubscript{3}AsS\textsubscript{4}) within the superlattice (Figure
S28). Washing the superlattice with acetonitrile removes a significant
portion of incorporated K\textsubscript{3}AsS\textsubscript{4}, as
confirmed by the elemental analysis (Table S1). Removal of
AsS\textsubscript{4}\textsuperscript{3-} triggers the oriented
attachment of edge-to-edge oriented PbS
-Sn\textsubscript{2}S\textsubscript{6}\textsuperscript{4-} NCs, which is
observed as gradual reduction of the lattice constant of \emph{fcc}
superlattice (Figures \ref{fig:5}B). It is known that
Sn\textsubscript{2}S\textsubscript{6}\textsuperscript{4-} ions can
dynamically rearrange in polar solvents and may form fragments
structurally compatible with PbS cores.\cite{54,55} The
dynamic nature of Sn-S-Sn bonding in polar solvents is expected to favor
both oriented attachment of NCs and disassembly of misbonded fragments,
gradually leading to the formation of permanent necks between NCs and
preventing redispersion of NCs into the colloidal phase.

\section{Conclusion}

In summary, our results demonstrate that PbS NCs capped with compact and
highly charged Sn\textsubscript{2}S\textsubscript{6}\textsuperscript{4-}
inorganic ligands can self-assemble into superlattices that exhibit
exceptionally high translational and orientational order, approaching
single-crystal precision. Structural and compositional analyses reveal
that the NCs adopt a distinctive edge-to-edge orientation, promoted by
the ionic layering in the electrolyte containing multi-charged ions. We
further reveal that the reversibility of this self-assembly is
influenced by the incorporation and partial removal of
K\textsubscript{3}AsS\textsubscript{4} salts during the washing process.
The superlattices with incorporated
K\textsubscript{3}AsS\textsubscript{4} can be completely disassembled
into their original colloidal state, whereas the partial removal of
salts results in permanent lattice contraction and epitaxial oriented
attachment of NCs. These findings highlight the interplay between ligand
chemistry, lattice stability, and reversibility, offering a tunable
strategy for control of NC assemblies. This ability to reversibly
assemble colloidal nanocrystals with atomic-level alignment opens
exciting opportunities for creating dynamic, reconfigurable materials
with single-crystal--like properties, holding potential for applications
in adaptive optoelectronics, self-healing materials, and nanostructured
devices.

\section{Methods/Experimental}

\textbf{Synthesis of PbS nanocrystals (NCs)}: The preparations of
Pb(OA)\textsubscript{2} and the thiourea precursor were carried out according to the
previously reported procedures.\cite{34} The synthesis of PbS
NCs follows the method adapted from the reported
procedures.\cite{34}

To obtain 5.7 nm PbS-OA NCs, Pb(OA)\textsubscript{2} (5.544 g, 7.2 mmol)
and 1-octene (60 mL) were added into a three-neck 250 mL round-bottom
flask under a N\textsubscript{2} atmosphere. The mixture was heated to
120 °C in an oil bath while maintaining nitrogen flow, during which
Pb(OA)\textsubscript{2} was completely dissolved in the solvent.
Separately, N-phenyl-N'-N'-dodecylthiourea (1.923 g, 6 mmol) was mixed
with diglyme (2 mL) in a 20 mL vial under nitrogen and sealed with a
septum cap. The vial was heated in an oil bath, and after the
temperature was stabilized, the thiourea solution was withdrawn using a
syringe and swiftly injected into the flask. After 10 min, the reaction
was quenched by removing the oil bath and cooling the flask with an air
stream. The resulting NCs were transferred into a nitrogen-filled
glovebox and precipitated by addition of anhydrous methyl acetate. The
NCs were then resuspended in anhydrous hexane and centrifuged to remove
insoluble residues. The supernatant was further purified by four cycles
of precipitation with methyl acetate and redispersion in hexane.
Finally, the washed NCs were redispersed in anhydrous hexane and stored
under N\textsubscript{2}.

\textbf{Ligand exchange:} The ligand exchange of PbS NCs follows the
previously reported method.\cite{32}

\textbf{Self-assembly:} Self-assembled superlattices were prepared by
mixing 20 µL of a stock PbS NC solution (50--100 mg/mL) with 40 µL of
\emph{N,N}-dimethylformamide (DMF). A 500 mM
K\textsubscript{3}AsS\textsubscript{4} solution was then added
incrementally in 5 µL aliquots at one-minute intervals, with continuous
stirring, until a total volume of 40 µL of
K\textsubscript{3}AsS\textsubscript{4} solution had been added. The
mixture was stirred for an additional 20 minutes to form a suspension of
superlattice in an NMF/DMF mixture. The resulting suspension was
centrifuged for 1 minute, and the supernatant was removed to obtain an
"as-prepared" superlattice. To produce a "washed" superlattice, the
material was resuspended in approximately 1 mL of acetonitrile (MeCN),
centrifuged, and the supernatant was discarded. This washing procedure
was repeated twice. The final superlattice samples were then collected
and stored.

\section{ACKNOWLEDGEMENTS}

We are grateful to Dr. A. Nelson for a critical reading and editing of
the manuscript. We thank Progna Banerjee for support for scanning
electron transmission microscopy measurements. We thank Dr. Nestor J.
Zaluzec for support in STEM-EDS measurements. We thank Dr. X. Zuo for
support for X-ray scattering experiments in Advanced Photon Source,
Argonne National Laboratory. Materials synthesis and detailed structural
characterization of nanocrystal superlattices were supported by the
National Science Foundation via the Center for Integration of Modern
Optoelectronic Materials on Demand (IMOD) Science and Technology Center
under Cooperative Agreement No. DMR-2019444. Simulations were supported
by the Office of Basic Energy Sciences (BES), US Department of Energy
(DOE) (award no. DE-SC0019375). Self-assembly experiments were supported
by the University of Chicago Materials Research Science and Engineering
Center, supported by National Science Foundation under award number
DMR-2011854. A.J. was partially supported by Kwanjeong Educational
Foundation. A.N.S. was supported by a Kavli ENSI Graduate Student
Fellowship from the Kavli Nanoscience Institute at UC
Berkeley. Y.A.T. was supported by University of
Chicago\textquotesingle s Quad Undergraduate Research Scholars Program.
Work performed at the Center for Nanoscale Materials and Advanced Photon
Source, U.S. Department of Energy Office of Science User Facilities, was
supported by the U.S. DOE, Office of Basic Energy Sciences, under
Contract No. DE-AC02-06CH11357. HAADF-STEM imaging was carried out at
the Facility for Electron Microscopy of Materials at the University of
Colorado Boulder (CU FEMM, RRID: SCR\_019306).

\section{Code availability}
The source code developed for this study, including the field-theoretic Hamiltonian implementation, CUDA-accelerated solvers and the Monte Carlo sampling routines, is available on GitHub at \url{https://github.com/ansingh1214/Rigid-Field}.

\section{Supporting Information}

Additional figures and tables are provided in the Supporting
Information, which includes detailed descriptions of the experimental
methods, nanocrystal and nanocrystal film characterization, SAXS, WAXS
and SAED analysis, Landau-Ginzburg field theory and numerical simulation
details.

\bibliographystyle{achemso} 
\bibliography{OrientationalAlignment}

\end{document}


\setcounter{page}{1}
\setcounter{figure}{0}
\setcounter{equation}{0}
\setcounter{table}{0}
\setcounter{section}{0}
\renewcommand{\thepage}{S\arabic{page}}     
\renewcommand{\thesection}{S\arabic{section}} 
\renewcommand{\thetable}{S\arabic{table}}   
\renewcommand{\thefigure}{S\arabic{figure}} 
\renewcommand{\theequation}{S\arabic{equation}} 

\setcounter{secnumdepth}{3} 
\setlength{\parskip}{0.5em}
\renewcommand{\thesection}{\arabic{section}}
\renewcommand{\thesubsection}{\thesection.\arabic{subsection}}
\renewcommand{\thesubsubsection}{\thesubsection.\arabic{subsubsection}}

\makeatletter
\renewcommand{\section}{\@startsection{section}{1}{\z@}%
  {0.8cm \@plus1ex \@minus.2ex}
  {0.5cm}
  {\normalfont\large\bfseries}}    

\renewcommand{\subsection}{\@startsection{subsection}{2}{\z@}%
  {0.6cm \@plus1ex \@minus.2ex}%
  {0.3cm}%
  {\normalfont\normalsize\bfseries}} 
\makeatother

\title{Supporting Information for Atomic Alignment in PbS Nanocrystal Superlattices with Compact
Inorganic Ligands via Reversible Oriented Attachment of Nanocrystals}

\author{Ahhyun Jeong}
\affiliation{Department of Chemistry, James Franck Institute, and Pritzker School of Molecular Engineering, University of Chicago, Chicago, Illinois 60637, United States}

\author{Aditya N. Singh}
\affiliation{Department of Chemistry, University of California, Berkeley, California 94720, USA}

\author{Josh Portner}
\affiliation{Department of Chemistry, James Franck Institute, and Pritzker School of Molecular Engineering, University of Chicago, Chicago, Illinois 60637, United States}

\author{Xiaoben Zhang}
\affiliation{Department of Chemistry, James Franck Institute, and Pritzker School of Molecular Engineering, University of Chicago, Chicago, Illinois 60637, United States}

\author{Saghar Rezaie}
\affiliation{Renewable and Sustainable Energy Institute, University of Colorado, Boulder, Colorado, USA}

\author{Justin C. Ondry}
\affiliation{Department of Chemistry, James Franck Institute, and Pritzker School of Molecular Engineering, University of Chicago, Chicago, Illinois 60637, United States}

\author{Zirui Zhou}
\affiliation{Department of Chemistry, James Franck Institute, and Pritzker School of Molecular Engineering, University of Chicago, Chicago, Illinois 60637, United States}

\author{Junhong Chen}
\affiliation{Department of Chemistry, James Franck Institute, and Pritzker School of Molecular Engineering, University of Chicago, Chicago, Illinois 60637, United States}

\author{Ye Ji Kim}
\affiliation{Department of Chemistry, James Franck Institute, and Pritzker School of Molecular Engineering, University of Chicago, Chicago, Illinois 60637, United States}

\author{Richard D. Schaller}
\affiliation{Center for Nanoscale Materials, Argonne National Laboratory, Argonne, Illinois 60439, United States}
\affiliation{Department of Chemistry, Northwestern University, Evanston, Illinois, USA}

\author{Youssef Tazoui}
\affiliation{Department of Chemistry, James Franck Institute, and Pritzker School of Molecular Engineering, University of Chicago, Chicago, Illinois 60637, United States}

\author{Zehan Mi}
\affiliation{Department of Chemistry, James Franck Institute, and Pritzker School of Molecular Engineering, University of Chicago, Chicago, Illinois 60637, United States}

\author{Sadegh Yazdi}
\affiliation{Renewable and Sustainable Energy Institute, University of Colorado, Boulder, Colorado, USA}

\author{David T. Limmer}
\affiliation{Department of Chemistry, University of California, Berkeley, California 94720, USA}
\affiliation{Chemical Sciences and Materials Sciences Divisions, Lawrence Berkeley National Laboratory, Berkeley, California 94720, USA}
\affiliation{Kavli Energy NanoSciences Institute, University of California, Berkeley, California 94720, USA}

\author{Dmitri V. Talapin}
\email[Corresponding author: ]{dvtalapin@uchicago.edu}
\affiliation{Department of Chemistry, James Franck Institute, and Pritzker School of Molecular Engineering, University of Chicago, Chicago, Illinois 60637, United States}
\affiliation{Center for Nanoscale Materials, Argonne National Laboratory, Argonne, Illinois 60439, United States}

\maketitle

\section{Materials}

\textbf{List of precursors:} Arsenic (V) sulfide (Santa Cruz
Biotechnology, 99.99\%), hexyl isothiocyanate (Sigma Aldrich, HPLC grade
\textgreater98\%), lead (II) oxide (Sigma Aldrich, 99.999\% trace metal
basis), \emph{N}-dodecylamine (Sigma Aldrich, 98\%), oleic acid (Fisher
Scientific, 99\%), oleylamine (Sigma Aldrich, technical grade 70\%),
phenyl isothiocyanate (Sigma Aldrich, 98\%), potassium sulfide (Strem,
anhydrous \textgreater95\%), sodium stannate trihydrate (Sigma Aldrich,
95\%), sodium sulfide nonahydrate (Sigma Aldrich, \textgreater98\%),
germanium disulfide (MP Biomedicals), tin (II) selenide (99.995\%,
Sigma-Aldrich), trifluoroacetic acid (Fisher Scientific,
\textgreater99\% for spectroscopy), trifluoroacetic anhydride (Sigma
Aldrich, 99\%), and triethylamine (Sigma Aldrich, \textgreater99.5\%)
were used as received.

\textbf{List of solvents:} Acetone (Fisher Scientific,
\textgreater99.5\%), acetonitrile (Sigma Aldrich, anhydrous 99.8\%),
diethylene glycol dimethyl ether (Sigma Aldrich, anhydrous 99.5\%),
dimethylsulfoxide (Sigma Aldrich, \textgreater99.9\%), hexanes (Fisher
Scientific, \textgreater98.5\%), 1-octadecene (Sigma Aldrich, technical
grade 90\%), 1-octene (Sigma Aldrich, for synthesis \textgreater97\%),
2-isopropanol (Fisher Scientific, \textgreater99.5\%), methanol (Fisher
Scientific, \textgreater99.9\%), ethanol (\textgreater99.5\%, anhydrous,
200 proof), methyl acetate (Sigma Aldrich, anhydrous 99.5\%),
\emph{N,N}-dimethylformamide (Sigma Aldrich, anhydrous 99.8\%),
\emph{N}-methylformamide (Sigma Aldrich, 99\%),
\emph{N}-methylpropionamide (Sigma Aldrich, 98\%), \emph{n}-hexane
(Sigma Aldrich, anhydrous 95\%), and toluene (Fisher Scientific,
\textgreater99.8\%) were used as received.

\section{Sample preparation}

\textbf{Synthesis of PbS nanocrystals (NCs):} Synthesis of oleate-capped
PbS (PbS-OA) NCs follows a previously reported procedure, including the
preparations of Pb(OA)\textsubscript{2} and the thiourea precursor.

To obtain 5.7 nm PbS-OA NCs, Pb(OA)\textsubscript{2} (5.544 g, 7.2 mmol)
and 1-octene (60 mL) were added into a three-neck 250 mL round-bottom
flask under a N\textsubscript{2} atmosphere. The mixture was heated to
120 °C in an oil bath while maintaining nitrogen flow, during which
Pb(OA)\textsubscript{2} was completely dissolved in the solvent.
Separately, N-phenyl-N'-N'-dodecylthiourea (1.923 g, 6 mmol) was mixed
with diglyme (2 mL) in a 20 mL vial under nitrogen and sealed with a
septum cap. The vial was heated in an oil bath, and after the
temperature was stabilized, the thiourea solution was withdrawn using a
syringe and swiftly injected into the flask. After 10 min, the reaction
was quenched by removing the oil bath and cooling the flask with an air
stream. The resulting NCs were transferred into a nitrogen-filled
glovebox and precipitated by the addition of anhydrous methyl acetate.
The NCs were then resuspended in anhydrous hexane and centrifuged to
remove insoluble residues. The supernatant was further purified by four
cycles of precipitation with methyl acetate and redispersion in hexane.
Finally, the washed NCs were redispersed in anhydrous hexane and stored
under N\textsubscript{2}.

Size-control of PbS-OA NCs was carried out by tuning the thiourea
precursor, temperature and solvents. To obtain 5 nm PbS-OA NCs, similar
procedure was used, except a solution of
\emph{N}-(\emph{p}-(trifluoromethyl)phenyl)-\emph{N}'-dodecylthiourea
(2.331 g, 6 mmol) in 2 ml of diglyme was injected into a solution of
Pb(OA)\textsubscript{2} (5.544 g, 7.2 mmol) and 1-octene (60 mL) at 120
\textsuperscript{o}C. To obtain 7.3 nm PbS-OA NCs, a solution of
\emph{N}-phenyl-\emph{N}'-\emph{N}'-dodecylthiourea (1.923 g, 6 mmol) in
2 ml of diglyme was injected into a solution of Pb(OA)\textsubscript{2}
(5.544 g, 7.2 mmol) in 1-octadecene (60 ml) at 120 \textsuperscript{o}C.
To obtain 8.3 nm PbS-OA NCs, a solution of
\emph{N}-\emph{N}-hexyl-\emph{N}'-dodecylthiourea (1.972, 6 mmol) in 2
ml of diglyme was injected to a solution of Pb(OA)\textsubscript{2}
(5.544 g, 7.2 mmol) in 1-octadecene (60 ml) at 150 \textsuperscript{o}C.
The same washing procedure was adopted across various PbS-OA NCs.

\textbf{Preparation of inorganic salt}The preparation of 200 mM
K\textsubscript{4}Sn\textsubscript{2}S\textsubscript{6}, 500 mM
K\textsubscript{3}AsS\textsubscript{4} and 200 mM
K\textsubscript{4}Sn\textsubscript{2}Se\textsubscript{6} solutions in
\emph{N}-methylformamide (NMF) follows the previously reported
method.\cite{1,2} A 500 mM
K\textsubscript{4}GeS\textsubscript{4} solution in NMF was prepared by
dissolving a stoichiometric ratio of anhydrous K\textsubscript{2}S and
GeS\textsubscript{2} in distilled NMF and stirring overnight.
Undissolved solids were removed by centrifugation.

\textbf{Ligand exchange:} The ligand exchange of PbS NCs follows
previously reported method.\cite{3}

\textbf{Self-assembly:} Self-assembled superlattices were prepared by
mixing 20 µL of a stock PbS NC solution (50--100 mg/mL) with 40 µL of
\emph{N,N}-dimethylformamide (DMF). A 500 mM
K\textsubscript{3}AsS\textsubscript{4} solution was then added
incrementally in 5 µL aliquots at one-minute intervals, with continuous
stirring, until a total volume of 40 µL of
K\textsubscript{3}AsS\textsubscript{4} solution had been added. The
mixture was stirred for an additional 20 minutes to form a suspension of
superlattice in an NMF/DMF mixture. The resulting suspension was
centrifuged for 1 minute, and the supernatant was removed to obtain an
"as-prepared" superlattice. To produce a "washed" superlattice, the
material was resuspended in approximately 1 mL of acetonitrile (MeCN),
centrifuged, and the supernatant was discarded. This washing procedure
was repeated twice. The final superlattice samples were then collected
and stored.

\textbf{Characterization methods}

\textbf{SAXS:} Small-angle X-ray scattering (SAXS) collected using a
SAXSLAB Ganesha Instrument with a Cu K-alpha source. Colloidal QDs were
characterized in glass capillaries that were flame-sealed in air.
Superlattices were suspended in a small volume of MeCN
(\textasciitilde20 µL) and were drop-cast on Kapton tape. The MeCN was
allowed to evaporate under vacuum before the measurement.

\textbf{TEM:} Transmission electron microscopy (TEM) images were
acquired using an FEI Tecnai G2 F30 X-TWIN operating at an accelerating
voltage of 300 kV. Superlattice samples were prepared by suspending the
superlattices in approximately 500 µL of MeCN and drop-casting the
suspension onto 400-mesh copper grids with an amorphous carbon support
film (Ted Pella). The MeCN was allowed to evaporate in air prior to
imaging. Dispersed nanocrystal (NC) samples were prepared by
drop-casting a dilute solution of colloidal NCs in NMF onto 400-mesh
copper grids with an amorphous carbon support (Ted Pella) that had been
treated with oxygen plasma (Harrick Plasma Cleaner, PDC-001) for 10
seconds. The NMF was evaporated under vacuum before imaging.

\textbf{TEM-SAED:} Transmission electron microscopy selected area
electron diffraction (TEM-SAED) images were collected using FEI Tecnai
G2 F30 X-TWIN at an accelerating voltage of 300 kV. The method for
sample preparation was identical to TEM.

\textbf{STEM-EDS:} STEM-EDS mapping were conducted under 300 kV using
the Analytical PicoProbe Electron Optical Beam Line, a prototype of the
ThermoFisher Spectra UltraX Illiad located at Argonne National
Laboratory. Microanalysis measurements of the elemental distribution
were obtained using the X-ray Perimeter Array Detector (XPAD) with a
capability over 4.5 sR was equipped, guaranteeing the ultrahigh
sensitivity for elemental analysis.\cite{4} For elemental
mapping shown in Figure 5E, Pb-L lines and Sn-L lines were used to map
Pb and Sn, respectively.

\textbf{HAADF-STEM:} High-angle annular dark field scanning transmission
electron microscopy (HAADF-STEM) images were acquired in an
aberration-corrected Thermo Fisher Scientific (TFS) Titan Themis
operated at an accelerating voltage of 300 kV. The imaging was performed
with the convergence semi-angle of 25 mrad, a HAADF detector collection
angle range of 52-200 mrad, a beam current of 41 pA, and a dwell time of
2 µsec per pixel. Samples were prepared by depositing dry powder on TEM
grids (Ted Pella, Prod\#01824; ultrathin carbon film on lacey carbon
support film, 400 mesh, Cu). A TFS low background double-tilt TEM holder
was used to tilt superlattices along the desired zone-axis.

\section{Analysis methods}

\subsection{Size distribution}

The size distribution of
PbS-Sn\textsubscript{2}S\textsubscript{6}\textsuperscript{4-} NCs was
determined from SAXS analysis of the corresponding NCs capped with oleic
acid ligands. Direct extraction of size distribution from the SAXS
patterns of
PbS-Sn\textsubscript{2}S\textsubscript{6}\textsuperscript{4-} NCs was
avoided because the
Sn\textsubscript{2}S\textsubscript{6}\textsuperscript{4-} ligands
consist of relatively heavy elements compared to the solvent
environment, scattering the X-ray and leading to an overestimation of NC
diameter. The size distribution was obtained in IgorPro9 using the
maximum entropy method via the Irena tool. The resulting number-weighted
size distributions were fitted with Gaussian functions to extract mean
diameters and standard deviations.

\subsection{Scherrer analysis of SAED patterns}

In this section, the selected area electron diffraction (SAED) pattern
shown in the inset of Figure 2A was analyzed to estimate the Scherrer
diameter of the
PbS-Sn\textsubscript{2}S\textsubscript{6}\textsuperscript{4-} NC
superlattice. Owing to the orientational alignment of the superlattice,
an extension of Scherrer analysis to the assembled structure was
anticipated. The center of the diffraction pattern was first determined
using the DiffTools plugin in Gatan DigitalMicrograph software. A linear
intensity profile passing through the center of the pattern (Figure
S12A) was then extracted to identify diffraction peaks. Scherrer
analysis was performed on the sharpest peak in the linear profile, as
indicated in Figure S12B.

The Scherrer diameter (\(D\)) is calculated from the full-width
half-maximum (FWHM) of the selected peak. The classical Scherrer
equation is given by:
\begin{equation}
D = \frac{K\lambda}{\beta\cos(\theta)}
\tag{4.1}
\end{equation}
Where \(K\) is the shape factor, \(\lambda\) is wavelength of the
incident beam, \(\beta\) is the FWHM of scattering angle in radians and
\(\theta\) is the Bragg angle.\cite{5} Since our measurement
is in reciprocal space in 1/\emph{d} scale rather than in angular space,
a modified form of the equation was employed, where \(\Delta q\) is the
FWHM in the reciprocal space. The shape factor, \(K\), can be estimated
as 0.93 for spherical particles.
\begin{equation}
D = \frac{K}{\Delta q}
\tag{4.2}
\end{equation}

From this analysis, a Scherrer diameter of 6.1 nm was obtained for the
PbS-Sn\textsubscript{2}S\textsubscript{6}\textsuperscript{4-} NC
superlattice. This value is larger than the estimated diameter of
individual PbS-Sn\textsubscript{2}S\textsubscript{6}\textsuperscript{4-}
NCs derived from powder X-ray diffraction shown in Figure S13 (PXRD)
(5.4 nm). The increase in Scherrer diameter upon self-assembly supports
the presence of oriented attachment of NCs within the superlattice. The
relatively small Scherrer diameter compared to the estimated
superlattice size (on the order of microns) may be attributed to
instrumental broadening, e-beam convergence, and to the emergence of
satellite peaks due to the periodic superlattice
structure.\cite{6} To confirm this, the diffraction patterns
of superlattices were modelled.

The Debye equation can be used to model the diffraction pattern of an
ensemble of identical randomly oriented particles. N atoms in each
particle will have its coordinates: \(r_{1}\), \(r_{2}\),\ldots{}
\(r_{N}\) and atomic form factor (structural amplitude) \(f_{1}\),
\(f_{2}\),\ldots{} \(f_{N}\). The amplitude of scattering is equal to
the sum of scattering amplitudes of each individual atom:

\begin{equation}
    F(\theta) = \sum_{j}^{N} f_j(\theta) e^{\frac{2\pi i}{\lambda}(\mathbf{s}-\mathbf{s}_0,\mathbf{r}_j)}
    \tag{4.3}
\end{equation}

Given that the intensity of scattering is $I(\theta) = F(\theta) \cdot F^*(\theta)$:

\begin{equation}
    I(\theta) = \sum_m \sum_n f_m(\theta) f_n(\theta) \exp \left( \frac{2\pi i}{\lambda} (\mathbf{s} - \mathbf{s}_0, \mathbf{r}_{mn}) \right)
    \tag{4.4}
\end{equation}

By integrating the wave vector in different directions, the intensity can be expressed as follows:

\begin{equation}
    \overline{I(\theta)} = \sum_m \sum_n f_m(\theta) f_n(\theta) \frac{\sin(2\pi S r_{mn})^*}{2\pi S r_{mn}}
    \tag{4.5}
\end{equation}

The angle-dependent atomic form factor in equation is calculated by Hartree-Fock wavefunctions by Cromer and Mann:

\begin{equation}
    f_m(\theta) = \exp \left( -B \sin^2\theta / \lambda^2 \right) \left[ \sum_{i=1}^{4} a_i \exp \left( -\frac{b_i \sin^2\theta}{\lambda^2} \right) + c \right]
    \tag{4.6}
\end{equation}

where $\lambda$ is wavelength of x-ray, B is Debye-Waller factor which is correlated with thermal vibrations of the atoms, $a_i$, $b_i$ \& $c$ are numerical fitting factors.\cite{7}

The lattice parameters of the simulated
PbS-Sn\textsubscript{2}S\textsubscript{6}\textsuperscript{4-} NC
superlattices were derived from experimental PXRD and SAXS data.
Scherrer analysis of the PXRD pattern (Figure S16) yielded NC dimensions
of 5.0 nm along the {[}100{]} direction and 4.6 nm along {[}111{]}
direction. SAXS analysis of the corresponding superlattice (Figure 5B,
blue line) indicated a lattice constant of 8.4 nm. These values suggest
that the edges of neighboring NCs are atomically precisely aligned,
consistent with the oriented order that were observed. The illustration
of the model is shown in Figure S17.

The diffraction pattern of the PbS-Sn\textsubscript{2}S\textsubscript{6}\textsuperscript{4-} NC superlattice was
simulated using the Debye equation, applied to a single face-centered
cubic (\emph{fcc}) unit cell. Figure S18 shows simulated scattering
patterns for NCs with dimensions of 5.0\,nm along the {[}100{]}
direction and 4.6\,nm along {[}111{]}, with lattice constants ranging
from 8.4\,nm to 20\,nm. Compared to the single-particle model, the
simulated pattern for a lattice constant of 8.4\,nm exhibits
significantly sharper peaks, indicating enhanced structural coherence.
As the lattice constant increases from 8.4\,nm to 8.7\,nm, the \{111\}
peak broadens and begins to split. At larger lattice constants, this
peak splitting becomes less pronounced, and no noticeable broadening is
observed at 20\,nm.

To validate the use of a single unit cell as a simulation model, the
effect of repeating units on the scattering pattern was examined. Figure
S19 shows simulated scattering patterns for one monolayer, one unit
cell, and two unit cells of the superlattice. The absence of significant
differences among the patterns supports the use of a single unit cell in
the simulations.

In summary, simulations of the PbS-Sn\textsubscript{2}S\textsubscript{6}\textsuperscript{4-} NC superlattice confirm that
self-assembly should lead to peak sharpening in the SAED pattern,
consistent with predictions from the Scherrer equation. However, the
emergence of satellite peaks in the superlattice may limit the extent of
peak sharpening.

\subsection{Wulff construction ratio from PXRD}

The Wulff construction ratio \emph{h}(111)/\emph{h}(100) for 5.7 nm
PbS-Sn\textsubscript{2}S\textsubscript{6}\textsuperscript{4-} NCs was
determined from the Scherrer size of the (111) and (200) peaks in the
PXRD pattern (Figure S6), calculated using SmartLab Studio II (Rigaku).
The Scherrer sizes were 55.7 Å and 54.4 Å, respectively, yielding a
Wulff ratio of 1.02.

\section{Field-theoretic simulations}

\subsection{Field theory for strong electrolytes}

\textbf{(A)} \textbf{Expression for the Hamiltonian.} The theoretical framework employed is a phenomenological Landau-Ginzburg
field theory, adapted for describing strongly correlated ionic systems.
\revision{The order parameter is the local charge density, defined
as~\(\phi\left( \mathbf{r} \right) = \left( \phi_{+}\left( \mathbf{r} \right) - \phi_{-}\left( \mathbf{r} \right) \right)/ \phi\),
where~\(\phi_{+}\left( \mathbf{r} \right)\)~and~\(\phi_{-}\left( \mathbf{r} \right)\)~are
the densities of cationic and anionic species weighted by their valency,
respectively, and $\phi$ is the bulk density}. This order parameter,$\phi (\mathbf{r})$, captures local
deviations from charge neutrality and its spatial fluctuations dictate
the system\textquotesingle s structural correlations. The effective
Hamiltonian, or free energy functional, for this order parameter is
given by
\begin{equation}
    \mathcal{H}\left\lbrack \{\phi(\mathbf{r})\} \right\rbrack = - a\int d\mathbf{r}\phi^{2}\left( \mathbf{r} \right) + m\int d\mathbf{r}\left| \nabla\phi\left( \mathbf{r} \right) \right|^{2} + Q^{2}\int\int d\mathbf{r}d\mathbf{r}'\frac{\phi\left( \mathbf{r} \right)\phi\left( \mathbf{r}' \right)}{\left| \mathbf{r} - \mathbf{r}' \right|}
 \tag{5.1}
\end{equation}

which incorporates three key contributions governing the behavior of the
electrolyte. The first two
terms~\(- a\int d\mathbf{r} \phi^{2}\left( \mathbf{r} \right)\) and
\(m\int d\mathbf{r} \left| \nabla\phi(\mathbf{r}) \right|^{2}\), with phenomenological
parameter~\(a\), \(m > 0\), represent the local free energy cost
associated with slow spatial variations in the charge density,
reflecting energetic costs associated with interfacial tension and other
entropic interactions that have correlations spanning short distances,
typically spanning the molecular diameter of the
electrolyte.\cite{8} The final
term,~\(Q^{2}\int\int d\mathbf{r}d\mathbf{r}'\frac{\phi\left( \mathbf{r} \right)\phi\left( \mathbf{r}' \right)}{\left| \mathbf{r} - \mathbf{r}' \right|}\)
, accounts for the long-range Coulomb interactions between charge
density fluctuations, where the coupling
strength~\(Q^{2} = q^{2}/\left( 4\pi\epsilon_{0}\epsilon \right)\)~depends
on the ionic charge~\(q\) and the solvent\textquotesingle s relative
dielectric constant~\(\epsilon\) \hspace{0pt}. Higher-order terms in the
order parameter expansion, are neglected, which is a valid approximation
assuming the system is studied above any bulk ordering temperature, such
that the harmonic description captures the essential fluctuations around
the disordered bulk liquid state. Consistent with the motivation
provided earlier, this Hamiltonian focuses on the interplay between
short-range correlations and electrostatics; explicit van der Waals
contributions are neglected in this analysis as the specific
orientational ordering is expected to be primarily dictated by the
longer-range forces that emerge from the anisotropic interactions of the
interfaces with the induced charge layering.

For analytical and computational convenience, the Hamiltonian is often
expressed in Fourier space:
\begin{equation}
\widehat{\mathcal{H}}\left\lbrack \{\phi\left( \mathbf{k} \right)\} \right\rbrack = \frac{1}{16\pi^{3}}\int d\mathbf{k\ }{\widehat{\chi}}^{- 1}(k)\left| \widehat{\phi}\left( \mathbf{k} \right) \right|^{2}
\tag{5.2}
\end{equation}

with susceptibility
\begin{equation}
\widehat{\chi}(k) = \left( - 2a + 2mk^{2} + \frac{8Q^{2}}{k^{2}} \right)^{- 1} = \left\lbrack 2a\left( l_{c}^{2}k^{2} + \frac{1}{l_{s}^{2}k^{2}} - 1 \right) \right\rbrack^{- 1}
\tag{5.3}
\end{equation}

where we note that phenomenological parameters \(a\), \(m\) and \(Q\)
can be related to physical length scales. In the long-wavelength limit
(\(k\) →0), the Coulomb term dominates, defining the Debye screening
length \(l_{s} = \sqrt{a/\left( 4\pi Q^{2} \right)}\) which
characterizes the exponential decay of correlations in the mean-field
limit. In the short-wavelength limit $k \rightarrow \infty$, the gradient term
dominates, defining a bare correlation length \(l_{c} = \sqrt{m/a}\),
representing the intrinsic length scale of short-range,
non-electrostatic correlations. The competition between these two length
scales,~\(l_{c}\)\hspace{0pt}~and~\(l_{s}\), dictates the nature of
charge correlations and screening in the system.

\textbf{(B)} \textbf{Susceptibility in real space.}

To understand the spatial structure of charge correlations predicted by
this model, we derive the real-space susceptibility,~\(\chi(r)\), that
describes the charge density response at position~\(r\)~due to a point
charge perturbation at the origin. This function can be computed by
performing an inverse Fourier transform
on~\(\widehat{\chi}(k)\ \)defined in Equation 5.3. We are interested in
a bulk theory that describes local order but not long-range
order,\revision{\cite{9,limmer2015interfacial, fredrickson1987surface}} which corresponds to the regime
\(l_{s}/l_{c} < \ 2\). Using the residue theory, the real-space
susceptibility~$\chi (r)$~characterized by~damped oscillations, takes the
form:
\begin{equation}
\chi(r) = A \exp( - r/\alpha) \cos(\gamma r/\alpha - \theta)\ /\ (4\pi r)\ 
\tag{5.4}
\end{equation}

Here, the parameters~\(\alpha\)~and~\(\gamma\)~govern the decay and
oscillation characteristics, respectively, while~\(\theta\)~is a phase
shift and~\(A\) is the amplitude. All these quantities are related to
the fundamental length scales \(l_{c}\) and \(l_{s}\) by:

\begin{equation}
\alpha = 2l_{c}l_{s}^{1/2}\left( 2l_{c} - l_{s} \right)^{- 1/2},\quad\gamma = \sqrt{\frac{2 + l_{s}/l_{c}}{2 - l_{s}/l_{c}}},\quad\theta = \sin^{- 1}(l_{s}/2l_{c}),\quad A = \frac{1}{al_{c}\sqrt{4l_{c}^{2} - l_{s}^{2}}}
\tag{5.5}
\end{equation}

This oscillatory form of \(\chi(r)\) mathematically represents the
phenomenon of charge layering where correlations decay with distance
\(r\) via an envelope function \(e^{- r/\alpha}\) while oscillating with
a characteristic wavelength given by \(2\pi\alpha/\gamma\). A
visualization of this oscillatory behavior and the influence of $\alpha$ and $\gamma$
is presented in \revision{Figure S18A and S18B}, which plots
\(4\pi rA^{- 1}\chi(r)\ \ \) to emphasize the structural oscillations
over the 1/\(r\) geometric decay. This susceptibility forms the basis
for understanding the effective interactions between the NCs mediated by
the electrolytes.

\subsection{Inclusion of Rigid body constraints}

To simulate the system in this work, the field theory described above
must be extended to incorporate the Pb NCs. We introduce terms to the
Hamiltonian representing the rigid NC bodies and their coupling to the
ionic environment mediated through the charged MCC ligands that are
grafted to the surface.

\textbf{(A)} \textbf{Expression for the modified Hamiltonian.} We assume
that the system comprises \(N\) NCs and denote \(\mathbf{r}_{j}\) and
\(\mathbf{R}_{j}\) as the position and orientation (expressed through
the rotation matrix) of the \(j^{th}\) NC respectively. The coupling
between the NC and the field theory is introduced through two additional
terms representing key physical eﬀects: (i) the asymmetric charge
density inside the nanocrystal is zero, that is imposed through a
quadratic penalty with Lagrange multiplier \(\lambda\):
\begin{equation}
\lambda\sum_{j}^{N}{\int_{\Omega_{j}}^{}d}\mathbf{r}\left| \phi\left( \mathbf{r} \right) \right|^{2}
\tag{5.6}
\end{equation}
where \(\int_{\Omega_{j}}^{}\) denotes the volume integral over the
\(j^{th}\) NC, and (ii) the interface of the nanocrystal attracts
negative charges, that is imposed through a linear term:
\begin{equation}
b\sum_{j}^{N}{\int_{\partial\Omega_{j}} {d \mathbf{r}\ }\phi}\left( \mathbf{r} \right)
\tag{5.7}
\end{equation}

where \(\int_{\partial\Omega_{j}}^{}\) denotes the surface integral over
the \(j^{th}\) NC and \(b > 0\) modulates the interaction strength. With
the inclusion of these two terms, the Hamiltonian in fourier space takes
the form:

\begin{equation}
\begin{split}
    \mathcal{H}[\{\phi(\mathbf{k})\}; \{\mathbf{r}_j\}, \{\mathbf{R}_j\}] = 
    & \quad \frac{1}{16\pi^3} \int d\mathbf{k} \, \hat{\chi}^{-1}(\mathbf{k}) \left| \hat{\phi}(\mathbf{k}) \right|^2 
    + \frac{\lambda}{64\pi^6} \iint d\mathbf{k}' d\mathbf{k} \, \hat{\phi}(\mathbf{k}) \hat{\phi}(\mathbf{k}') \sum_{j}^{N} \hat{\eta}_j (-\mathbf{k}-\mathbf{k}'; \mathbf{r}_j, \mathbf{R}_j) \\
    & + \frac{b}{8\pi^3} \int d\mathbf{k} \, \hat{\phi}(\mathbf{k}) \sum_{j}^{N} \hat{\Sigma}_j (-\mathbf{k}; \mathbf{r}_j, \mathbf{R}_j)
\end{split}
\tag{5.8}
\end{equation}

where \(\widehat{\chi}\left( \mathbf{k} \right)\) is the inverse of the
susceptibility in Fourier space defined in Equation 5.3 and
\({\widehat{\eta}}_{j}\left( \mathbf{k} \right)\) and
\({\widehat{\Sigma}}_{j}\left( \mathbf{k} \right)\) is the volume and
surface form-factor of the \(j^{th}\) NC respectively, computed via the
fourier transform of the volume and surface integrals defined in
Equation 5.6 and 5.7 respectively. By taking the functional derivative
of the Hamiltonian, the corresponding Euler-Lagrange equation for fixed
positions and orientations of the NCs is given by

\begin{equation}
{\widehat{\chi}}^{- 1}\left( \mathbf{k} \right)\widehat{\phi}\left( \mathbf{k} \right) + \sum_{j}^{N}\frac{\lambda}{4\pi^{3}}\left\lbrack \widehat{\phi} \circledast {\widehat{\eta}}_{j} \right\rbrack\left( \mathbf{k} \right) = - \sum_{j}^{N}{b{\widehat{\Sigma}}_{j}\left( \mathbf{k} \right)\ }
\tag{5.9}
\end{equation}

where \(\circledast\) denotes the convolution operator, and net
neutrality is imposed by disregarding the \(k = 0\) term. While the
Euler-Lagrange equations can be solved through a linear solver to get
the free-energies, the accuracy of numerical calculations strongly
resides in the resolution of the interface and the volume form factors.
The Hamiltonian is highly sensitive to the form factors, resulting in
strong artifacts for different orientations when standard
interpolation-based methods are used. To resolve this issue, we derive
analytical equations for the form factors of truncated octahedra.

\textbf{(B)} \textbf{Interpolation free representation using analytical
form factors}. Accurately representing the truncated octahedral NC
geometry within the Fourier space Hamiltonian (Equation 5.8) and the
corresponding Euler-Lagrange equation (Equation 5.9) requires
computation of the volume and surface form factors,
\({\widehat{\eta}}_{j}\left( \mathbf{k} \right)\) and
\({\widehat{\Sigma}}_{j}\left( \mathbf{k} \right)\). Direct calculation
of these factors on a discrete grid can introduce interpolation
artifacts, particularly when representing arbitrarily oriented NCs that
are not aligned with the grid's axes. To overcome this, we employ
analytical expressions for the form factors derived from the NC's
geometry.

The surface of the truncated octahedron is composed of 6 square faces
and 8 hexagonal faces. The total surface form factor
\(\widehat{\Sigma}_j\left( \mathbf{k} \right)\) for an \revision{NC indexed by $j$ centered at the
origin is computed by summing the analytical Fourier transforms of each of its individual faces $l$, denoted
\({\widehat{\Sigma}}_{j}\left( \mathbf{kl} \right)\):}
\begin{equation}
\widehat{\Sigma_{j}}\left( \mathbf{k} \right) = \sum_{l \in \partial\Omega_{j}}^{}{\widehat{\Sigma}}_{jl}\left( \mathbf{k} \right)\widehat{G}\left( \mathbf{k} \right)
\tag{5.10}
\end{equation}

The specific analytical form of
\({\widehat{\Sigma}}_{jl}\left( \mathbf{k} \right)\) can be obtained
analytically for both squares and hexagons\revision{\cite{wuttke2021numerically, lee1983fourier}} and depend on the distance
from the center of the face from the origin and the relative orientation
of the face with respect to the grid's principal axis.
\(\widehat{G}\left( \mathbf{k} \right)\) is a Gaussian smearing
function, defined as:
\begin{equation}
\widehat{G}\left( \mathbf{k} \right) = \exp\left( - \sigma^{2}\left| \mathbf{k} \right|^{2}/2 \right)
\tag{5.11}
\end{equation}

where \(\sigma = 2\mathring{\mathrm{A}}\) is the standard deviation in real space
chosen to avoid numerical instabilities due to Gibbs oscillations. All
these terms are calculated by first computing the relative distance to
the truncated octahedra's center and the relative orientation of the
faces with respect to the principal axis of the truncated octahedra, and
then by computing the relative position \(\mathbf{r}_{j}\) and
orientation \(\mathbf{R}_{\mathbf{j}}\) of the \(j^{th}\) octahedra with
respect to the grid.

The volume form factor,
\({\widehat{\eta}}_{j}\left( \mathbf{k} \right)\), is obtained from the
surface form factors using the divergence theorem:\revision{\cite{wuttke2021numerically}}
\begin{equation}
\widehat{\eta_{j}}(k) = \frac{i}{\left| \mathbf{k} \right|^{2}}\sum_{l \in \partial\Omega_{j}}^{}\left( \mathbf{k} \cdot {\widehat{\mathbf{n}}}_{jl} \right){\widehat{\Sigma}}_{jl}\left( \mathbf{k} \right)\widehat{G}\left( \mathbf{k} \right)
\tag{5.12}
\end{equation}

where \({\widehat{\mathbf{n}}}_{jl}\) is the outward unit normal vector
of face \(l\), that depends on the position and orientation of the of
the NC. For \(\mathbf{k} = 0\), \({\widehat{\eta}}_{j}(0)\) corresponds
to the volume of the truncated octahedron. This analytical,
interpolation-free approach allows for accurate and efficient evaluation
of the rigid body constraints for any NC position and orientation within
the simulation framework.

\subsection{Computational methods}

To explore the thermodynamic landscape governed by the Hamiltonian, our
computational approach involves the minimization of the charge density
field and the sampling of nanoparticle configurations.

\textbf{(A)} \textbf{Minimization of charge density field.} For a given
configuration of nanocrystals, defined by their positions
\(\{\mathbf{r}_{j}\}\) and orientations \(\{\mathbf{R}_{j}\}\), the
equilibrium charge density field
\(\widehat{\phi}\left( \mathbf{k} \right)\) is determined by numerically
solving the Euler-Lagrange equation in Equation 5.9. The system is
discretized onto a periodic cubic grid, with a box size \(L\)
corresponding to the desired physical dimensions. The grid spacing
\(dx = L/N_{g}\) is chosen to be approximately
\(0.8\mathring{\mathrm{A}} - 1\mathring{\mathrm{A}}\), which is more than sufficient to
converge free energy estimates up to \(0.01\%\) for the smallest
wavector chosen.

We solve the Euler-Lagrange equation, Equation 5.9, efficiently in
Fourier space, which allows for the straightforward application of the
analytical form factors described previously and the use of Fast Fourier
Transforms (FFTs) for evaluating convolution terms. We employ the CUDA
accelerated matrix free Conjugate Gradient (CG) algorithm implemented in
CuPy to solve the resulting linear set of equations. To accelerate
convergence, a preconditioner based on the Fourier space susceptibility,
\(\widehat{\chi}\left( \mathbf{k} \right)\), is applied during the CG
iterations . The linear operator within the CG method incorporates the
inverse susceptibility
\({\widehat{\chi}}^{- 1}\left( \mathbf{k} \right)\) and the volume
constraint term, which involves the convolution
{[}\(\widehat{\phi} \circledast {\widehat{\eta}}_{j}\rbrack(\mathbf{k)}\)
implemented via FFTs .

The iterative process continues until the residual norm falls below a
specified tolerance threshold, set to \(10^{- 6}\) for all calculations.
The resulting field \({\widehat{\phi}}^{*}\left( \mathbf{k} \right)\)
represents the equilibrium charge density distribution for the specified
nanocrystal arrangement and is subsequently used for calculating the
system's free energy at a fixed position and orientation of the NCs. The
code for the field theory implementation is available online.

\textbf{(B)} \textbf{Free energy computation for 2 nanocrystals.} To
characterize the effective interaction potential between a pair of NCs
mediated by the strong electrolytes, we performed calculations using the
field theory described above. The system consisted of two truncated
octahedral NCs, each with a diameter of 57 Å, placed within a cubic
simulation box of side length \(L = 200\) Å, discretized onto a
\(200 \times 200 \times 200\) grid. For the field parameters, we used
\(\alpha = 3.0\mathring{\mathrm{A}}\) and \(\gamma = 2.1\), which allows us to
uniquely define the physical length scales \(l_{c}\) and \(l_{s}\). The
amplitude of the density waves is dictated by the ratio between the
strength of the source term at the interface \(b\) to the harmonic term
\(a\) in the Hamiltonian. For all simulations, we use
\(b/a = 100\mathring{\mathrm{A}}\). Additionally, the magnitude of the Lagrange
multiplier is chosen to be \(\lambda/a = 500\), which is sufficient to
prevent charge penetration into the volume of the NC up to \(2 - 4\%\)
of the maximum density observed. The energy scale is set by defining the
reduced temperature \(k_{\mathrm{B}}T = 1\), and the strength of the source term
\(b{/k}_{B}T = 100{\mathring{\mathrm{A}}}^{- 2}\).

For this two-particle system, the center-to-center separation distance,
\(r\), was systematically varied between \(62\mathring{\mathrm{A}}\) to
\(95\mathring{\mathrm{A}}\). At each separation distance, the free energy of the
system was computed for different relative orientations of the NCs,
focusing specifically on comparing configurations where the NCs were
aligned versus non-aligned. The equilibrium charge density field
\({\widehat{\phi}}^{*}\left( \mathbf{k} \right)\) for each specific
configuration (separation \(r\), relative orientation
\(\mathbf{R}_{1},\mathbf{R}_{2}\)) was obtained using the preconditioned
CG solver. The total free energy for both orientation was then
calculated from this field, summing the relevant contributions
(harmonic, gradient, Coulomb, interface, and constraint terms) and
subtracting it from the total free energy at the largest distance of
separation for the aligned case. At the largest distance of separation
considered, the absolute free energy difference between the aligned and
nonaligned system was found \(0.003\) \(k_{\mathrm{B}}T\), \(< 0.02\%\) of the
free energy difference between the free energy at the minima of the
aligned state, correctly reflecting the physical result that the
Hamiltonian of non-interacting NCs is invariant to the orientation of
the NC.

\textbf{(C)} \textbf{Phase diagram for orientational order.} To
determine the thermodynamically preferred orientational phases of the
NCs within an FCC superlattice structure, we performed Monte Carlo (MC)
simulations as a function of the electrolyte screening parameters
\(\alpha\) and reduced wavevector \(\gamma\). The simulations modeled a
system of four truncated octahedral NCs placed on the sites of a single
FCC unit cell with lattice constant \(90\mathring{\mathrm{A}}\). The parameter
space scanned covered \(\alpha\) from \(1.8\mathring{\mathrm{A}}\) to
\(4.0\mathring{\mathrm{A}}\) in steps of \(0.1\mathring{\mathrm{A}}\) and \(\gamma\) from
1.1 to 4.0 in steps of 0.1. The grid size \(N_{g}\) was chosen between
100 to 120, depending on the value of \(\gamma/\alpha\) that sets the
Nyquist limit. As before, the relative strength of the source term and
the Lagrange multiplier is set through the ratios
\(b/a = 100\mathring{\mathrm{A}}\) and \(\lambda/a = 500\). Finally, the energy
scale is set through the reduced temperature, \(k_{\mathrm{B}}T = 50\) and the
magnitude of the source term \(b/k_{\mathrm{B}}T = 2{\mathring{\mathrm{A}}}^{- 2}\),
chosen to allow for efficient sampling of the system in a regime where
thermodynamics governing the orientational order is driven by the free
energetics of charge layering, while enabling efficient sampling of the
phase space without getting stuck in trapped states.

For each \((\alpha,\gamma)\) pair, an independent MC simulation was
performed to sample the orientational configurations of the four NCs,
where the initial orientations were drawn from the Haar distribution. At
each MC step, a single NC with index \(j\) was chosen at random, and its
orientation matrix \(\mathbf{R}_{j}\) was perturbed to generate a trial
orientation \(\mathbf{R}_{j}'\). This rotational move was implemented by
generating a \(3 \times 3\) skew-symmetric matrix \(\mathbf{W}\) from a
random 3D vector drawn from an isotropic Gaussian distribution with
standard deviation \(\epsilon = 0.2\), and then applying the rotation
\(\mathbf{R}_{j}' = e^{\mathbf{W}}\mathbf{R}_{j}\). Each simulation
consisted of 40,000 such rotational MC steps. The trial move was
accepted based on the standard Metropolis criterion using the change in
total free energy, \(\Delta F\), calculated via the field theory solver.
The acceptance rate was found to be 0.3 at the start of the simulation
and decreased to \textless0.1 as the system approached an
orientationally ordered state.

To quantify the degree of alignment, we defined a single-particle order
parameter, \(s_{j}\), for the \(j^{th}\) NC based on its orientation
matrix \(R_{j}\) relative to the simulation box axes

\begin{equation}
s_{j} = \left( \sum_{l = 1}^{3}\max_{m = 1,2,3}\left| \left( \mathbf{R}_{j} \right)_{lm} \right| \right) - 2
\tag{5.13}
\end{equation}

This parameter measures the alignment of the NC's primary axes
(represented by the columns of \(R_{j}\)) with the Cartesian axes of the
simulation frame, accounting for all possible symmetry elements of the
octahedral group \(O_{h}\). The average order parameter for the system,
\(\langle s\rangle = (1/N)\sum_{j = 1}^{N}s_{j}\), was calculated at
each MC step. After discarding the initial portion of the trajectory for
equilibration, the equilibrium value \(\langle s\rangle\) for each
\((\alpha,\gamma)\) point was obtained by averaging \(s\) over the final
10,000 MC steps.

The orientational phase diagram (presented as Figure 4B in the main
text) was constructed by plotting the equilibrium average order
parameter \(\langle s\rangle\) as a function of \(\alpha\) and
\(\gamma\). For all cases of (\(\alpha\) and \(\gamma\)) considered, we
found that the order parameter plateaued around 3 different values of
\(\langle s\rangle\), given by \(\langle s\rangle \approx 0.45\),
\(\langle s\rangle \approx 0.65\) and \(\langle s\rangle \approx 0.95\).
This is observed from Figure S19A, where we plot the average order
parameter as a function of \(\alpha\) and \(\gamma\). By visualizing the
orientations at different values of \((\gamma,\alpha)\), we found that
the typical orientations centered around the three different values of
\(s\) were similar to each other, as shown in Figure 4C in the main
text. From these results, we chose the phase-boundaries through
\(0.6 \leq \langle s\rangle \leq 0.85\) (diagonally aligned),
\(\langle s\rangle \leq 0.6\) (face-to-face aligned) and
\(\langle s\rangle > 0.85\) (edge-to-edge aligned). An additional step
of Gaussian smoothing was performed to get the final phase diagram shown
in Figure 4B.

The low effective temperature used for performing the MC simulation
ensured that the resulting phase diagram primarily reflects the
energetically preferred orientational order dictated by the free energy
of the charge-layering. As a final step of validation, we compared the
average free energies of some representative configurations sampled
within each identified phase region and then compared the relative free
energies between them. The phase diagram constructed by choosing the
phase with the minimum free energy, plotted in Fig. S19B.

\textbf{(D)} \textbf{Orientational order in a 32-site FCC Lattice.} To
ensure that our results, especially the emergence of edge-to-edge
alignment at large screening length is not affected by finite size
effects, we consider a \(2 \times 2 \times 2\) FCC lattice with the same
lattice constant of \(90\mathring{\mathrm{A}}\) as before that contains 32 NCs.
For this system, we chose \(\alpha = 3.0\mathring{\mathrm{A}}\) and
\(\gamma = 1.1\), for which the 4-site results predicted the
edge-to-edge aligned state to be favorable. The remaining parameters are
chosen as \(N_{g} = 180\), \(b/a = 100\mathring{\mathrm{A}}\),
\(\lambda/a = 500\), \(k_{\mathrm{B}}T = 50\) and
\(b/k_{\mathrm{B}}T = 2{\mathring{\mathrm{A}}}^{- 2}\). As before, the orientations of all
NCs are drawn randomly from the Haar distribution, and 50,000 Monte
Carlo steps are performed to obtain the orientational order of the NCs
in the super-lattice. Representative snapshots illustrating the
system\textquotesingle s evolution are presented chronologically in Fig.
S19B. The simulation progresses from an initial random configuration
towards states with increased orientational order and pronounced charge
layering between the NCs. The final configuration after 50,000 MC steps,
shown in Fig. S19B, clearly exhibits the edge-to-edge alignment
predicted by the 4-site simulations and observed experimentally (Figure
4C main text), validating that our results are unaffected by finite size
effects. The quantitative evolution towards this state is shown in Fig.
S19B and S19B, which plot the average order parameter $\langle s \rangle$ (computed
using Equation 5.13 averaged over 32 NCs) and the system free energy
\({F/k}_{B}T\) versus MC steps, respectively. The order parameter
rapidly increases towards $\langle s\rangle \approx 1$, while the free energy decreases
and plateaus, indicating relaxation into the stable edge-to-edge aligned
structure.

Finally, to ensure the stability of the observed phase is robust against
the specific choice of numerical parameters, we performed additional
validation simulations. Fig. S20E and S20F show the evolution of $\langle s \rangle$ and
\({F/k}_{B}T\) for an identical simulation where the Lagrange multiplier
enforcing the volume constraint was doubled ($\lambda/a=1000$). The
system clearly relaxes to the same high-order edge-to-edge state,
confirming the result is not sensitive to the strength of this
constraint.

\section{Supporting Figures}

\includegraphics[width=6.5in,height=2.88958in,alt={A graph of normalized number of numbers AI-generated content may be incorrect.}]{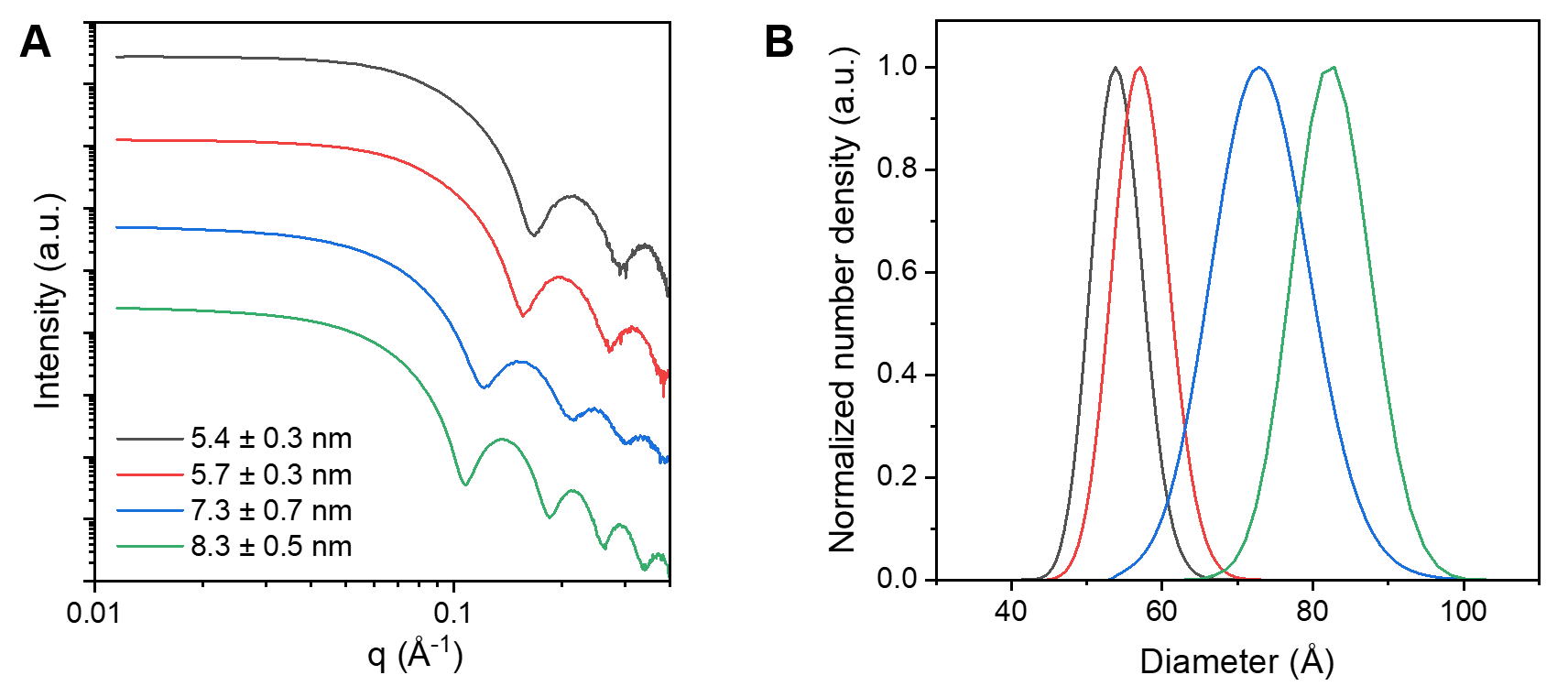}

\includegraphics[width=3.6132in,height=2.85118in,alt={A graph of different colored lines AI-generated content may be incorrect.}]{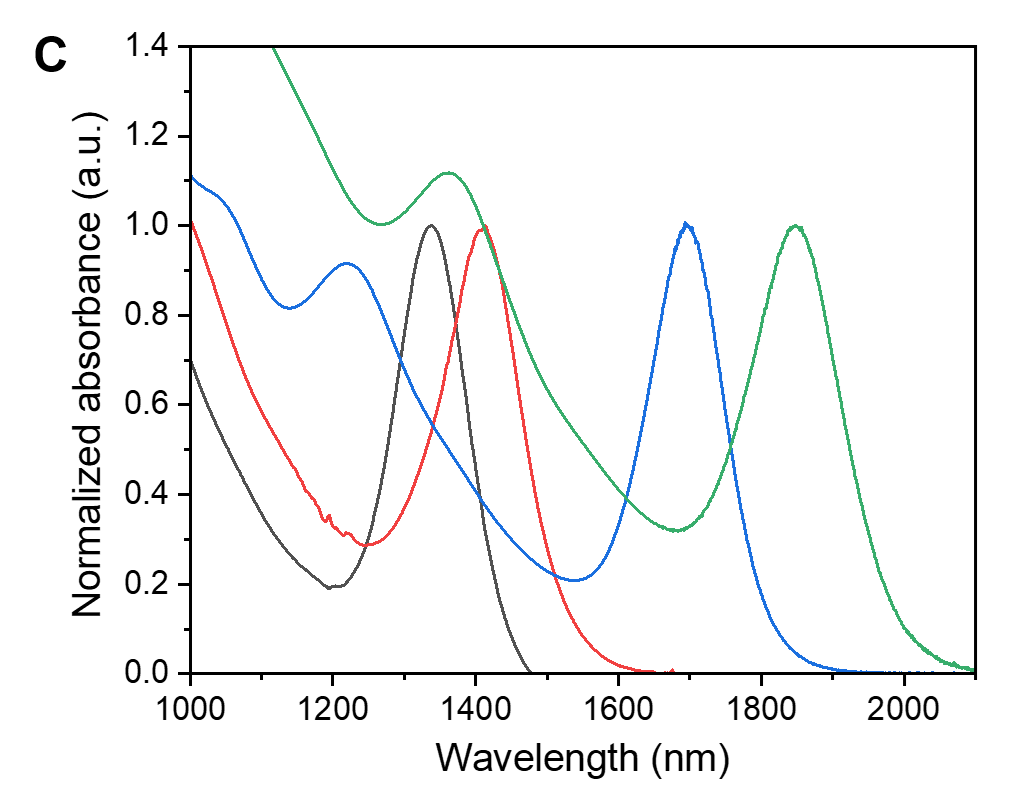}

\textbf{Figure S1. (A)} SAXS patterns and \textbf{(B)} size distribution
of 5.4, 5.7, 7.3 and 8.3 nm PbS-OA NCs. The standard deviations of the
diameter are 0.3, 0.3, 0.7 and 0.5 nm, respectively. Each size
distribution was obtained by fitting the form factor of the
corresponding SAXS pattern (see Section 4.1). \textbf{(C)} UV-Vis
spectra of the PbS-OA NCs.

\includegraphics[width=6.5in,height=3.10139in,alt={A comparison of different colored lines AI-generated content may be incorrect.}]{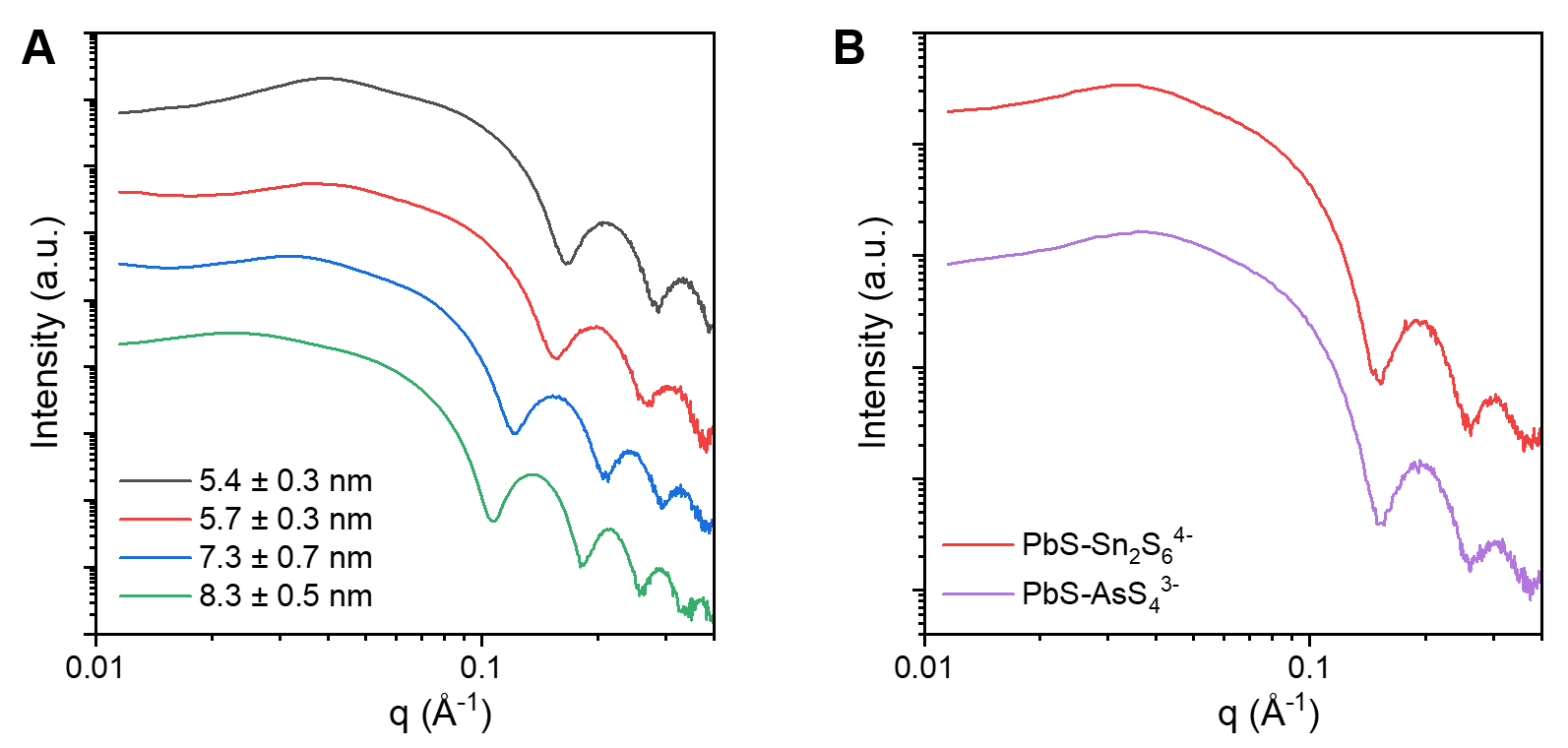}

\textbf{Figure S2.} (A) SAXS patterns of
PbS-Sn\textsubscript{2}S\textsubscript{6}\textsuperscript{4-} NCs with
various sizes and (B) SAXS patterns of 5.7 nm PbS NCs with various metal
chalcogenide complex (MCC) ligands.

\includegraphics[width=6.68954in,height=2.73085in,alt={A graph of different colored lines AI-generated content may be incorrect.}]{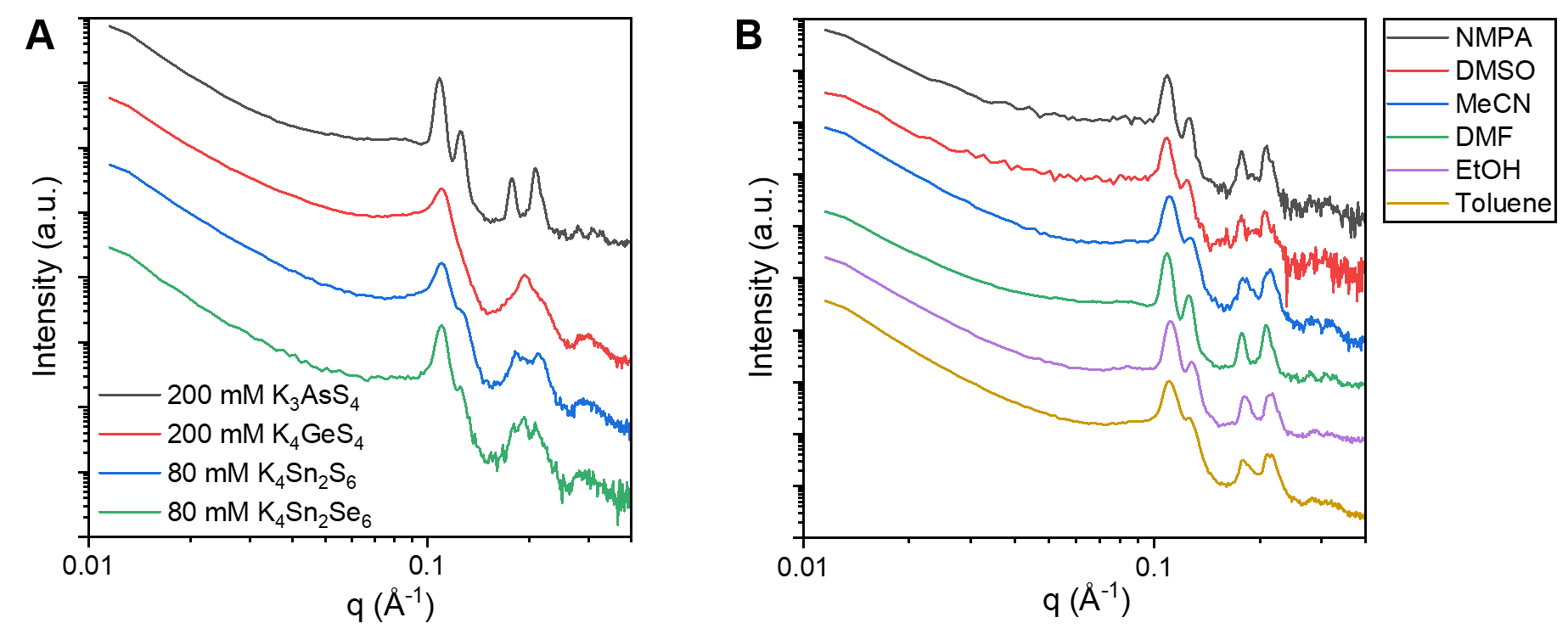}

\textbf{Figure S3. (A)} SAXS patterns of 5.7 nm
PbS-Sn\textsubscript{2}S\textsubscript{6}\textsuperscript{4-} NCs that
were self-assembled by adding 40\% volume fraction of DMF as non-solvent
and various salts as flocculants. \textbf{(B)} SAXS patterns of 5.7 nm
PbS-Sn\textsubscript{2}S\textsubscript{6}\textsuperscript{4-} NCs that
were self-assembled by adding 200 mM of
K\textsubscript{3}AsS\textsubscript{4} as flocculant and 40\% volume
fraction of various organic solvents as non-solvents.

\includegraphics[width=3.38719in,height=3.00323in,alt={A graph of a graph showing a number of light AI-generated content may be incorrect.}]{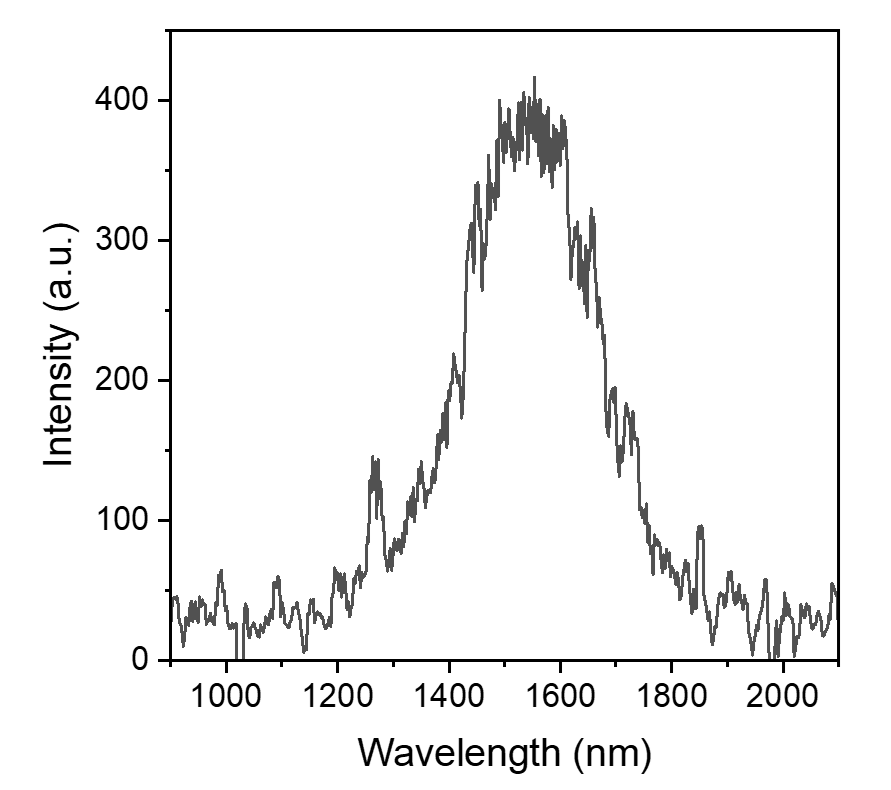}

\textbf{Figure S4.} PL spectrum of 5.7 nm
PbS-Sn\textsubscript{2}S\textsubscript{6}\textsuperscript{4-} NC
superlattice.

\includegraphics[width=6.5in,height=3.01042in,alt={A graph of different colored lines AI-generated content may be incorrect.}]{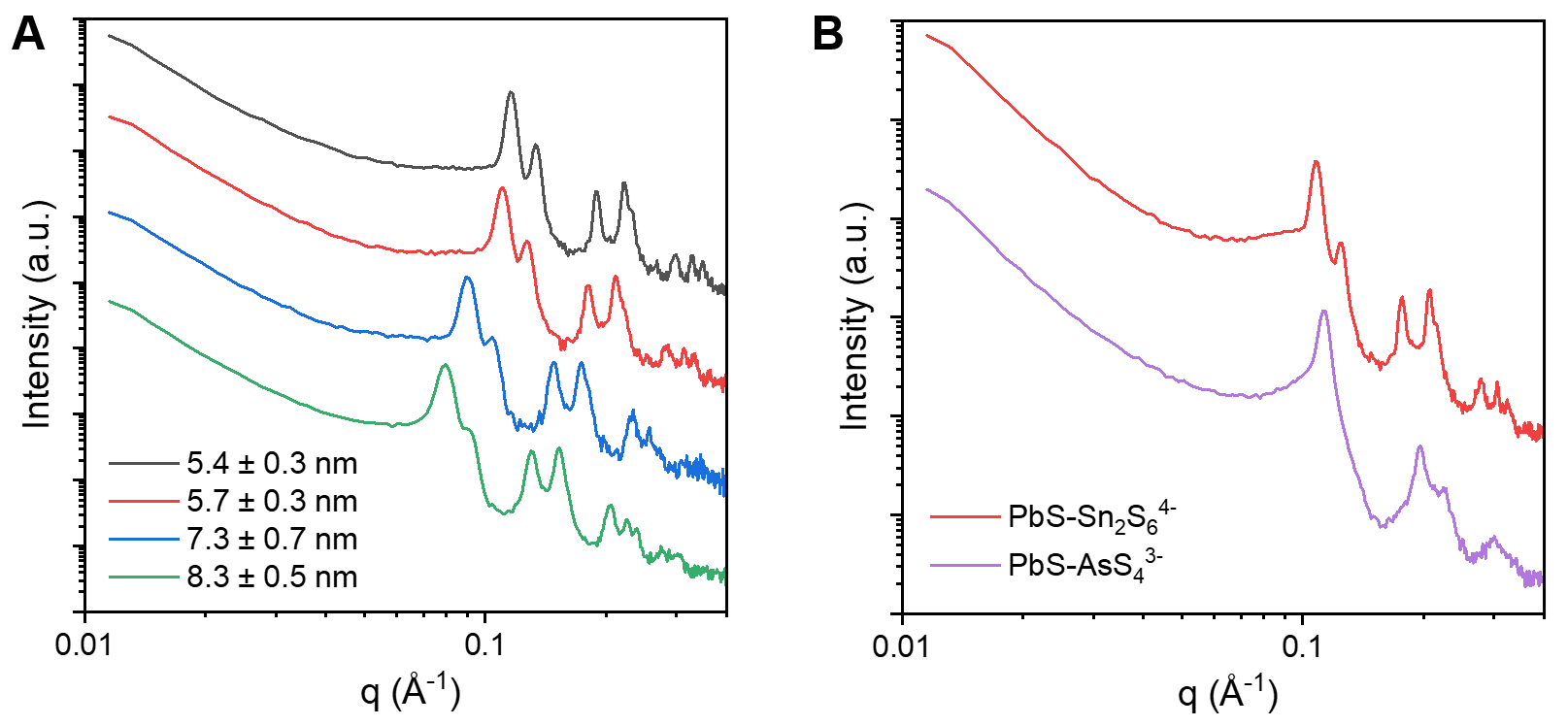}

\textbf{Figure S5. (A)} SAXS pattern of superlattices of
PbS-Sn\textsubscript{2}S\textsubscript{6}\textsuperscript{4-} NCs with
various sizes. \textbf{(B)} SAXS pattern of superlattices of 5.9 nm PbS
NCs with various inorganic ligands.

\includegraphics[width=6.5in,height=2.87639in,alt={A graph of a graph and a graph of a graph AI-generated content may be incorrect.}]{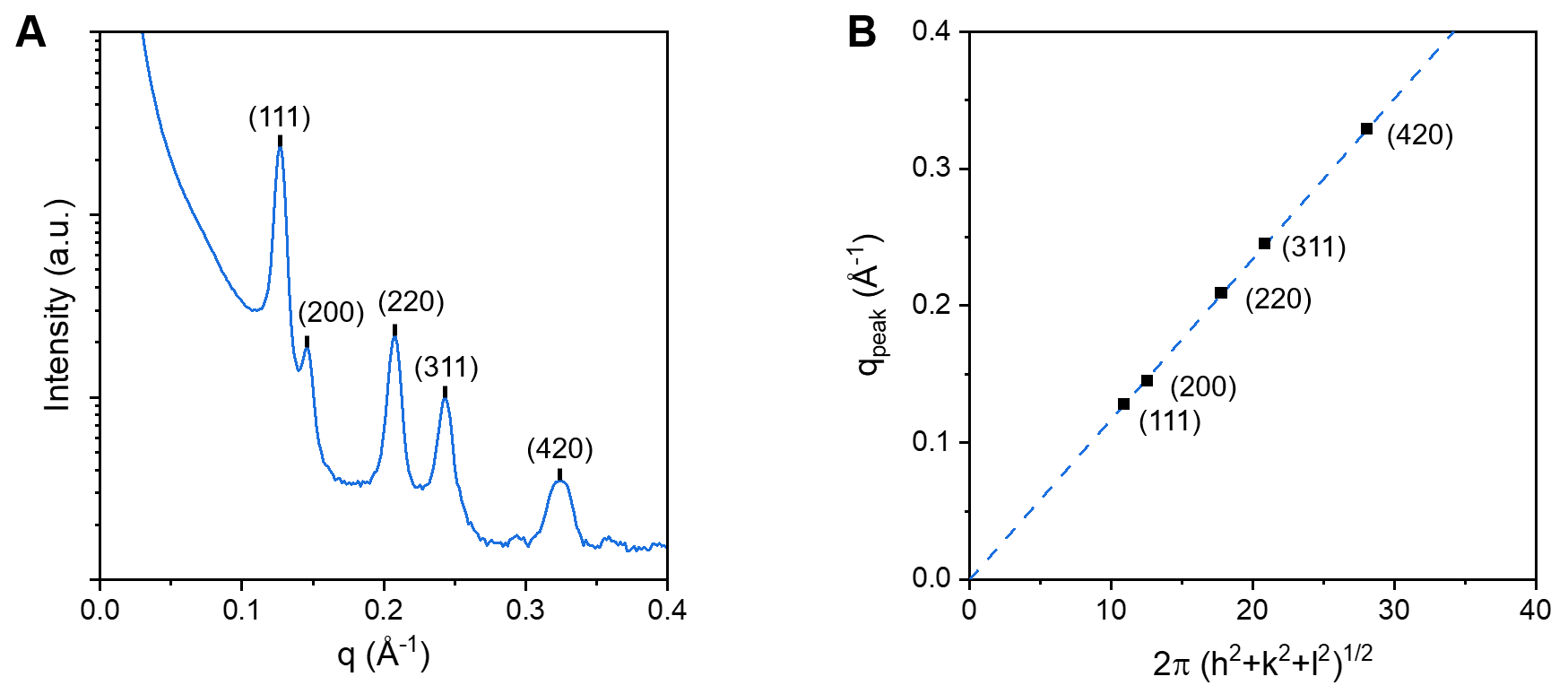}

\textbf{Figure S6.} Plot of SAXS pattern (A) and the plot of
\emph{q}\textsubscript{peak} versus
2$\pi$(\emph{h}\textsuperscript{2}+\emph{k}\textsuperscript{2}+\emph{l}\textsuperscript{2})\textsuperscript{1/2}
of the superlattice of 5.7 nm
PbS-Sn\textsubscript{2}S\textsubscript{6}\textsuperscript{4-} NCs after
washing with MeCN. The assigned Miller indices satisfy the selection
rule for an \emph{fcc} crystal structure (where~\emph{h},~\emph{k},
and~\emph{l}~are either all odd or all even). The data points align
closely with a straight line passing through the origin, confirming that
the superlattices exhibit an \emph{fcc} structure.

\includegraphics[width=3.08666in,height=3.13678in,alt={A pen with numbers and lines AI-generated content may be incorrect.}]{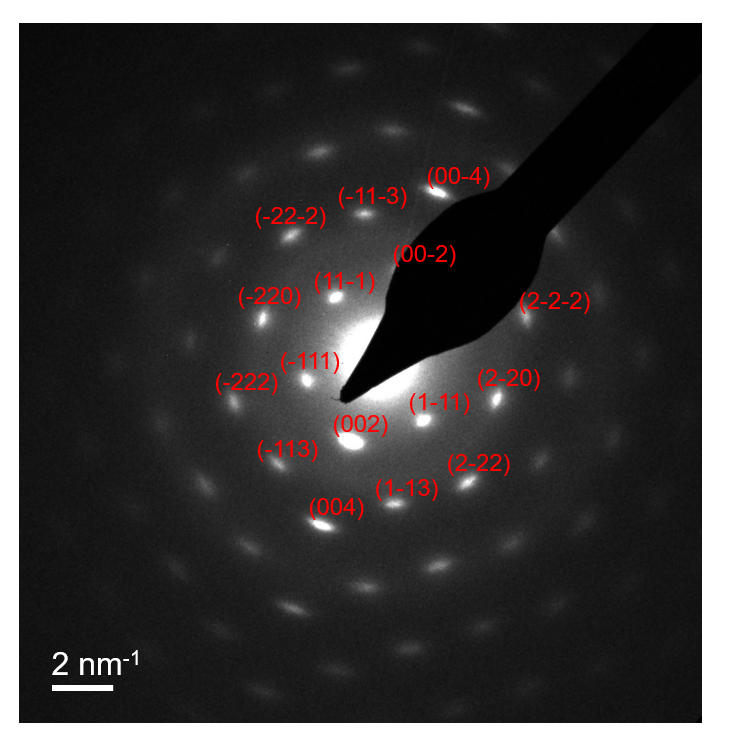}

\textbf{Figure S7.} Spot assignment of an SAED image of 5.7 nm
PbS-Sn\textsubscript{2}S\textsubscript{6}\textsuperscript{4-} NC
superlattice, shown in main text Figure 1C.

\includegraphics[width=6.38354in,height=2.98103in,alt={A close-up of a black and white photo AI-generated content may be incorrect.}]{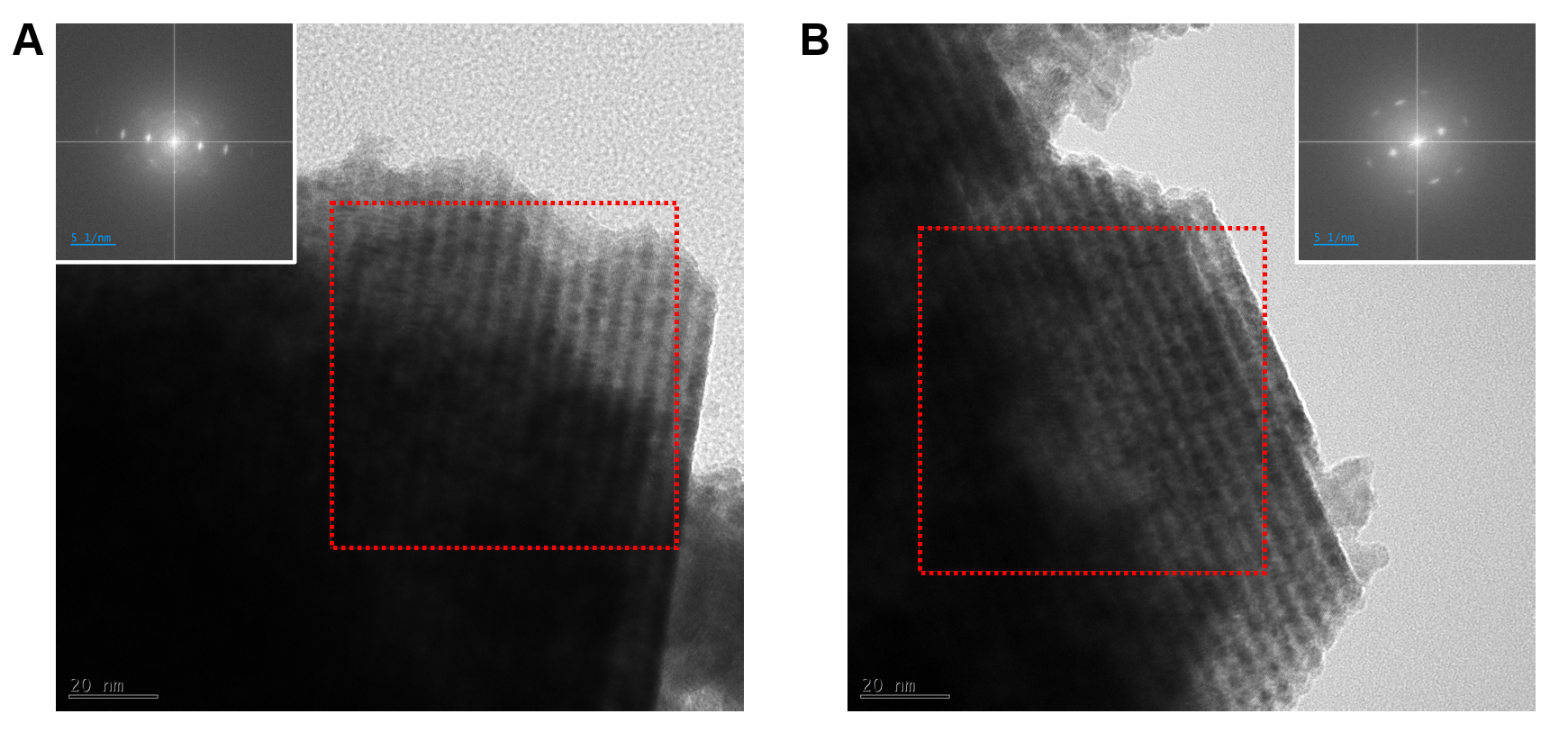}

\textbf{Figure 8.} TEM image of \textbf{(A)} as-prepared and
\textbf{(B)} washed 5.7 nm
PbS-Sn\textsubscript{2}S\textsubscript{6}\textsuperscript{4-} NC
superlattices. Fourier transformed images of the areas highlighted in
red boxes are shown in the inset.

\includegraphics[width=6.5in,height=2.16111in,alt={A close-up of a greyscale image AI-generated content may be incorrect.}]{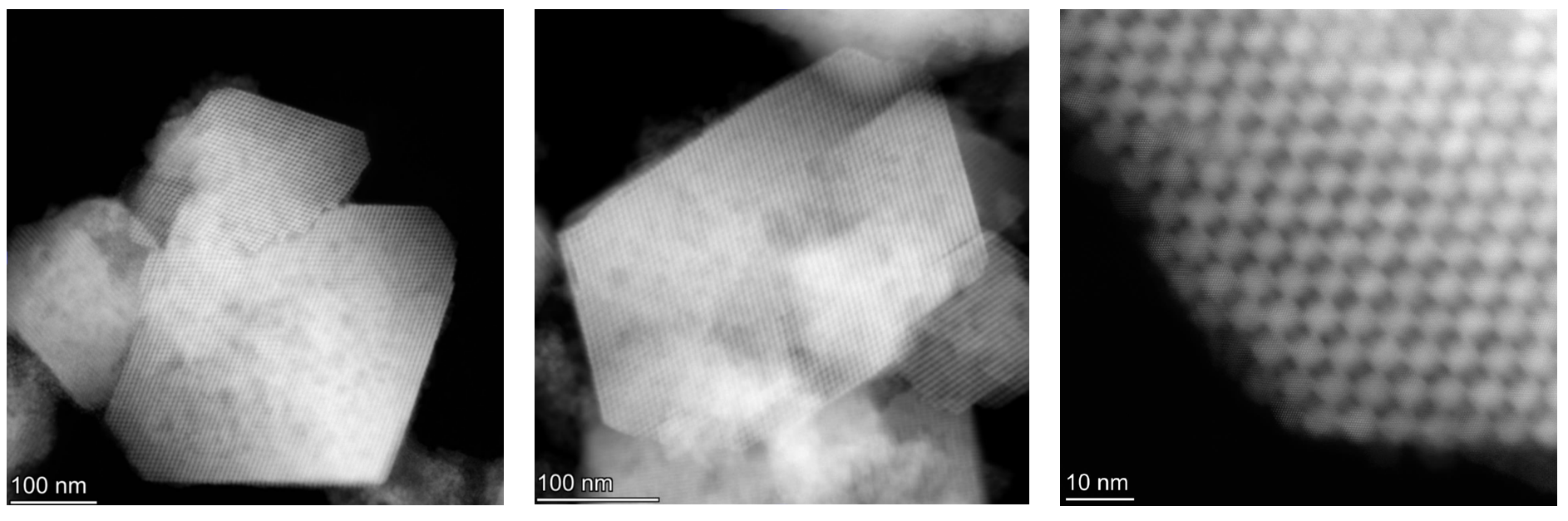}

\includegraphics[width=6.5in,height=2.15208in,alt={A close-up of a black and white image AI-generated content may be incorrect.}]{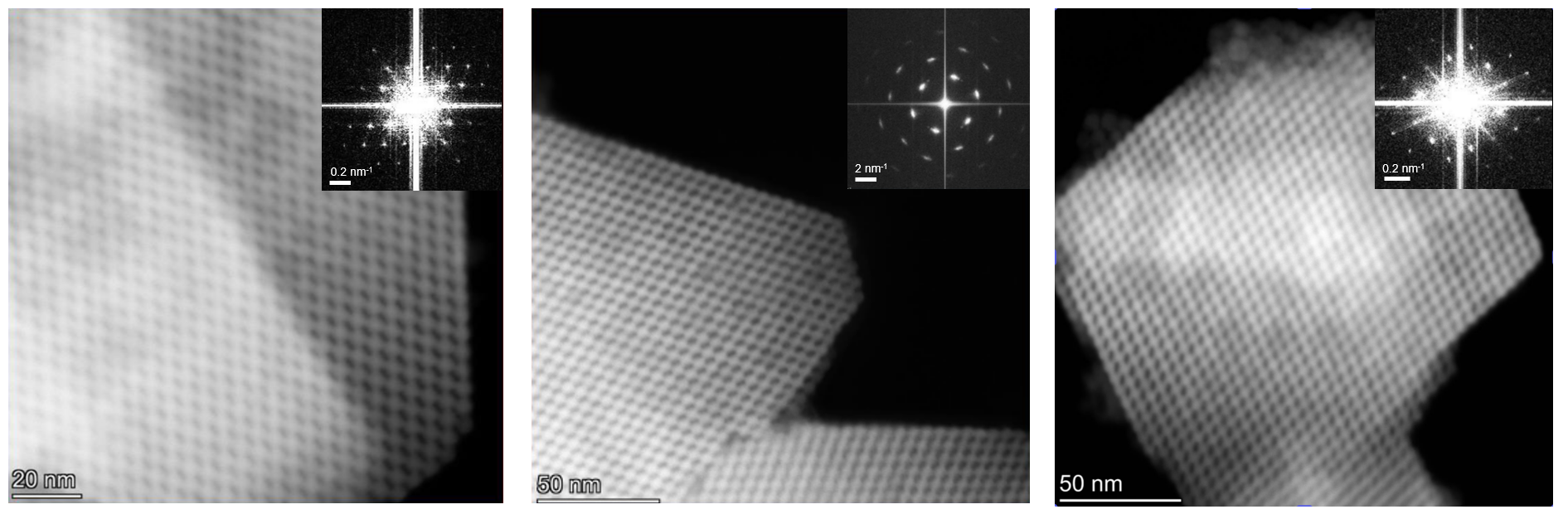}

\textbf{Figure S9.} HAADF-STEM images of 5.7 nm
PbS-Sn\textsubscript{2}S\textsubscript{6}\textsuperscript{4-} NC
superlattices. Insets: Fourier transformed image of the corresponding
STEM images.

\includegraphics[width=6.5in,height=2.99375in,alt={A close-up of a microscope Description automatically generated}]{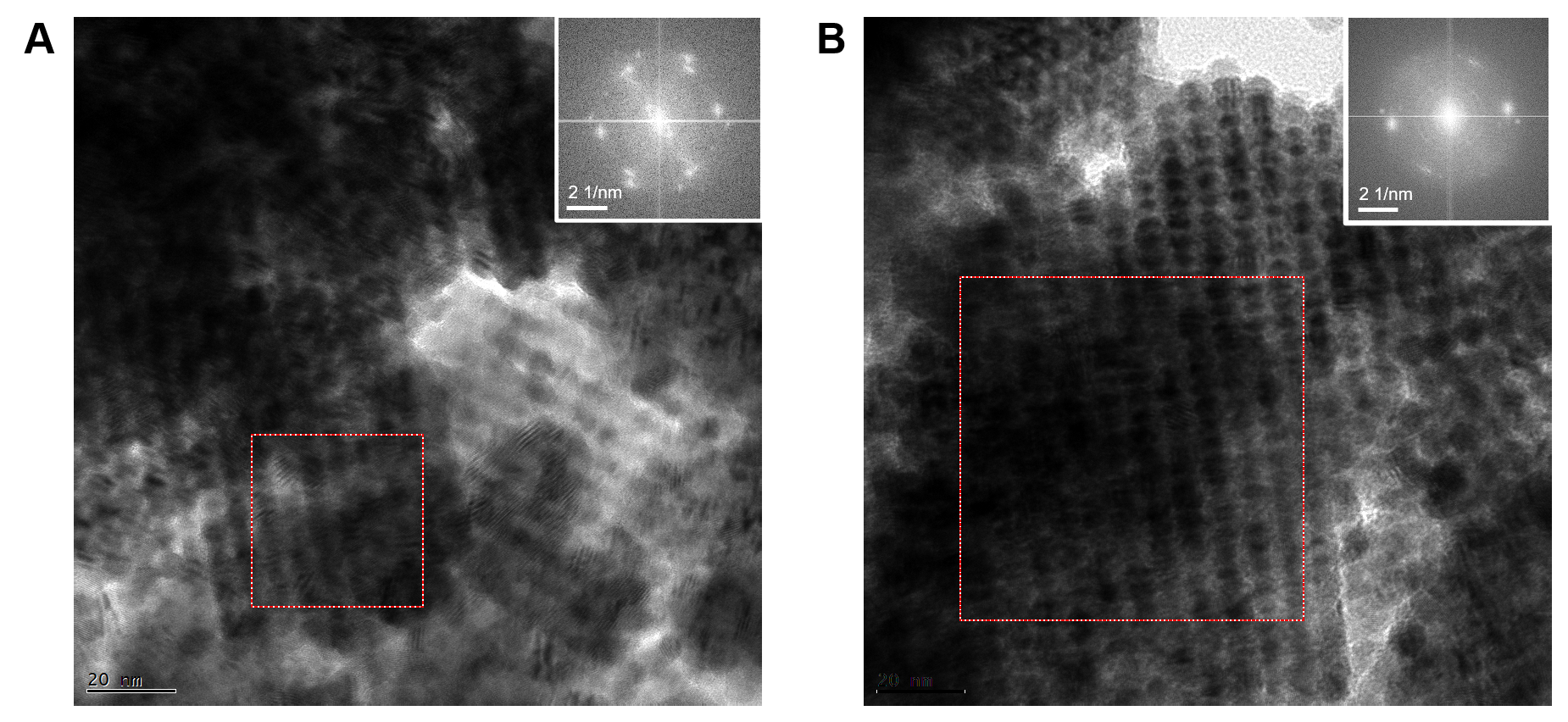}

\textbf{Figure S10.} TEM images of \textbf{(A)} 7.3 nm and \textbf{(B)}
8.3 nm PbS-Sn\textsubscript{2}S\textsubscript{6}\textsuperscript{4-} NC
superlattices. Inset: Fourier transformed image of the region enclosed
by red dashed box. The distinct spot patterns in wide-angle FT-TEM
images shown in the inset indicate that there is a strong preference for
the NCs to be orientationally aligned within the superlattice.

\includegraphics[width=6.5in,height=2.99236in,alt={A close-up of a microscope AI-generated content may be incorrect.}]{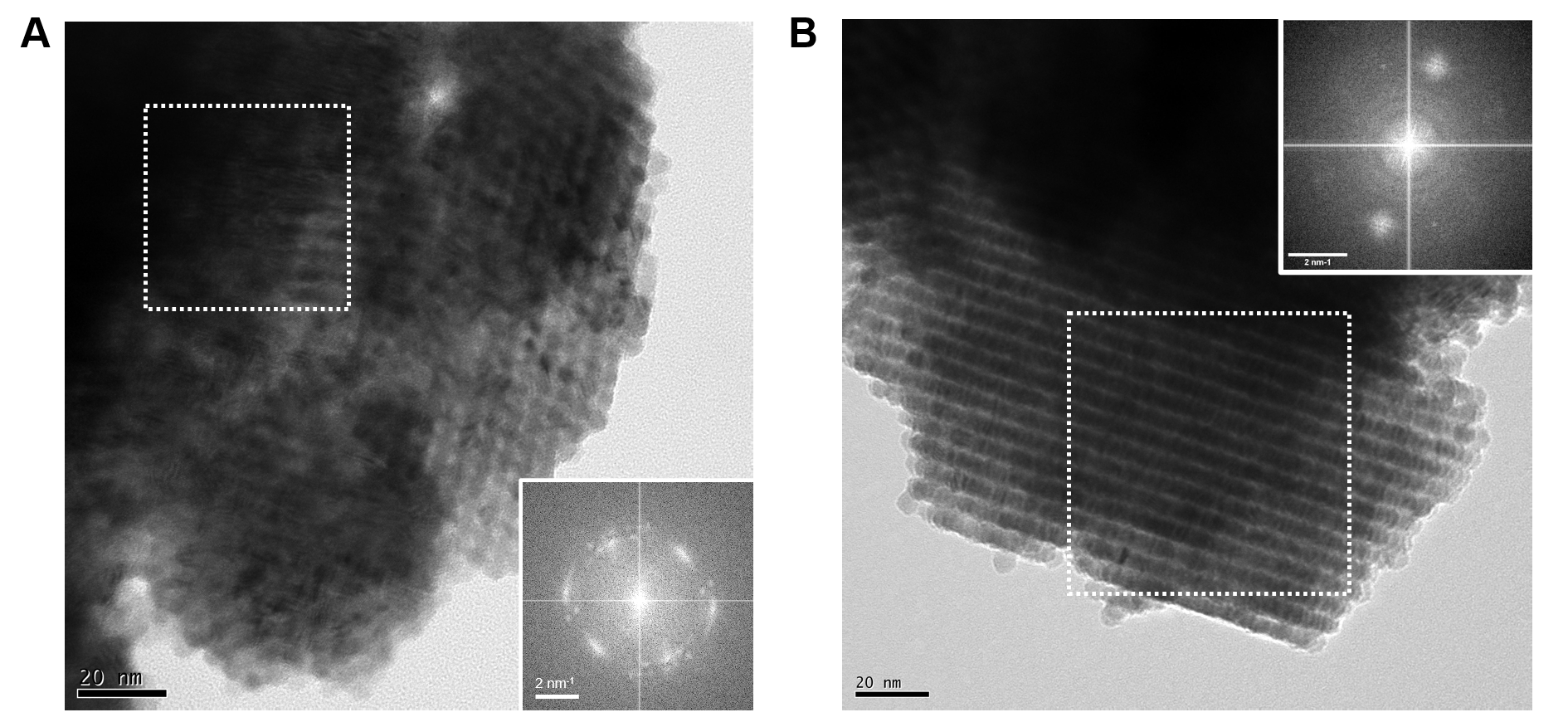}

\textbf{Figure S11.} HRTEM image of the superlattice of \textbf{(A)} 5.7
nm PbS-AsS\textsubscript{4}\textsuperscript{3-} and \textbf{(B)} 6.4 nm
PbSe-Sn\textsubscript{2}S\textsubscript{6}\textsuperscript{4-} NCs.
Inset: Fourier transformed image of the region enclosed by white dashed
box.

\includegraphics[width=6.5in,height=2.90556in,alt={A close-up of a graph and a diagram AI-generated content may be incorrect.}]{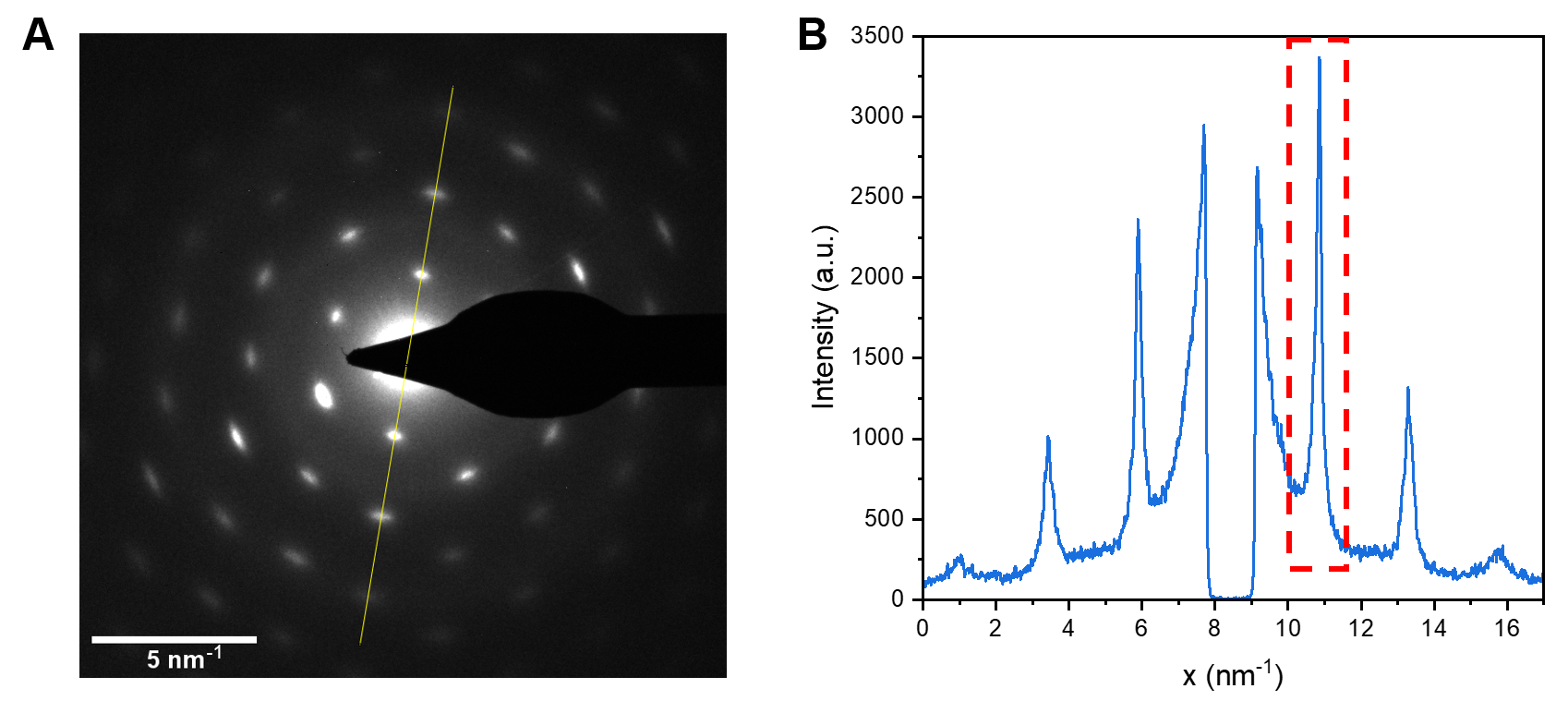}

\textbf{Figure S12.} (A) SAED image of the 5.7 nm
PbS-Sn\textsubscript{2}S\textsubscript{6}\textsuperscript{4-} NC
superlattice. The yellow line indicates the location where the linear
profile shown in (B) is obtained. The peak indicated by the dashed red
box in (B) is the peak where the full-width half-maximum (FWHM) was
obtained for the Scherrer analysis. The detailed description of the
Scherrer analysis is given in Section 4.2.

\includegraphics[width=3.47073in,height=2.94975in,alt={A graph of a graph Description automatically generated}]{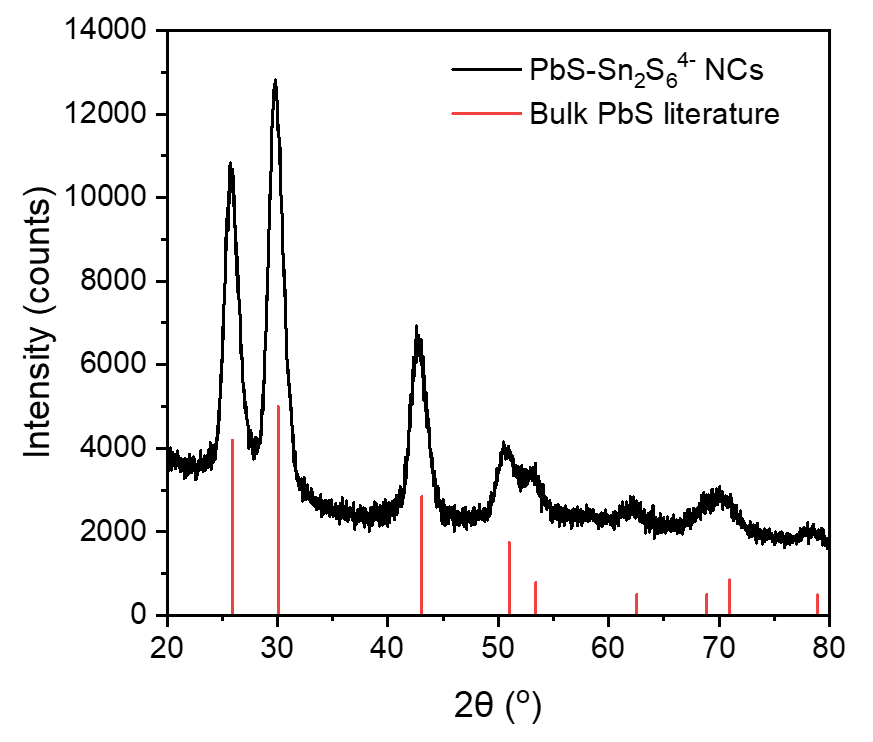}

\textbf{Figure S13.} PXRD pattern of PbS NCs. The lines are the PXRD
pattern of bulk PbS obtained from the literature (rock-salt structure,
unit cell length 5.9 Å). Details of the Wulff construction analysis are
presented in Section 4.2.

\includegraphics[width=6.25357in,height=2.14465in,alt={A diagram of a hexagon with yellow dots AI-generated content may be incorrect.}]{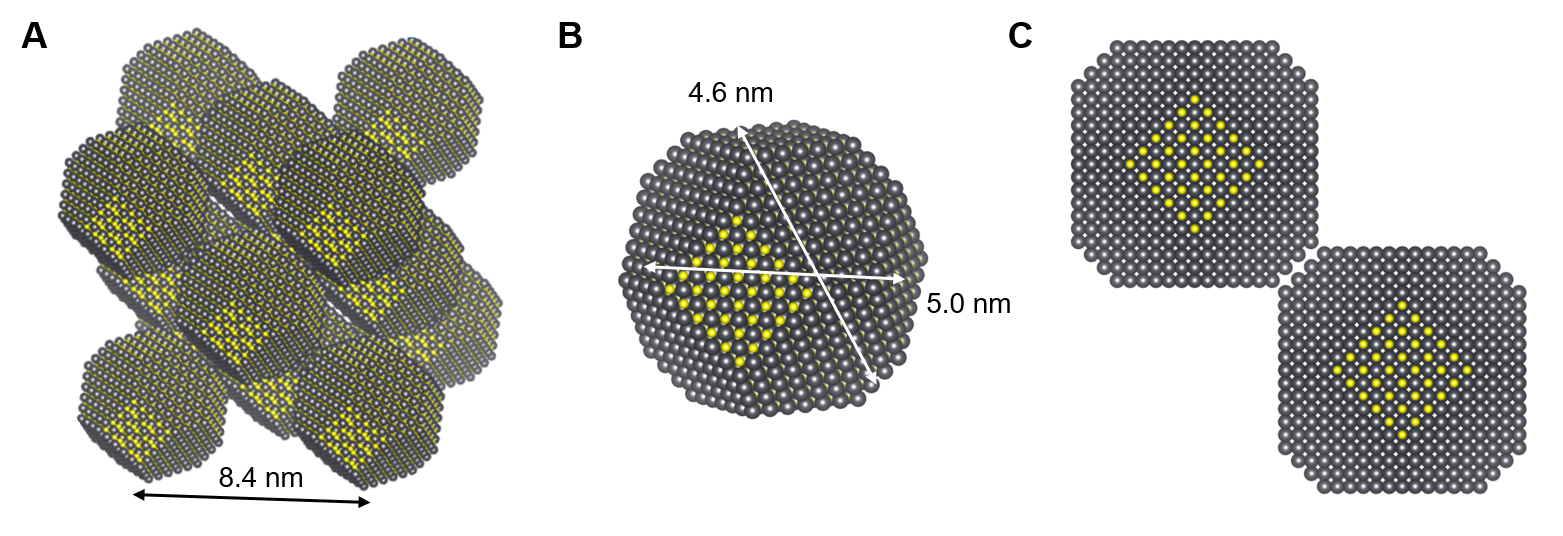}

\textbf{Figure S14.} Illustration of
PbS-Sn\textsubscript{2}S\textsubscript{6}\textsuperscript{4-} NC (A)
superlattice and (B) nanocrystal for simulating the scattering pattern.
The atomic edges of the NCs are precisely aligned, as illustrated in
(C). Details of the simulation are in Section 4.2 of the SI.

\includegraphics[width=5.54342in,height=4.22407in,alt={A diagram of different types of objects AI-generated content may be incorrect.}]{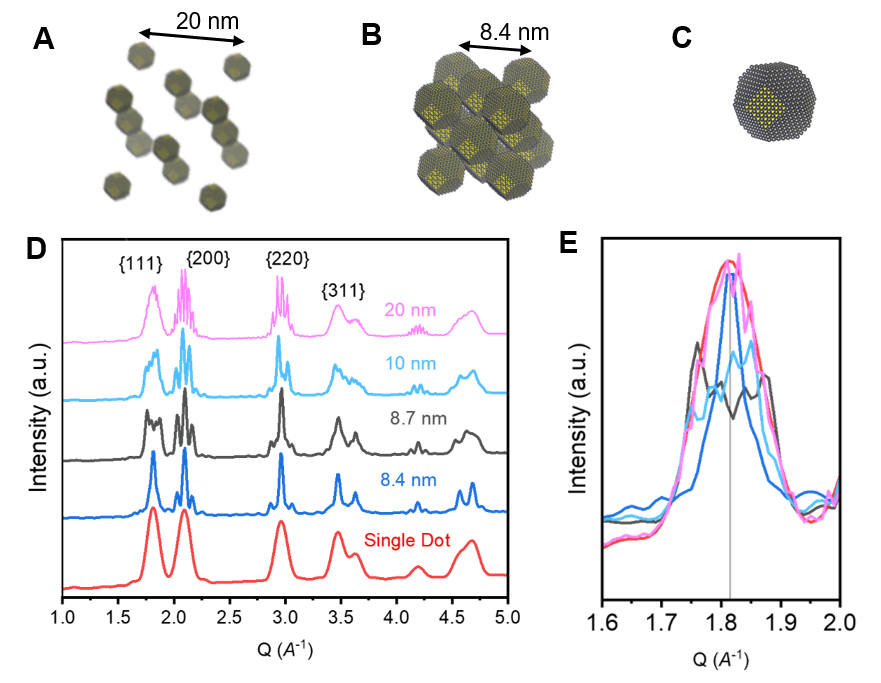}

\textbf{Figure S15.} Illustration of
PbS-Sn\textsubscript{2}S\textsubscript{6}\textsuperscript{4-} NC
superlattice with the lattice constant of (A) 20 nm and (B) 8.4 nm. (C)
presents the illustration of a single
PbS-Sn\textsubscript{2}S\textsubscript{6}\textsuperscript{4-} NC
particle. (D) Simulated scattering pattern of
PbS-Sn\textsubscript{2}S\textsubscript{6}\textsuperscript{4-} NC
superlattices with varying lattice constant and a single NC particle.
(E) Enlarged image of {[}111{]} peak. Details of the simulation are in
Section 4.2.

\includegraphics[width=6.40695in,height=3.66141in,alt={A diagram of a cell AI-generated content may be incorrect.}]{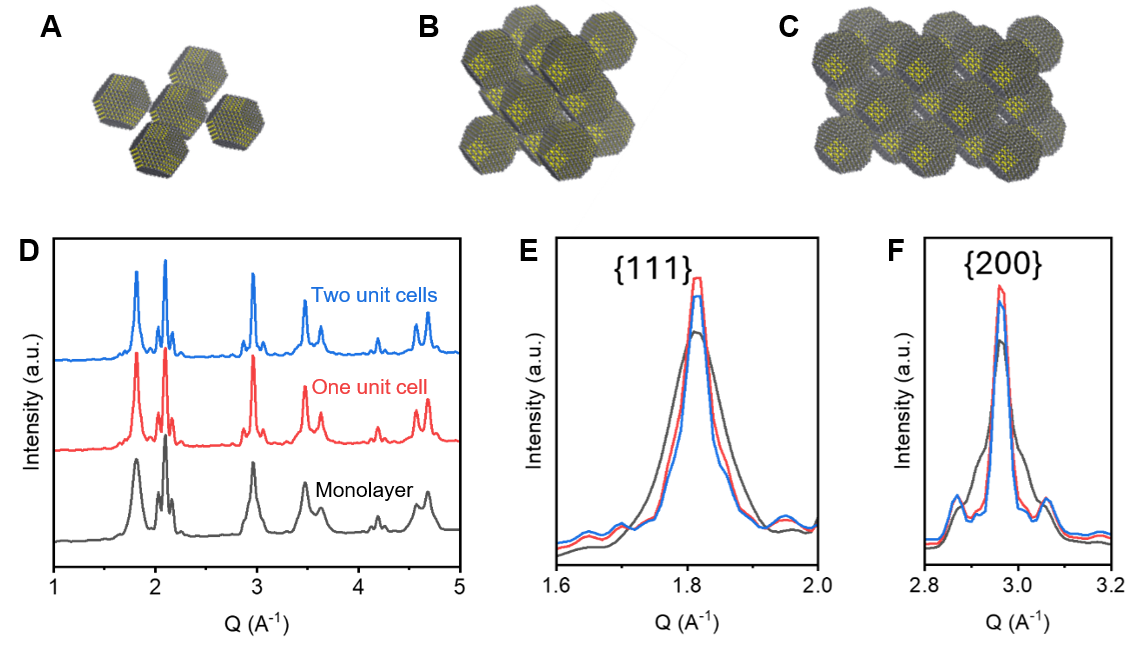}

\textbf{Figure S16.} Illustration of (A) one monolayer, (B) one unit
cell and (C) two unit cells of
PbS-Sn\textsubscript{2}S\textsubscript{6}\textsuperscript{4-} NC
superlattice. (D) Simulated scattering pattern of
PbS-Sn\textsubscript{2}S\textsubscript{6}\textsuperscript{4-} NC
superlattices with different numbers of repeating units. Also shown are
enlarged scattering patterns of (E) the {[}111{]} peak and (F) the
{[}200{]} peak. Details of the simulation are in Section 4.2.

\includegraphics[width=6.13219in,height=2.89903in,alt={A close-up of a microscope AI-generated content may be incorrect.}]{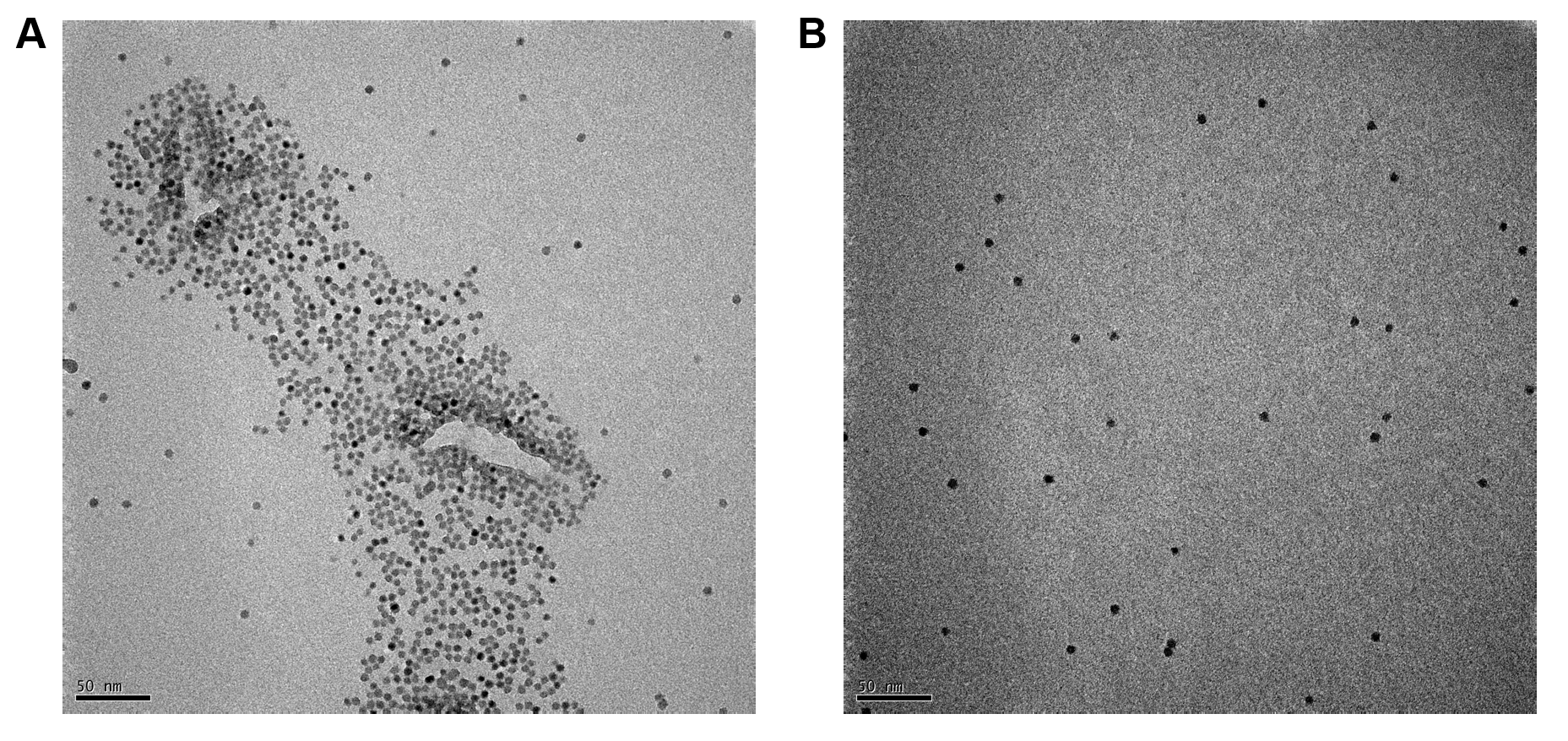}

\textbf{Figure S17.} TEM images of 5.7 nm PbS-Sn2S64- NCs. (A) and (B)
are taken from a same sample but at different spots.

\includegraphics[width=6.5in,height=2.18611in,alt={A graph of a function AI-generated content may be incorrect.}]{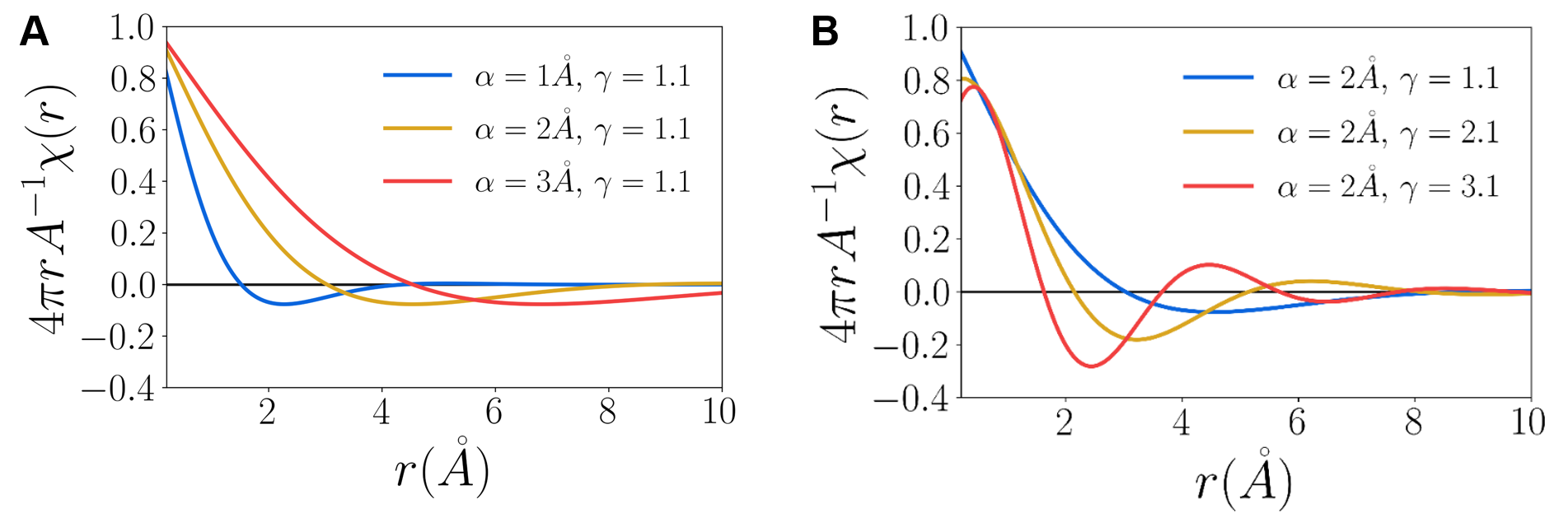}

\textbf{Figure S18.} Plots of theoretically calculated susceptibility of
5.7 nm nanoparticles with varied values of \textbf{(A)} $\alpha$ and
\textbf{(B)} $\gamma$.

\includegraphics[width=5.76596in,height=2.40187in]{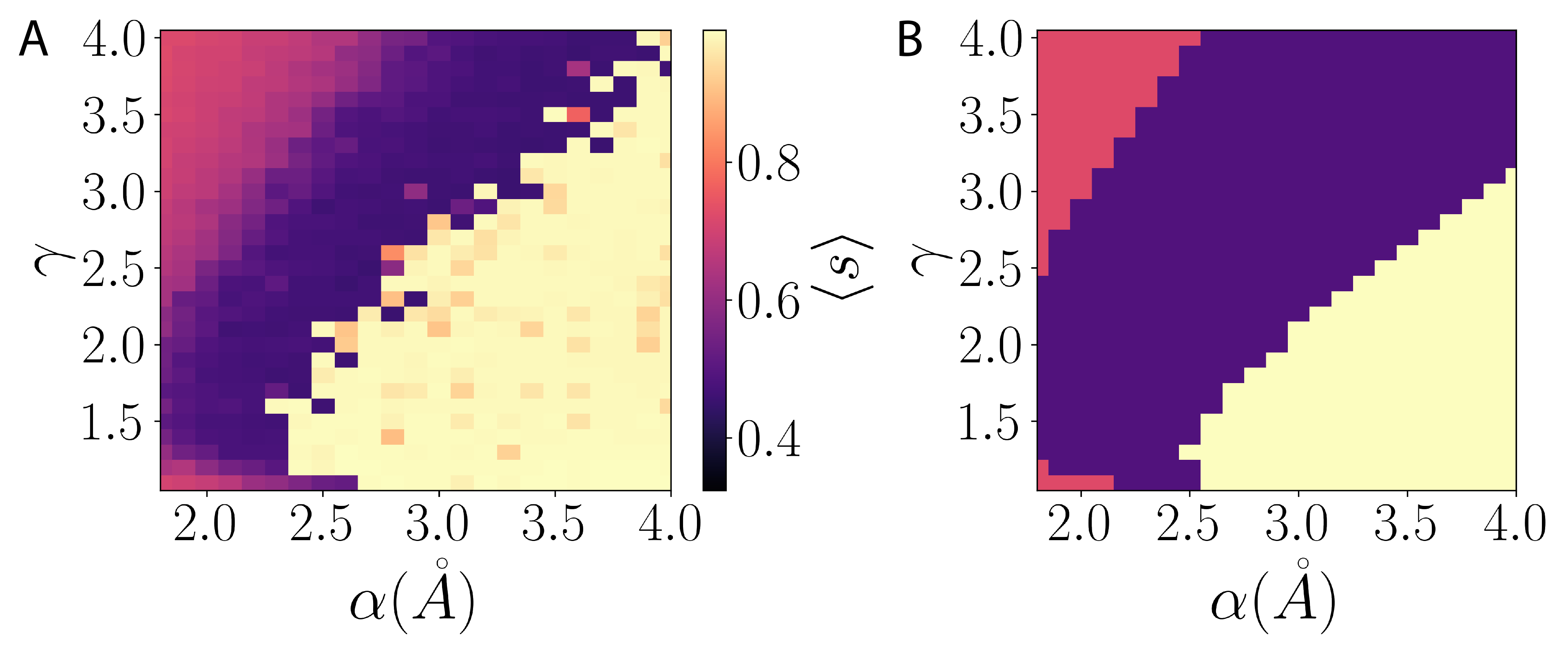}

\textbf{Figure S19.} (A) Average values of the order parameter $\langle s \rangle$
obtained from Monte Carlo production runs. (B) Phase diagram constructed
by comparing the free energy of representative orientations in each of
the three phases, exhibiting consistency with Figure 4B in the main
text. The pink, purple and yellow regions correspond to diagonal
orientation, face-to-face orientation and edge-to-edge orientation,
respectively. Illustrations of these orientational phases can be found
in Figure 4C.

\includegraphics[width=6.5in,height=4.15069in]{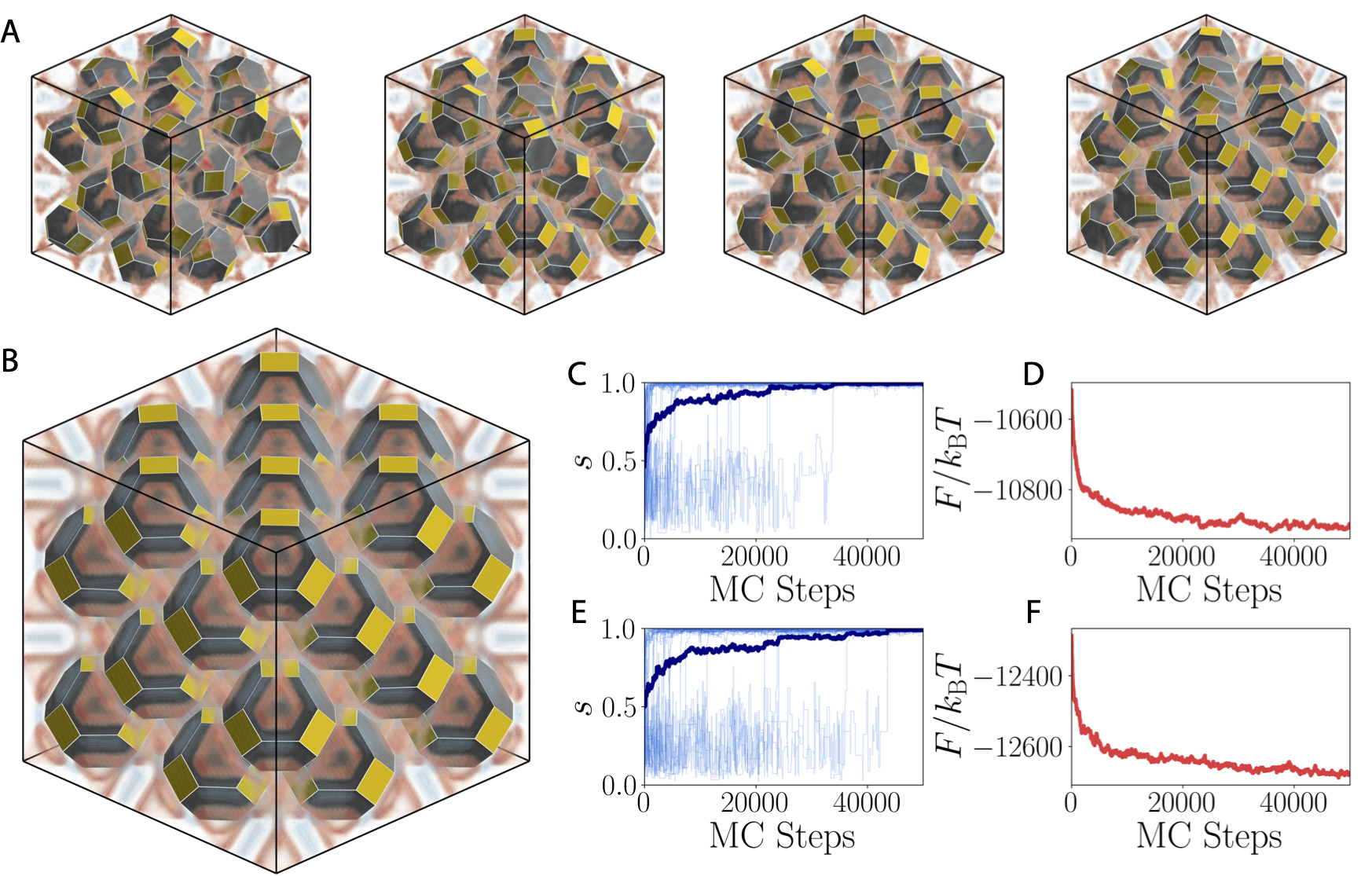}

\textbf{Figure S20.} Validation of orientational ordering in a~2×2×2~FCC
lattice (32 NCs) via Monte Carlo simulation (using~$\alpha=3.0\mathring{\mathrm{A}}$, $\gamma=1.1$). (A)
Snapshots from the Monte-Carlo simulation for the 2x2x2 FCC lattice,
arranged chronologically from left to right. (B) Final snapshot of the
Monte-Carlo run, with NCs showing edge-to-edge alignment, consistent
with the charge layering depicted in Fig. 3E in the main text. (C) Value
of the order parameter s defined in Eq. 16 for the 32 NC
(semi-transparent) and the average (opaque) as a function of Monte-Carlo
steps. (D) Corresponding evolution of the system\textquotesingle s free
energy. (F,G) Equivalent plots to (D,E) from a simulation where the
Lagrange multiplier $\lambda$~(enforcing volume constraints) was doubled,
confirming the robustness of the relaxation to the ordered state.

\includegraphics[width=6.30902in,height=2.99881in,alt={A graph of a graph of a graph of a graph of a graph of a graph of a graph of a graph of a graph of a graph of a graph of a graph of a graph of AI-generated content may be incorrect.}]{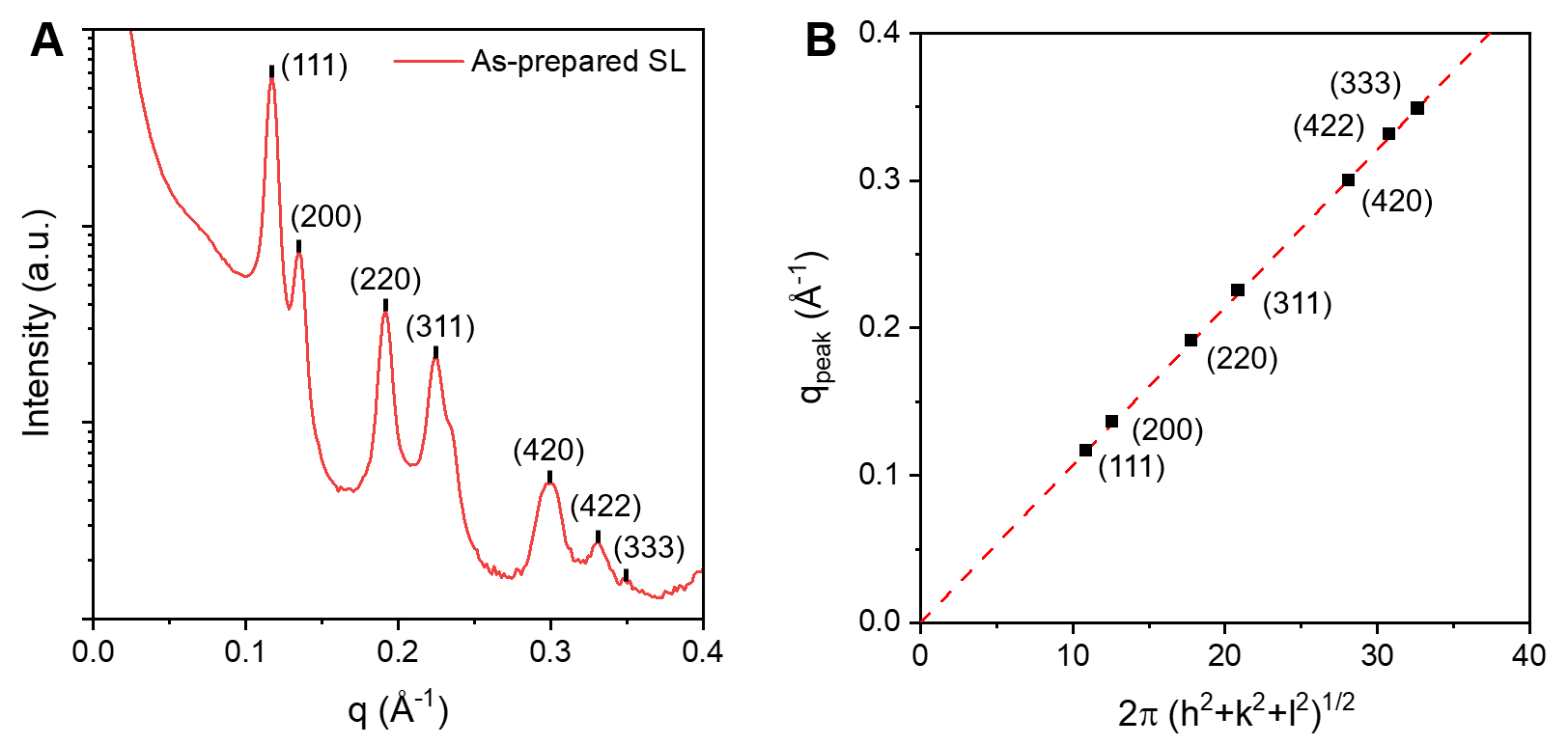}

\textbf{Figure S21. (A)} SAXS pattern of as-prepared superlattice of 5.7
nm PbS-Sn\textsubscript{2}S\textsubscript{6}\textsuperscript{4-} NCs
with peak assignment. The superlattice was deposited on Kapton tape and
dried before the SAXS measurement. \textbf{(B)} Peak analysis of the
superlattice.

\includegraphics[width=3.17523in,height=1.81132in,alt={Image}]{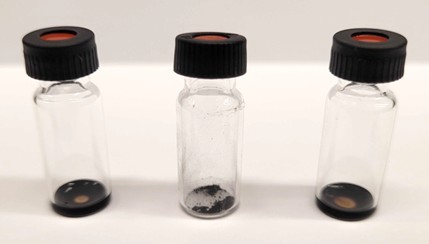}

\textbf{Figure S22.} Photograph of colloidal 5.7 nm
PbS-Sn\textsubscript{2}S\textsubscript{6}\textsuperscript{4-} NCs
(left), superlattices in powdered form (center), and colloidal NCs that
are recovered by redispersing superlattices in NMF (right).

\includegraphics[width=6.52867in,height=3.04881in,alt={A comparison of a graph AI-generated content may be incorrect.}]{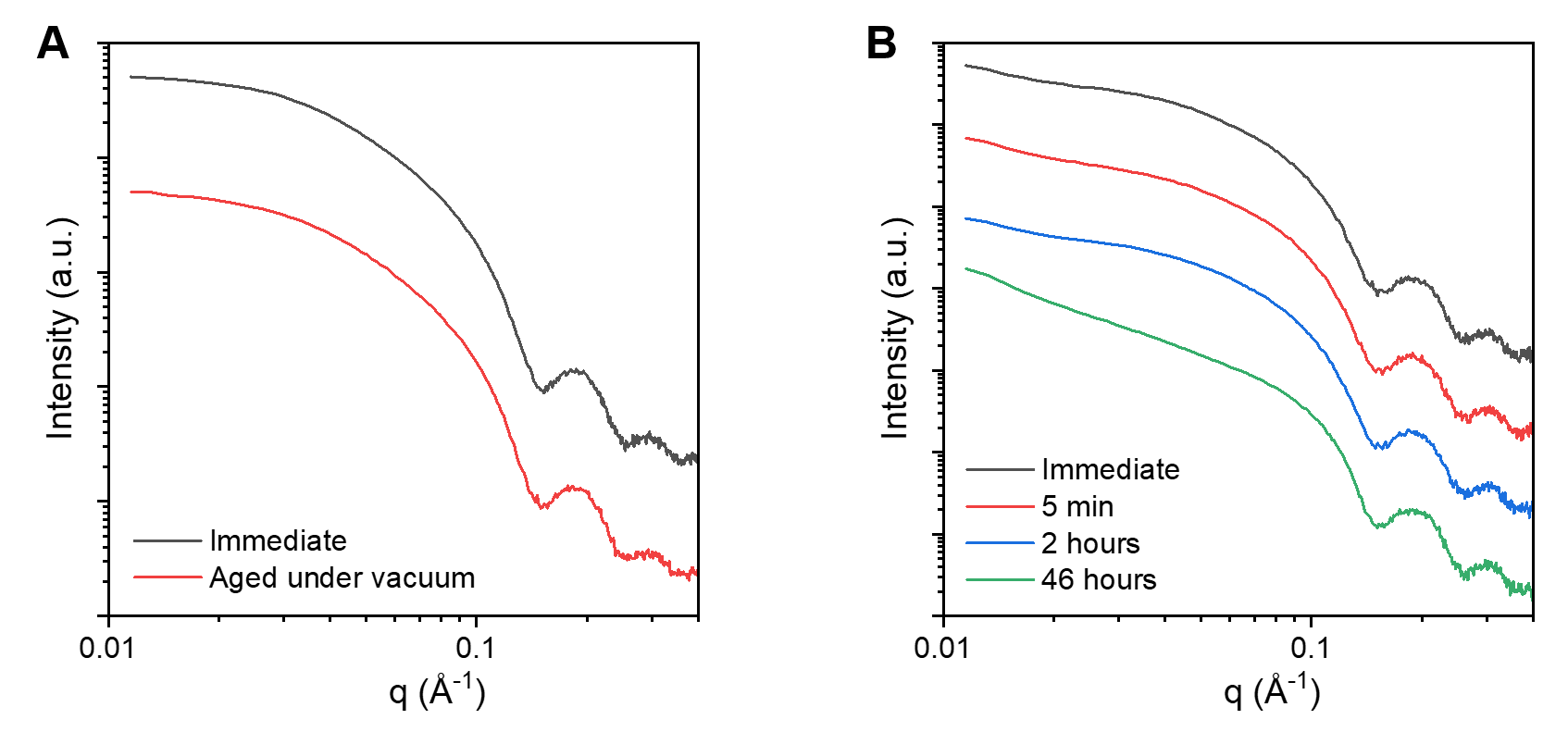}

\textbf{Figure S23.} SAXS patterns of as-prepared 5.7 nm
PbS-Sn\textsubscript{2}S\textsubscript{6}\textsuperscript{4-} NC
superlattices redispersed in NMF. \textbf{(A)} Superlattices were
redispersed in NMF immediately after solvent removal or after drying
under vacuum for 2 hours. \textbf{(B)} Superlattices were stored under a
N\textsubscript{2} atmosphere for the indicated durations following
solvent removal and subsequently redispersed in NMF.

\includegraphics[width=6.5in,height=2.08403in,alt={A black cloud in a white background AI-generated content may be incorrect.}]{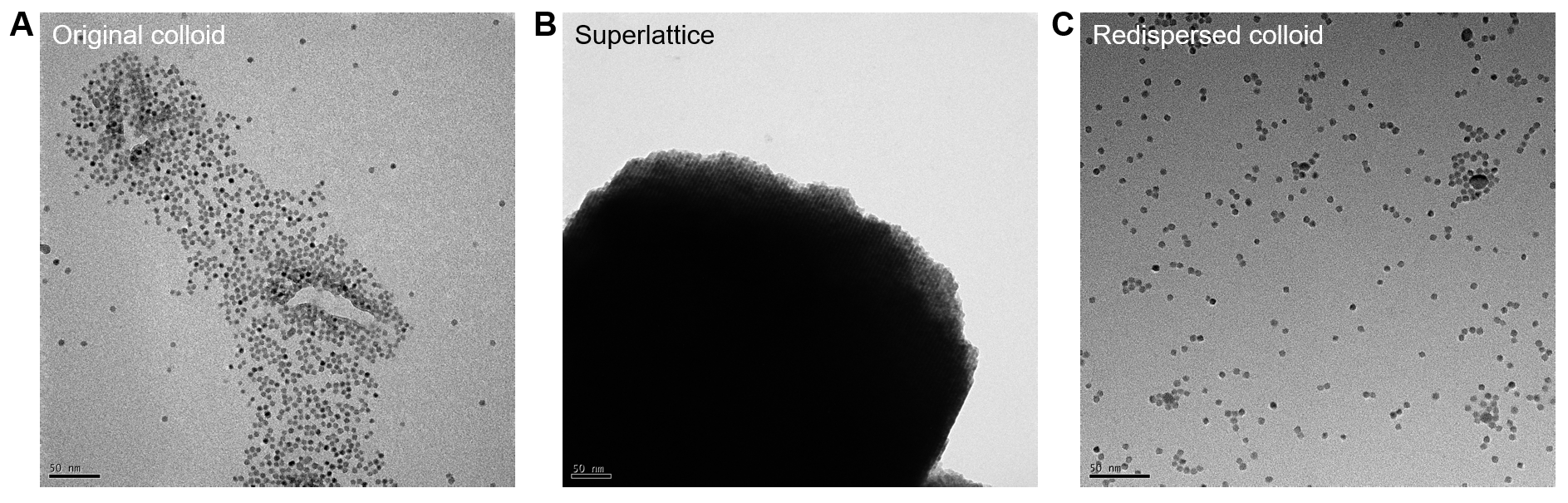}

\textbf{Figure S24.} TEM images of \textbf{(A)} original colloid,
\textbf{(B)} superlattice and \textbf{(C)} redispersed colloid of 5.7 nm
PbS-Sn\textsubscript{2}S\textsubscript{6}\textsuperscript{4-} NCs.

\includegraphics[width=6.5in,height=2.96042in,alt={A comparison of a graph AI-generated content may be incorrect.}]{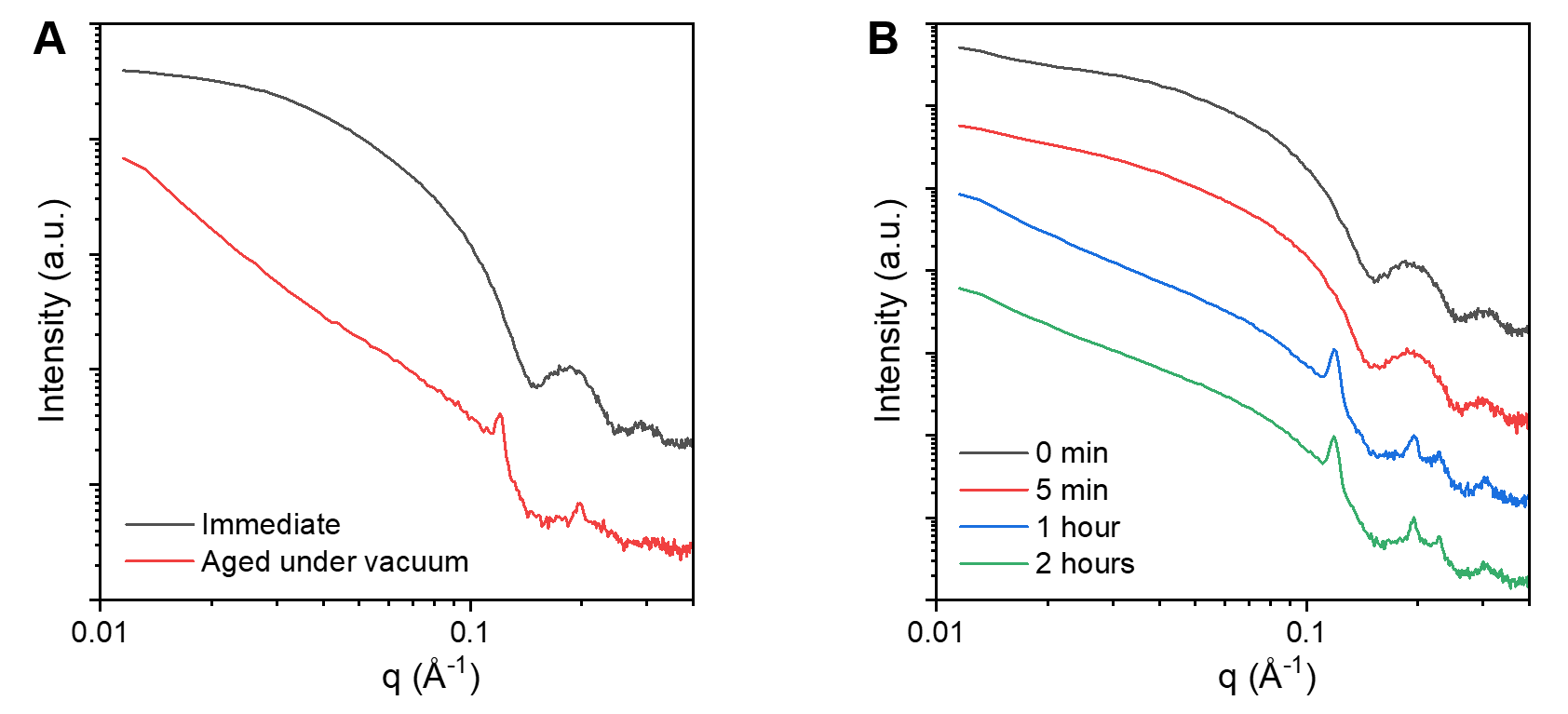}

\textbf{Figure S25.} SAXS patterns of washed 5.7 nm
PbS-Sn\textsubscript{2}S\textsubscript{6}\textsuperscript{4-} NC
superlattices redispersed in NMF. \textbf{(A)} Superlattices were
redispersed in NMF immediately after solvent removal or after drying
under vacuum for 2 hours. \textbf{(B)} Superlattices were stored under a
N\textsubscript{2} atmosphere for the indicated durations following
solvent removal and subsequently redispersed in NMF.

\includegraphics[width=4.18225in,height=2.44992in,alt={A collage of different colors AI-generated content may be incorrect.}]{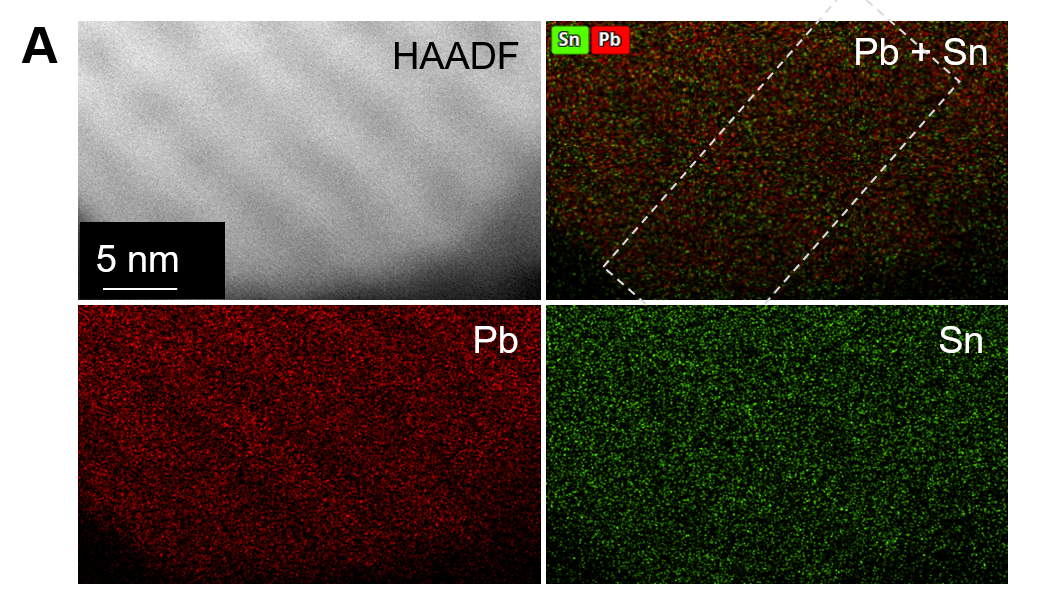}

\includegraphics[width=3.49676in,height=1.98187in,alt={A graph of a graph of a person\textquotesingle s position AI-generated content may be incorrect.}]{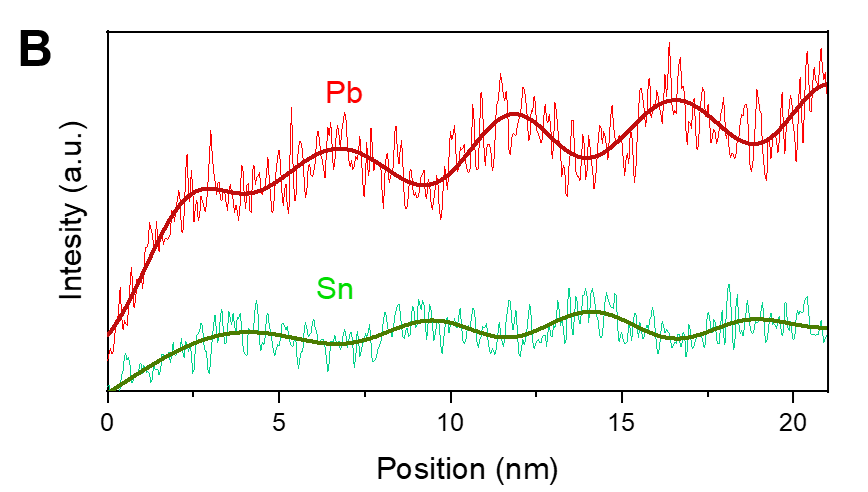}

\textbf{Figure S26.} \textbf{(A)} HAADF-STEM image and STEM-EDS mapping
of 5.9 nm PbS-Sn\textsubscript{2}S\textsubscript{6}\textsuperscript{4-}
NC superlattice. \textbf{(B)} Line scan of the STEM-EDS mapping. The
area where the line scan was carried out is marked with a dashed grey
box. \textbf{(C)} Intensity counts of S, As and Sn relative to the
intensity count of Pb on a line scan of STEM-EDS of 5.9 nm
PbS-Sn\textsubscript{2}S\textsubscript{6}\textsuperscript{4-} NC
superlattice.

\includegraphics[width=6.5in,height=4.20556in,alt={A green tower with a white background AI-generated content may be incorrect.}]{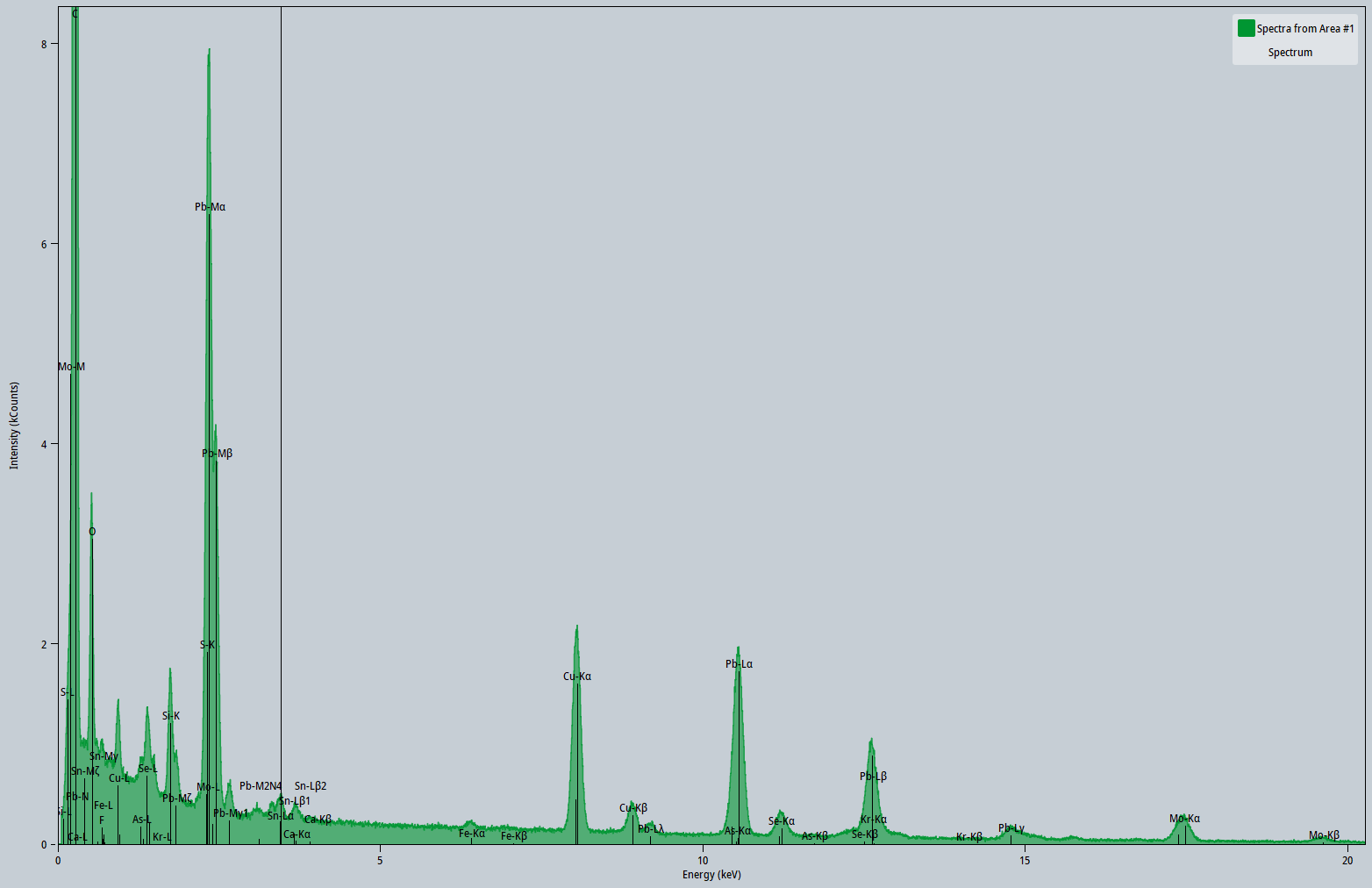}

\textbf{Figure S27.} Accumulative STEM-EDS spectrum of 5.9 nm
PbS-Sn\textsubscript{2}S\textsubscript{6}\textsuperscript{4-} NC
superlattice. The HAADF-STEM image and STEM-EDS mapping of the region
where this EDS spectrum is taken can be found in Figure 5E in the main
text and Figure S26. The origin of other elements is as follows: Cu from
TEM grid, Fe from TEM, Mo from TEM holder and Se from EDS detector. Ca
and Si are commonly found impurities from air dust.

\includegraphics[width=6.5in,height=2.18542in,alt={A black and pink background AI-generated content may be incorrect.}]{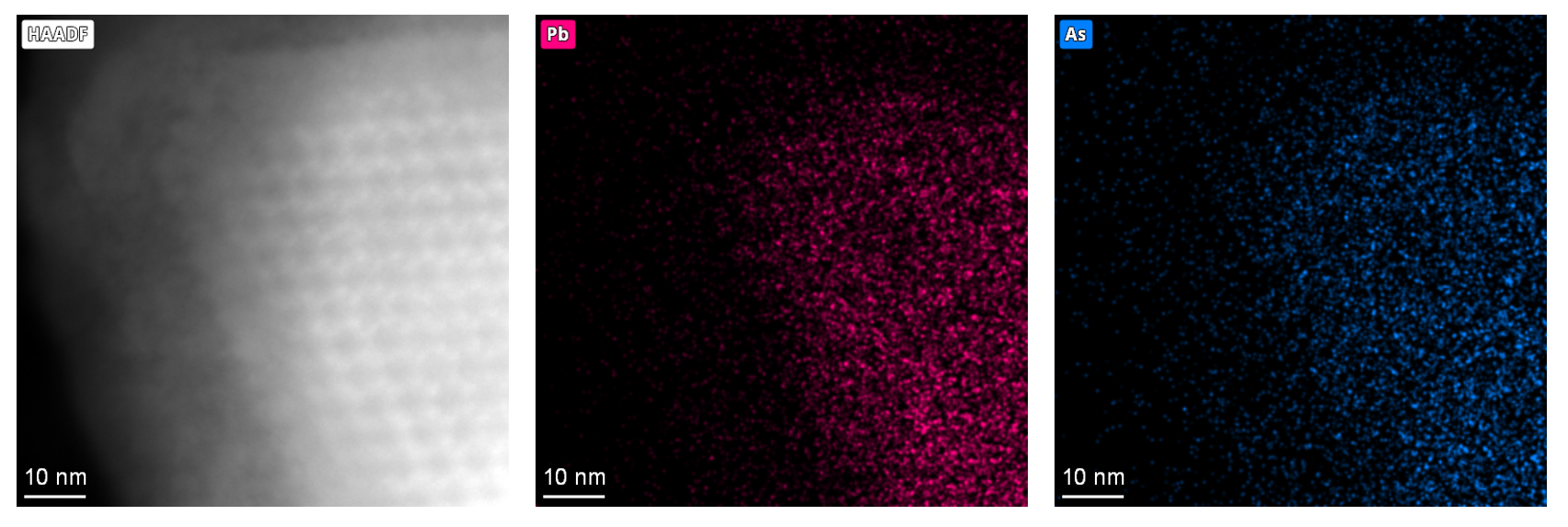}

\textbf{Figure S28.} HAADF-STEM image and STEM-EDS images of unwashed
5.9 nm PbS-Sn\textsubscript{2}S\textsubscript{6}\textsuperscript{4-} NC
superlattices.

\section{Supporting Tables}

\textbf{Table S1.} Relative mole ratio of five elements (Pb, S, K, Sn,
As) in 5.7 nm
PbS-Sn\textsubscript{2}S\textsubscript{6}\textsuperscript{4-} NC
superlattices before and after washing with MeCN, obtained through the
ICP-OES measurements. \begin{table}[h]
    \centering
    \label{tab:s1}
    \begin{tabular}{|l|l|l|l|}
        \hline
        \textbf{Element} & \textbf{NC solution} & \textbf{Before wash} & \textbf{After wash} \\ \hline
        Pb               & 1.000                & 1.000                & 1.000               \\ \hline
        S                & 1.131                & 1.190                & 1.122               \\ \hline
        K                & 0.137                & 0.277                & 0.222               \\ \hline
        Sn               & 0.094                & 0.074                & 0.072               \\ \hline
        As               & None                 & 0.086                & 0.065               \\ \hline
    \end{tabular}
\end{table}

\bibliographystyle{achemso} 
\bibliography{OrientationalAlignmentSI}